\documentclass[a4paper,fleqn,usenatbib]{mnras}

\usepackage[T1]{fontenc}
\usepackage{ae,aecompl}
\usepackage{newtxtext,newtxmath}

\usepackage{graphicx}	
\usepackage{amsmath}	
\usepackage{amssymb}	
\usepackage{natbib}
\usepackage[flushleft]{threeparttable}
\usepackage{siunitx}

\title[Stellar mass and luminosity functions at $z=6-9$]{Evolution of the galaxy stellar mass functions and UV luminosity functions at $z=6-9$ in the Hubble Frontier Fields}

\author[Bhatawdekar et al.]{Rachana Bhatawdekar,$^{1}$\thanks{E-mail: ppxrb2@nottingham.ac.uk}
Christopher J. Conselice$^{1}$, Berta Margalef-Bentabol$^{2}$, \newauthor Kenneth Duncan$^{3}$
\\
$^{1}$School of Physics and Astronomy, University of Nottingham, Nottingham NG7 2RD, UK\\
$^{2}$Sorbonne Universit\'e, Observatoire de Paris, Universit\'e PSL, CNRS, LERMA, F-75014, Paris, France\\
$^{3}$Leiden Observatory, Leiden University, PO Box 9513, NL-2300 RA Leiden, The Netherlands\\
\\
}

\date{Accepted XXX. Received YYY; in original form ZZZ}

\pubyear{2018}

\begin{document}
\label{firstpage}
\pagerange{\pageref{firstpage}--\pageref{lastpage}}
\maketitle

\begin{abstract}
We present new measurements of the evolution of the galaxy stellar mass functions (GSMF) and UV luminosity functions (UV LF) for galaxies from $z=6-9$ within the Frontier Field cluster MACSJ0416.1-2403 and its parallel field. To obtain these results, we derive the stellar masses of our sample by fitting synthetic stellar population models to their observed spectral energy distribution with the inclusion of nebular emission lines. This is the deepest and farthest in distance mass function measured to date and probes down to a level of M$_{*} = 10^{6.8}M_{\odot}$. The main result of this study is that the low-mass end of our GSMF to these limits and redshifts appears to become steeper from $-1.98_{-0.07}^{+0.07}$ at $z=6$ to $-2.38_{-0.88}^{+0.72}$ at $z=9$, steeper than previously observed mass functions at slightly lower redshifts, and we find no evidence of turnover in the mass range probed. We furthermore demonstrate that the UV LF for these system also appears to show a steepening at the highest redshifts, without any evidence of turnover in the luminosity range probed. Our $M_{\mathrm{UV}}-M_{*}$ relation exhibit shallower slopes than previously observed and are in accordance with a constant mass-to-light ratio. Integrating our GSMF, we find that the stellar mass density increases by a factor of $\sim15_{-6}^{+21}$ from $z=9$ to $z=6$. We estimate the dust-corrected star formation rates (SFRs) to calculate the specific star formation rates ($\mathrm{sSFR}=\mathrm{SFR/M_{*}}$) of our sample, and find that for a fixed stellar mass of $5\times10^{9}M_{\odot}$, sSFR $\propto(1+z)^{2.01\pm0.16}$. Finally, from our new measurements, we estimate the UV luminosity density ($\rho_{\textrm{UV}}$) and find that our results support a smooth decline of $\rho_{\textrm{UV}}$ towards high redshifts.
\end{abstract}

\begin{keywords}
galaxies: evolution -- galaxies: luminosity function, mass function -- galaxies: formation --galaxies: high-redshift
\end{keywords}



\section{Introduction}
Exploring the very first galaxies is one of the major contemporary problems in astronomy. We do not know when the first galaxies formed, nor how their formation occurred -- two related but distinct questions. Until the James Webb Space Telescope (JWST) launches, these problems can be best addressed through deep imaging observations, particularly with the Hubble Space Telescope (HST). Hubble allows us to view the Universe back to within 500 million years of the Big Bang, and perhaps earlier. This allows us to address the question of how much of the current galaxy population was in place at these early times, and perhaps to also investigate the deeper question of how that formation occurred. The \textit{HST} has spent several 1000s of orbits over two decades on the various Hubble (Ultra) Deep Fields, to understand the formation and evolution of galaxies, which existed when the universe was 500 million years old until today. This has led to interesting results on early galaxy formation and cosmic dawn at redshifts $z\sim6-10$ (e.g., \citealt{Bouwens2011,Bouwens2015,Finkelstein2012,McLure2013,Schenker2013,Schmidt2014,Duncan2014,Ishigaki2017}). 

To extend its reach even farther beyond its native technical capabilities, \textit{HST} observed six massive clusters of galaxies as gravitational lenses as a part of Hubble Frontier Fields (HFF) program \citep{Lotz2017}. Using Director's Discretionary (DD) observing time, the HFF program (FF program 13495; P.I. Lotz, Co-PI: Mountain) imaged six massive clusters, Abell 2744 (z$\sim$0.308), MACSJ0416.1-2403 (z$\sim$0.396), MACSJ0717.5+3745 (z$\sim$0.545), MACSJ1149.5+2223 (z$\sim$0.543), Abell S1063 (z$\sim$0.543), Abell 370 (z$\sim$0.543), and their parallel fields, for 140 orbits each with the Advanced Camera for Surveys (ACS), and Wide Field Camera 3 (WFC3) onboard \textit{HST} in three optical (F435W, F606W, and F814W) and four near-infrared (F105W, F125W, F140W, and F160W) bands over two epochs. These clusters are being used as gravitational lenses to magnify faint distant galaxies, while the flanking fields are for deep observations of otherwise blank areas. 

In fact, we now have a good understanding of the amount of light emitted by these early galaxies in the rest-frame UV.  We are able to measure this for galaxies at a range of luminosities and masses back to this epoch of $z \sim 7$ (e.g., \citealt{Bouwens2007,Oesch2010,Grazian2011,Schenker2013,McLure2013}).  What this reveals is that there is a gradual steepening of the UV luminosity function as one goes back in time, such that there are more fainter galaxies per brighter galaxies as one probes from $z\sim 3$ to $z \sim 7$.  This is such that there are in fact more galaxies at these earlier times than there are today, as there are so many fainter and lower mass systems (e.g., \citealt{Conselice2016}).  However, the issue with UV photometry and the UV luminosity function is that these are a combination of a few physical processes. Namely, a higher star formation rate will produce a higher UV luminosity, but this does not necessarily correlate with the underlying stellar mass of the galaxy. It is easy to imagine scenarios whereby the underlying mass is high, but because of dust or a low star formation rate, the UV luminosity is not as high. Therefore, it is essential to measure the stellar mass functions as well as luminosity functions as they can be substantially different.  

Whilst there is a significant amount of work done on the Ultraviolet Luminosity Function (UV LF) in these and other fields (e.g., \citealt{Bouwens2015, Finkelstein2015,Mcleod2016,Laporte2016,Livermore2017,Atek2018}), this reveals the ongoing star formation, while the stellar masses reveal the past formation, and is thus complementary.    The stellar mass, unlike the UV luminosity, is a measure of the integrated formation and merging history of a galaxy. It includes all processes such as star formation and mergers, which contribute to building up the mass of a galaxy. In this sense, it is an excellent indicator for how galaxy formation has progressed over the epochs before it is measured, as it is the integral of all galaxy formation processes previous to the time in which we observe it. This is in contrast to the UV luminosity which is a good, albeit affected by dust, measure of the instantaneous star formation rate of a galaxy. This is in a sense the differential of the formation process, albeit with uncertainties due to dust. UV luminosities, therefore, as such do not tell us about the past history of a galaxy or how/when it formed.

There are many other reasons to search for these early galaxies and to study their stellar masses. The first is that by pushing back in time, we can find when the very first galaxies formed and therefore understand the physical causes of that early formation.  Whilst this is still not possible due to technical limitations, we are able to see galaxies when they are forming at high star formation rates back to this early epoch of around $z \sim 7-8$ and measure their stellar masses (e.g., \citealt{Labbe2010,McLure2011,Oesch2014,Duncan2014,Song2016,Laporte2016}).

Just as for the UV luminosity function, the stellar mass function also evolves with time, and becomes steeper with higher redshift, such that there are more lower mass galaxies per massive/bright galaxy at higher redshift than at lower redshift (e.g., \citealt{Gonzalez2011,Duncan2014,Grazian2015,Song2016}). The stellar masses of these galaxies allows us to probe the past star formation history to reveal, and place constraints, on when the first epochs of star formation occurred and to probe whether galaxies could have caused reionization (e.g., \citealt{Duncan2015}).   

As there are so many more faint and low mass galaxies compare to massive ones, and the fact that the stellar mass function continues to be occupied by progressively lower mass galaxies, it is worth asking if, and how, this continues. At some point we may expect the mass function to turn over, such that at some mass limit we see a natural decline in the number of galaxies lower than some stellar mass limit. Stellar masses also provide information about the mass-to-light ratio of the lowest mass galaxies, which can only be probed through deep observations.  Reaching these galaxies will be routine with the JWST (e.g., \citealt{Kalirai2018,Gardner2006}), but until then our best chance to study these systems is through deep observations of lensing clusters (e.g.,  \citealt{Kneib2011,Bouwens2009,Coe2013,Zheng2012,Bradley2014,Zitrin2014}). In this paper, we therefore study the MACSJ0416.1-2403 cluster and its parallel field, and combine these data to derive the first galaxy stellar mass function (GSMF) at $z=6-9$ for the Frontier Fields program, by combining the \textit{HST} imaging with \textit{Spitzer} and ground-based VLT data.

Currently there are inconsistencies, and large uncertainties, in the best available measurements of galaxy number densities and luminosity functions at $z>6$. One of the most intriguing recent results from the deepest existing \textit{HST} data is that there is an apparent decline of the number of galaxies at the highest redshifts, $z\sim6-11$ (e.g., \citealt{Bouwens2016}). Based on extrapolating the UV LF at lower redshifts, we should have found more systems than the candidates discovered in the deepest \textit{HST} data (e.g., \citealt{Oesch2013}). On the other hand, deep \textit{HST} imaging of lensing clusters in the CLASH and HFF clusters have found a significant number of $z>6$ candidates (e.g., \citealt{Atek2014, Livermore2017}). This includes several lensed candidates at $z\sim9$ behind the HFFs (e.g., \citealt{Zitrin2014}). Another intriguing result from HFF studies is that the UV LF continues to remain unbroken to magnitudes as faint as $M\mathrm{_{UV}=-12.5}$ at $z\sim6$ (e.g., \citealt{Livermore2017}) and $M\mathrm{_{UV}=-15}$ at $z\sim9$ (e.g., \citealt{Ishigaki2017}), whereas more recently \citet{Bouwens2017b} and \citet{Atek2018} have reported a possible turnover in the faint-end of the UV LF beyond an absolute magnitude of $M\mathrm{_{UV}=-15}$ at $z\sim6$. In this paper, we therefore revisit this issue by deriving the UV luminosity function from $z=6-9$ with our new measurements.

The structure of this paper is as follows: In Section~\ref{sec:dataset} we describe the properties of the data used in this study. Section~\ref{sec:methods} describes in detail the method we developed to remove the massive cluster galaxies on the critical line of MACSJ0416.1-2403 cluster and the construction of multiwavelength catalog, from 0.4 to 4.5$\mu$m, using \textit{HST}, \textit{Spitzer} and ground-based VLT data. In this section, we also describe the photometric redshift determination and the selection criteria used to construct our sample of high-redshift galaxies at $z=6-9$, along with the completeness simulations undertaken to take into account the effects of completeness and selection functions. We conclude this section describing the SED fitting method used to estimate the stellar masses for our sample. In Section~\ref{sec:results}, we present our results by including a derivation of the GSMF, UV LF and an analysis of mass-to-light ratio of our sample of galaxies. We also present and discuss the estimated total stellar mass density (SMD), the specific star formation rates (sSFR) and the UV luminosity density in this section. Finally, we summarize our results and present the conclusions of this work in Section~\ref{sec:summary}.
Throughout this paper, we adopt a $\Lambda$CDM cosmology with $H_{0}$ = 70 km s$^{-1}$ Mpc$^{-1}$, $\Omega_{M} = 0.3$, and $\Omega_{\Lambda} = 0.7$.  All magnitudes are quoted in the AB system \citep{Oke1983}, and a \citet{chabrier2003} stellar initial mass function is used.

 \section{The Data}
 \label{sec:dataset}
 
\subsection{HST data}
Observations of MACSJ0416.1-2403, hereafter MACSJ0416, (RA: 04:16:08.9, Dec: -24:04:28.7) and its parallel field (RA: 04:16:33.1, Dec: -24:06:48.7) were carried out between Jan 2014--Feb 2014 (Epoch 1) and July 2014--September 2014 (Epoch 2) as a part of the HFF program. In this study, we use the final reduced and calibrated v1.0 mosaics and their associated weight and rms maps provided by the Space Telescope Science Institute (STScI) on the HFF website \footnote[1]{http://www.stsci.edu/hst/campaigns/frontier-fields/FF-Data}, drizzled at 60 mas pixel-scale. For a detailed description of data release, including the data reduction pipeline, exposure times, and calibration procedures, we refer the reader to the STScI data release documentation \footnote[2]{https://archive.stsci.edu/pub/hlsp/frontier/}.  

We compute the depth of these \textit{HST} images by placing 100s of $0\farcs2$ radius apertures in random positions on the images and measuring fluxes in them. Table~\ref{tab:dataset_table} lists the resulting 5$\sigma$ limiting magnitudes for the seven bands using our detection methods. We find that the 5$\sigma$  limiting magnitudes of the cluster are lower than the parallel field (See Table~\ref{tab:dataset_table}). This is the result of the cluster field being dwarfed by the light from the massive foreground galaxies, making the effective raw depths shallower than the parallel field.

\subsection{VLT data}
\label{sec:vltdata}
HFF data by itself is inadequate for characterizing the galaxies we are interested in at $z\gtrsim6$. Typically, the way to study the stellar masses and stellar populations of $z\gtrsim6$ galaxies is through the use of \textit{Spitzer}/IRAC data at $> 3 \mu$m. This is often included and used along with \textit{HST} imaging by the use of the $K_{s}$ band, the reddest ground based filter we can obtain deep imaging data from. In fact, good quality $K_{s}$ band data has been crucial to fully exploit \textit{Spitzer}/IRAC data, which has poor resolution and is affected by blended sources. Finally, $K_{s}$ band at 2.2 $\mu$m also fills in the gap between the 1.6$\mu$m F160W band and the IRAC 3.6$\mu$m band. 

Because of the aforementioned reasons, and in order to put better constraints on redshift estimates, we include longer wavelength $K_{s}$ band data in our analysis. $K_{s}$ band observations of MACSJ0416 fields were obtained between October 2013 and February 2014 with the High Acuity Wide-field K-band Imager (HAWK-I) on the 8.2 m UT4 telescope at the ESO Very Large Telescope. We use the fully reduced $K_{s}$ band images that are made available to the public through the Phase 3 infrastructure of the ESO Science Archive Facility (ESO program 092.A-0472, P.I. Brammer). The full HAWK-I 7$^{'}$.5 $\times$ 7$^{'}$.5 field of view covers both the cluster and the parallel fields in a single pointing.  

Similar to the \textit{HST} bands, we calculate the depth of the image by placing 100s of $0\farcs4$ radius apertures in random positions on the image and measuring fluxes in them. Table~\ref{tab:dataset_table} describes the 5$\sigma$ limiting magnitudes for the $K_{s}$ band. We refer the reader to \citet{Brammer2016} for a detailed description of observations and data reduction process.

\subsection{Spitzer data}
\label{sec:spitzerdata}

The Balmer break/4000 angstrom break is crucial to estimate the age of stellar populations and subsequently in the measurement of galaxy stellar mass. This break is at observed wavelengths beyond 2.4 $\mu$m at $z>5$, and only the IRAC camera on board \textit{Spitzer} can identify the break. IRAC data is also essential to put better constraints on redshift estimates in addition to obtaining robust stellar mass estimates. We therefore include \textit{Spitzer} data in our analysis. In addition to \textit{HST}, the \textit{Spitzer} Space Telescope has dedicated $\sim$ 1000 hours of Director's Discretionary time to observe the Frontier Fields at 3.6 $\mu$m and 4.5 $\mu$m and has observed each field for $\sim$50 hours in each channel. We use the final reduced mosaics made available to the public on IRSA website \footnote[3]{http://irsa.ipac.caltech.edu/data/SPITZER/Frontier/} (Program ID 90258, P.I T. Soifer). The depth of \textit{Spitzer} data was again, similar to the \textit{HST} and $K_{s}$ band data, computed by using a similar method of placing 100s of $1\farcs4$ apertures in random positions in the images.  The 5$\sigma$ limiting magnitudes for both the channels are as listed in Table~\ref{tab:dataset_table}. 

 \begin{table}
	 \centering
	  \caption{Description of dataset for the MACS0416 cluster and its parallel field. The 5$\sigma$ depths are calculated using 100s of $0\farcs2$ radius apertures in random positions for \textit{HST} images, $0\farcs4$ radius apertures for HAWK-I image and $1\farcs4$ radius apertures for IRAC images.}
	 \label{tab:dataset_table}
	 \begin{tabular}{lccr} 
	         \hline
	 	 \hline
		& MACS0416 Cluster & MACS0416 Parallel\\
		 \hline
		 Filter & Depth (5$\sigma$) & Depth (5$\sigma$) & Instrument\\
		 \hline
		 F435W &28.87 &28.91 &ACS\\
		 F606W &28.95 &29.01 &ACS\\
		 F814W &29.35 &29.40 &ACS\\
		 F105W &29.22 &29.30 &WFC3\\
		 F125W &28.95 &28.02 &WFC3\\
		 F140W &28.85 &28.93 &WFC3\\
		 F160W &28.65 &28.75  &WFC3\\
		 Hawk-I $K_{s}$  &26.25 &26.35  &HAWK-I\\
		 IRAC 3.6 &25.10 &25.16 &IRAC\\
		 IRAC 4.5 &25.13 &25.20 &IRAC\\
		 \hline
	 \end{tabular}
 \end{table}

\section{Methods}
\label{sec:methods}
\subsection{Divide and Conquer}
 
One aim of the Hubble Frontier Fields (HFF) program is to find the faintest and the earliest galaxies in the Universe, $\sim$10-100 times fainter than any previously studied, and examine their properties, using clusters as gravitational lenses. While the clusters provide a magnified boost to the light from background galaxies, providing us with a deeper view of the early Universe than we would obtain from just examining blank fields, the overwhelming luminosity of the brightest galaxies in the cluster impedes the detection of the faint galaxies. For this reason, it was imperative that we first model and subtract the foreground galaxies from the cluster. In this study, we choose to model and subtract galaxies on the critical line since the greatest magnification of distant objects occurs along the line of sight of the densest areas of the cluster.

Two other studies \citet{Livermore2017} and \citet{Merlin2016} have attempted to subtract the foreground light using wavelet decomposition method and GALFIT, respectively, delivering promising results. More recently, \citet{Shipley2018} have sought to model the light from the brightest cluster members that contribute significant light to the cluster using IRAF. 

In this work, to subtract the massive foreground galaxies, we build upon the strategy used by \citet{Gu2013} to detect faint substructures in NGC 4889. As a first attempt, we try to model the brightest galaxies in the reddest band F160W, which is our detection image, using GALAPAGOS \citep{Barden2012} as a stand alone application. GALAPAGOS is a IDL based software that uses SExtractor \citep{Bertin1996} to detect the sources, and then fits a single S\'ersic profile \citep{Sersic1968} to the detected sources using GALFIT \citep{Peng2002}. Unfortunately, the resulting residuals from this procedure were unsatisfactory. This is primarily because GALAPAGOS uses only a single S\'ersic component to model the galaxies. The galaxies in the MACS0416 cluster are very bright/massive and also lie in a crowded field. They therefore cannot be modelled with simply one or two S\'ersic components. As the massive galaxies are embedded in a crowded field, we also need to fit and remove all the neighbouring galaxies at the same time. 

To overcome these difficulties, we developed an iterative procedure, dubbed "Divide and Conquer", in which we split the image into small regions, with the target bright galaxy in the center, and make use of GALAPAGOS and GALFIT on those regions to model the small neighbouring galaxies first before attempting to model the big galaxies. This procedure is essentially a series of iterations between fitting away the small objects first in a small region, one by one, using one or more S\'ersic models to simulate the galaxy of interest as well as the local sky background (whose parameters are thrown away afterwards as they are only needed to approximate the sky).  We keep increasing the complexity until we get a reasonable residual (e.g. there are no over-subtracted regions in the residual). This process is repeated on all the small galaxies until we are left with the central galaxy, with all the neighbours subtracted away and the process is then repeated on the central galaxy.  After we get a good fit, we do a massive simultaneous fit with all the neighbours together. We then create an image with these objects subtracted out from the original image and move on to fit the next bright galaxy. 

The basic steps of our method are as follows:
\begin{enumerate}
  \item To subtract the bright galaxies, we select a rectangular patch covering the central bright galaxy, and its nearby region with smaller galaxies, and use GALAPAGOS to create postage stamps to acquire the first guess of the model parameters of each source in the stamp.
  \item We then model the small neighbouring galaxies first using GALFIT, and once the small galaxies are fitted away cleanly, we model the central bright galaxy.
  \item After all the galaxies in the rectangular patch are modelled accurately, we subtract the final model from the original image and move to the next rectangular patch and repeat the process in iteration.
\end{enumerate} 
In the following sections we describe our procedure in detail.

\subsection{Subtraction of brightest cluster galaxies}

\subsubsection{Creating postage stamp and obtaining initial guess of model parameters}
Frontier Fields images are sky subtracted and in units of counts/s. The very first step that we perform, before we run GALAPAGOS, is to add sky background to the detection image (F160W band) and convert the image to counts. This step was necessary in order for GALFIT to produce a reliable sigma image. We then select a rectangular patch on the detection image that consists of the target central bright galaxy and the nearby region with smaller galaxies and run GALAPAGOS on it to get the postage stamp as well as the first guess of model parameters for all the sources in the stamp. For each object that was contributing light to the stamp, we obtain the initial guesses for all the model parameters such as the position within the stamp, magnitude, effective radii, S\'ersic index, axis ratio and position angle required to produce a S\'ersic model. 

\subsubsection{Modelling brightest cluster galaxies using GALFIT}
Once the postage stamp and the initial guess of model parameters are obtained, the next step is to model the galaxies. As described above, we first model the small neighbouring galaxies, along with the local background, as accurately as possible (and eventually all the bright galaxies on the critical line). 

Every galaxy is modelled using a S\'ersic profile described by the following expression for the surface brightness $\mathrm{\sum(r)}$ at radius \textsl{r},

\begin{equation}
\sum(r)=\sum_{e}\exp\Bigg\{-k\left[(\frac{r}{r_{e}})^{\frac{1}{n}}) - 1\right]\Bigg\},
\end{equation}

\noindent where \textit{$r_e$} is the half-light radius, $\sum_e$ is the effective surface brightness, \textit{k} is a normalisation coefficient and \textit{n} is the S\'ersic index. The de Vaucouleurs profile with index $n=4$ is a good fit for massive elliptical galaxies, and exponential profiles with $n=1$ to Gaussian $n=0.5$ tend to fit spiral galaxy discs and dwarf ellipticals. We start with the smallest/faintest galaxy that is contributing to the light in the stamp (with the neighbouring sources masked), with a single component S\'ersic model and the initial guesses obtained from GALAPAGOS. We run GALFIT on this system and visually inspect the residual.  Although the reduced $\chi^{2}$ value is the most reliable method to judge the goodness of fit, for our application, since we are not striving to obtain any physical information from the parameters, we judge the goodness of fit merely by looking at the quality of the residuals; the smoothness of the residual gives a good indication of how successful the model has been. To refine the fit, we slowly build the complexity by adding additional components and visually examining the fit and the model parameters after each trial. For example, in addition to the inadequate components required to model a galaxy, we also found that GALAPAGOS was unable to accurately estimate the radius for almost all the sources, defaulting to either a very small value or a very large value, resulting in poor residuals. As such, we found that GALFIT converged faster when we estimated the radius from the stamp image and changed this value manually. We note that this would be unsatisfactory were we interested in the properties of these galaxies. But since the aim is only to remove the foreground galaxies, this tuning by hand is not critical. This was similarly done for the other model parameters. We let all the fitting parameters free to vary but we fix them once we get a reasonable residual before moving on to the next galaxy. 

\begin{figure*}
\centering
\begin{minipage}{0.33\textwidth}
\centering
\includegraphics[width=1\textwidth, height=0.19\textheight]{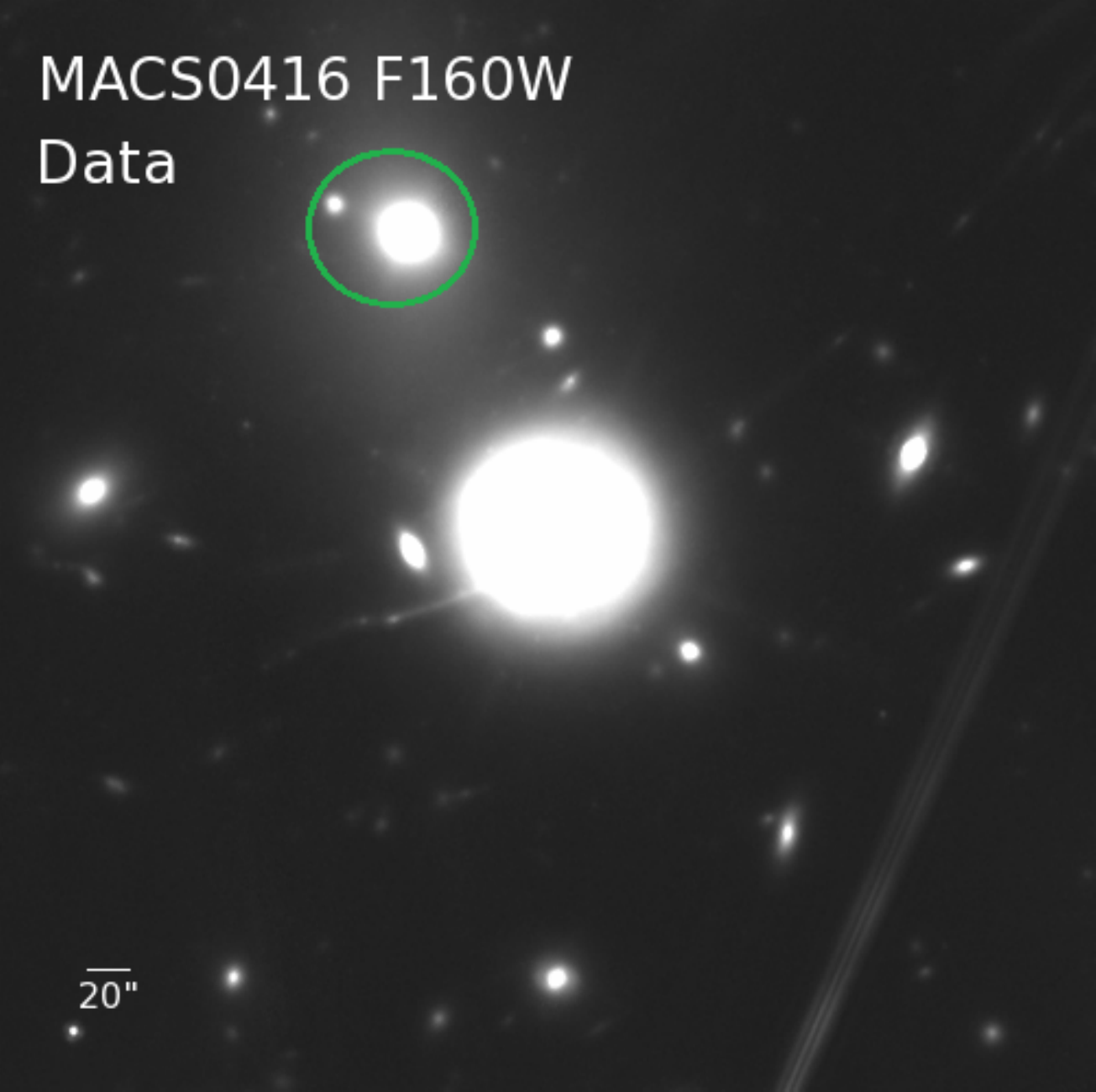}
\end{minipage}
\begin{minipage}{0.33\textwidth}
\centering
\includegraphics[width=1\textwidth, height=0.19\textheight]{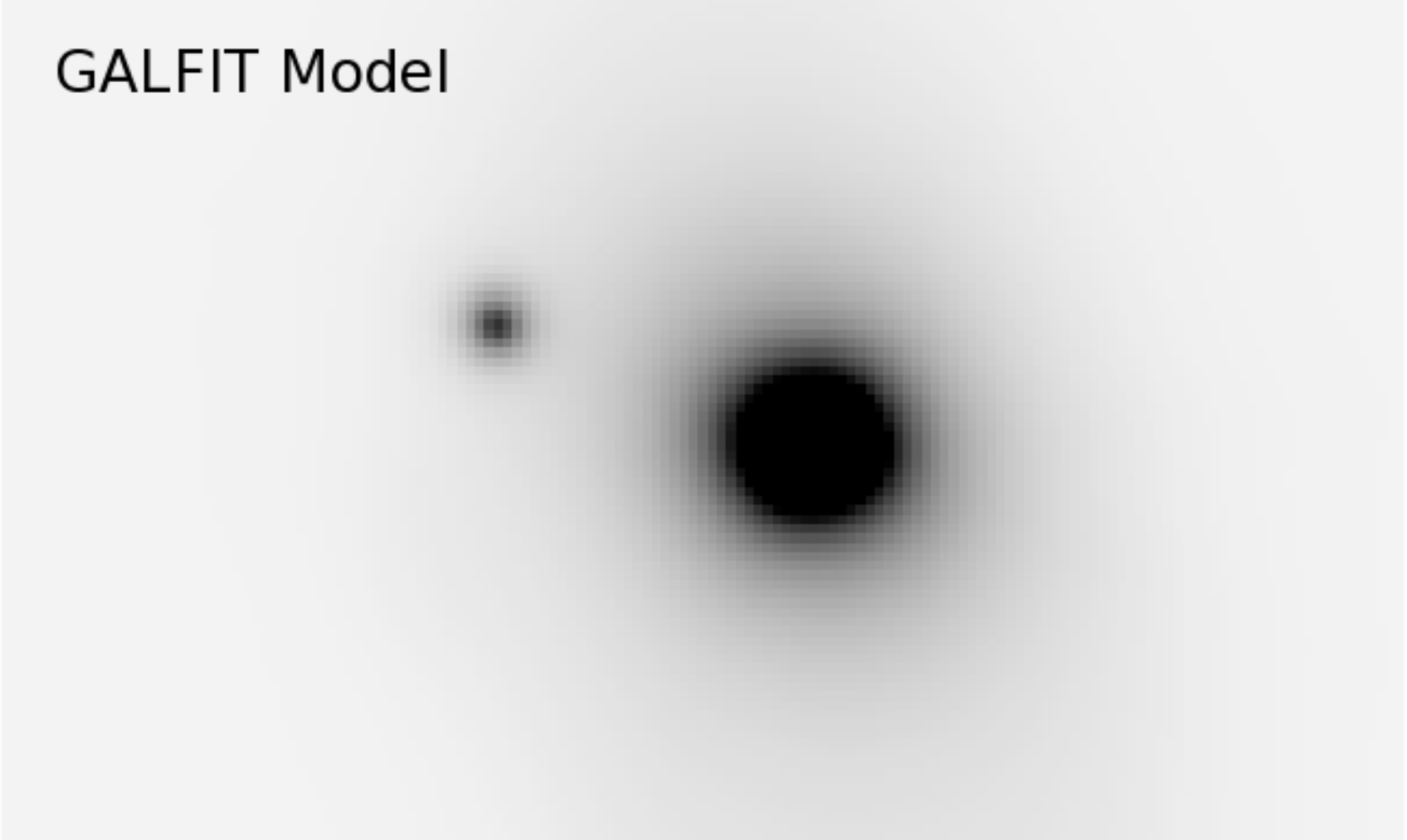}
\end{minipage}
\begin{minipage}{0.33\textwidth}
\centering
\includegraphics[width=1\textwidth, height=0.19\textheight]{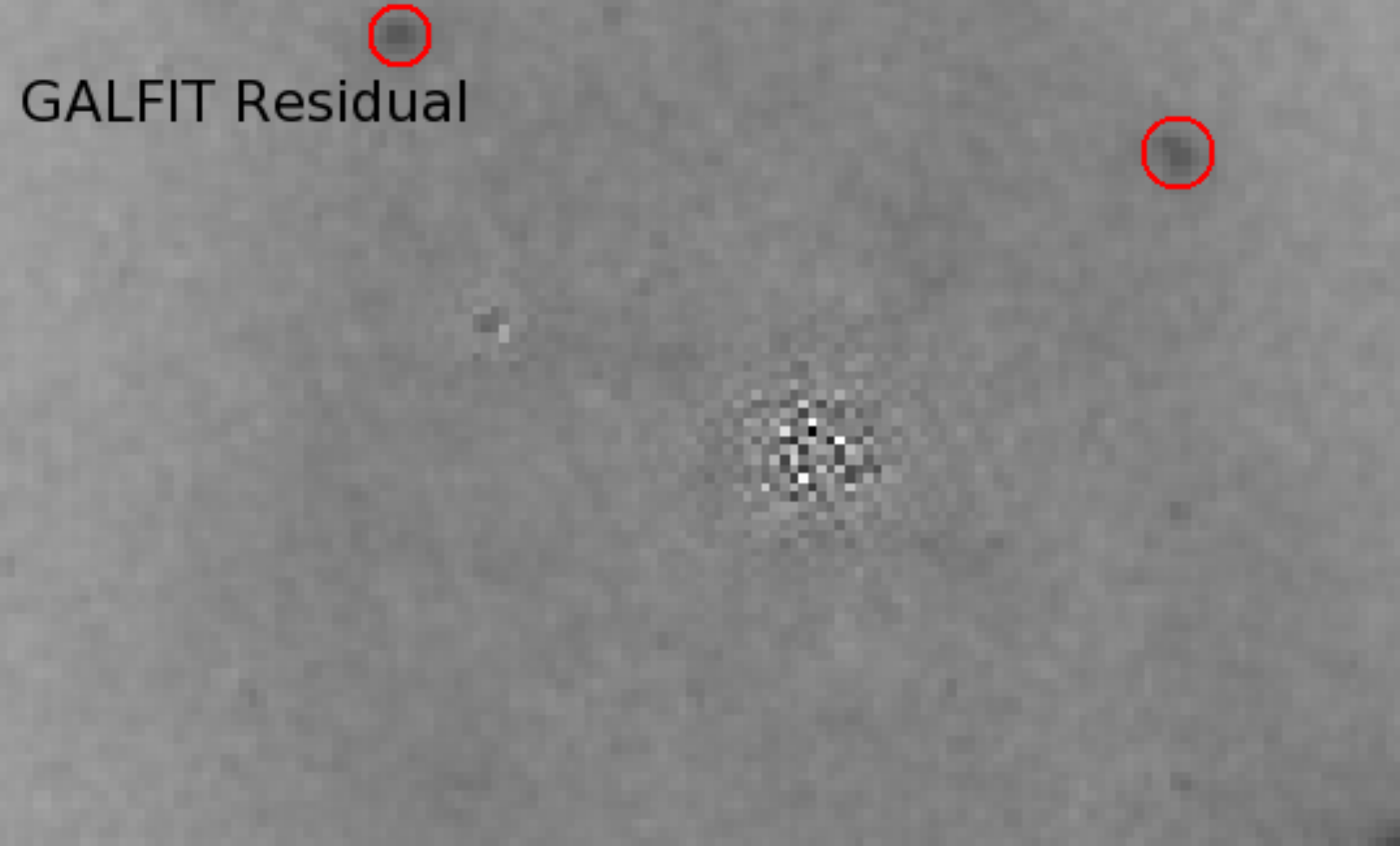}
\end{minipage}
\begin{minipage}{0.33\textwidth}
\centering
\includegraphics[width=1\textwidth, height=0.19\textheight]{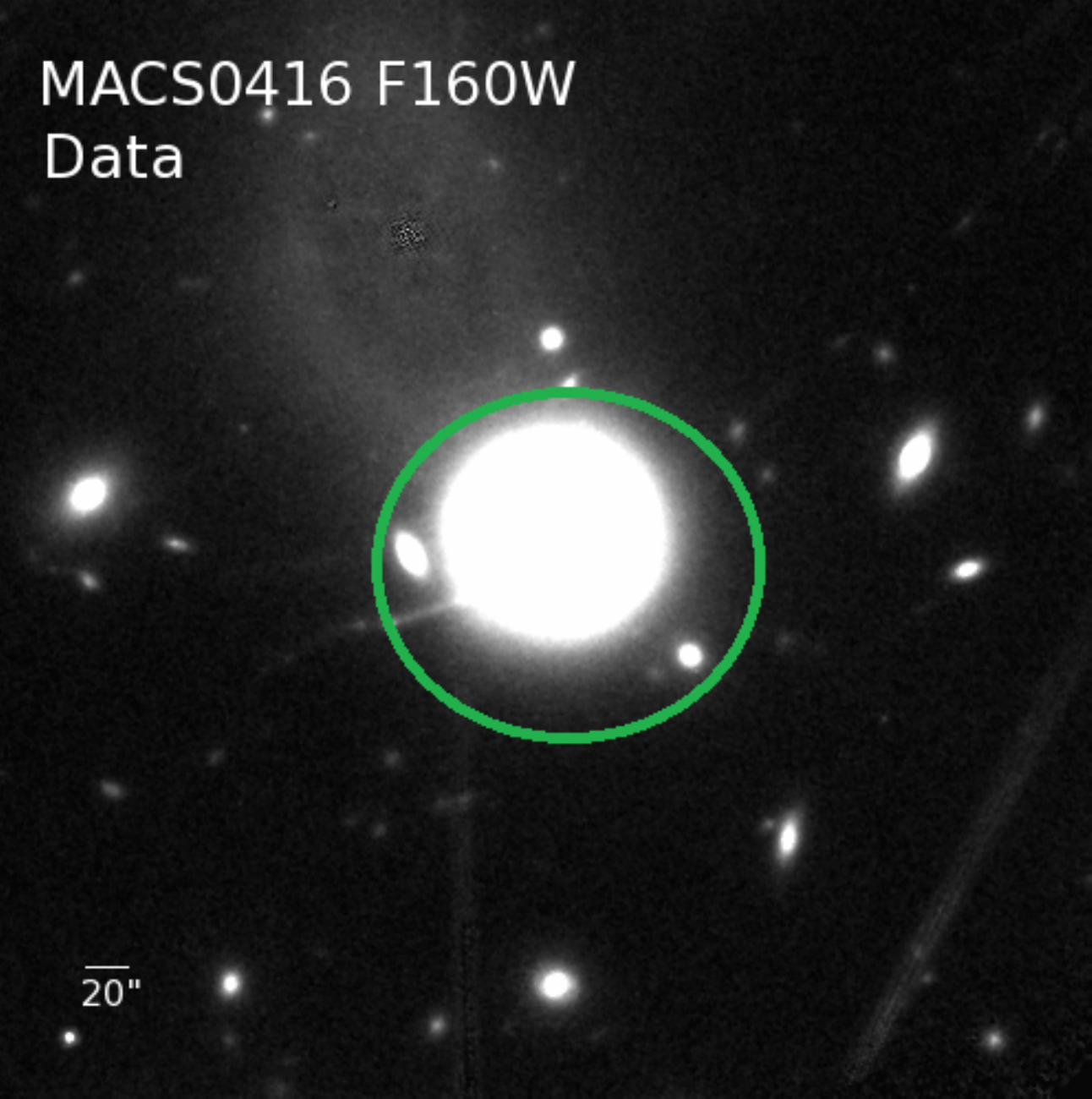}
\end{minipage}
\begin{minipage}{0.33\textwidth}
\centering
\includegraphics[width=1\textwidth, height=0.19\textheight]{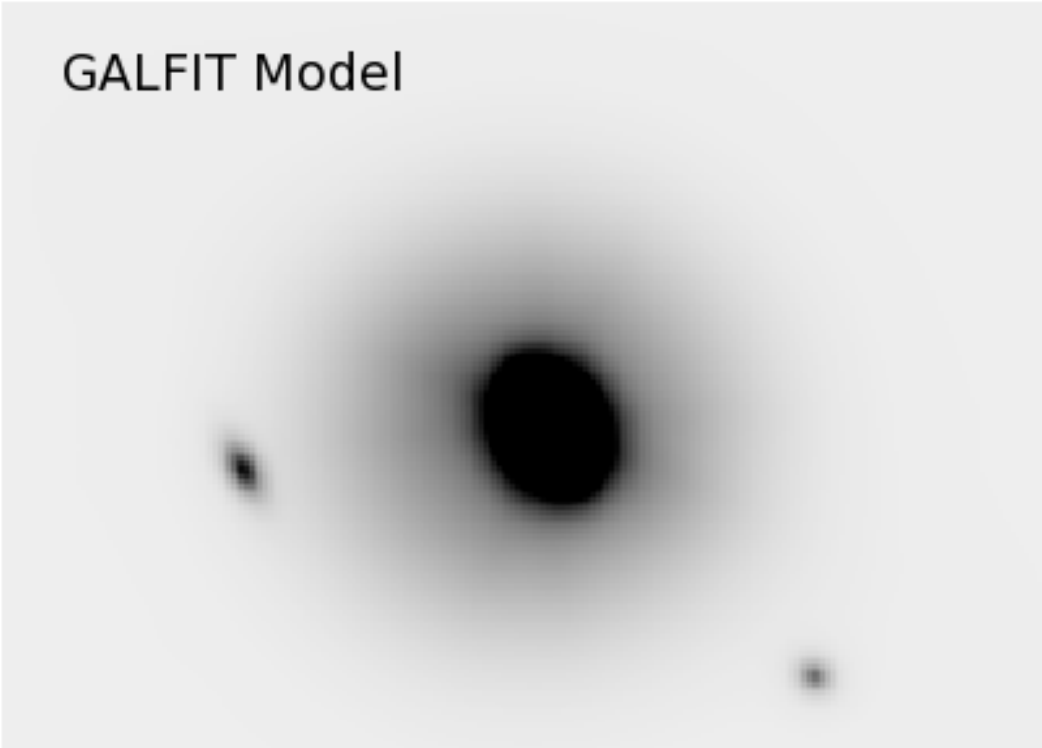}
\end{minipage}
\begin{minipage}{0.33\textwidth}
\centering
\includegraphics[width=1\textwidth, height=0.19\textheight]{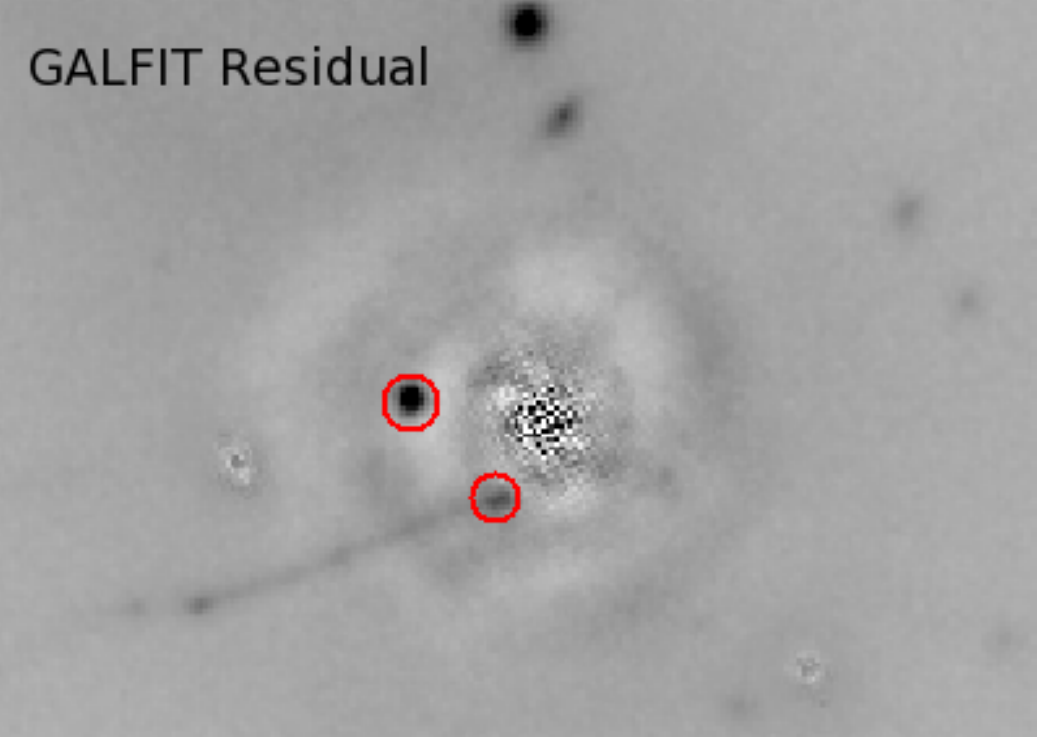}
\end{minipage}
\caption{Illustration of our multi-component fitting procedure on the F160W image. Top row left: Postage stamp showing the target central bright galaxy to be subtracted along with neighbouring smaller galaxies. Middle: GALFIT models of the smaller systems on the top (shown by green circle in the top left panel) that we attempt to subtract first before subtracting the brightest galaxy in the center. Two components were required to model the smallest galaxy at the top, and four components were required to model the brighter galaxy. Right: Residual after models subtraction, revealing two faint sources shown by red circles. Bottom row left: Postage stamp with the smaller systems on the top subtracted out. Middle: GALFIT model of the brightest central galaxy (shown by green circle in the bottom left panel) we want to subtract. Six components were required to model the brightest galaxy in the center. Right: Residual image, clearly revealing the faint sources that were obscured behind it, shown by red circles. }
\label{fig:sub_method}
\end{figure*}

This step is performed on all the neighbouring galaxies that contribute light to the stamp until we are left with the central target bright galaxy. We then repeat the process on the target bright galaxy once again until we get a reasonable residual. Finally, we do a massive simultaneous fit with all the neighbours together. Fig.~\ref{fig:sub_method} illustrates our method, showing the postage stamp for our target brightest cluster galaxy in the center, surrounded by smaller neighbouring galaxies. For example, two components were required to model the smallest galaxy on the top and four components to model the bright galaxy below it (enclosed in green circle in the top left panel)), revealing two faint sources (shown by red circles in the residual image). Similarly, for the massive galaxy in the middle (enclosed in green circle in the bottom left panel), six components were required, once again revealing two faint sources (shown by red circles in the residual image) that were obscured by the light of this massive galaxy. As shown in Fig.~\ref{fig:sub_method}, our procedure clearly allows us to detect objects that are obscured by the bright galaxies once the bright galaxies are subtracted from the cluster. We note that the number of S\'ersic components that were required to model some of the galaxies are higher than that are studied usually. However, we attribute no physical significance to the total number of components; they were solely needed for the purpose of subtraction. In this work, we obtained GALFIT models in the range $0.5<n<6.55$, indicating that the brightest cluster galaxies in MACSJ0416 cluster vary greatly in profile, and hence structure and morphology.

\subsubsection{Subtracting models from the original image}
\begin{figure*}
\centering
\begin{minipage}{0.33\textwidth}
\centering
\includegraphics[width=1\textwidth, height=0.3\textheight]{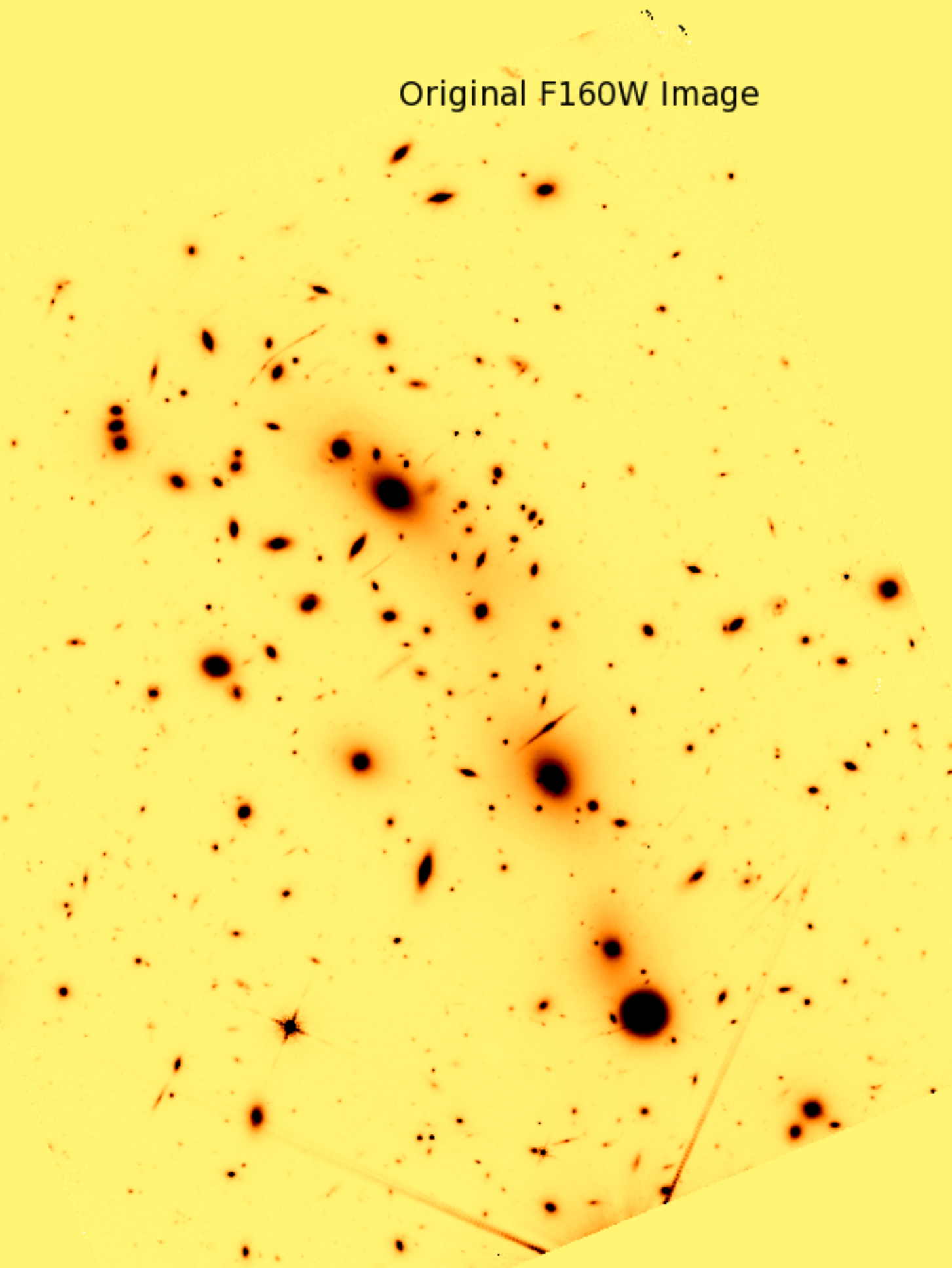}
\end{minipage}
\begin{minipage}{0.33\textwidth}
\centering
\includegraphics[width=1\textwidth, height=0.3\textheight]{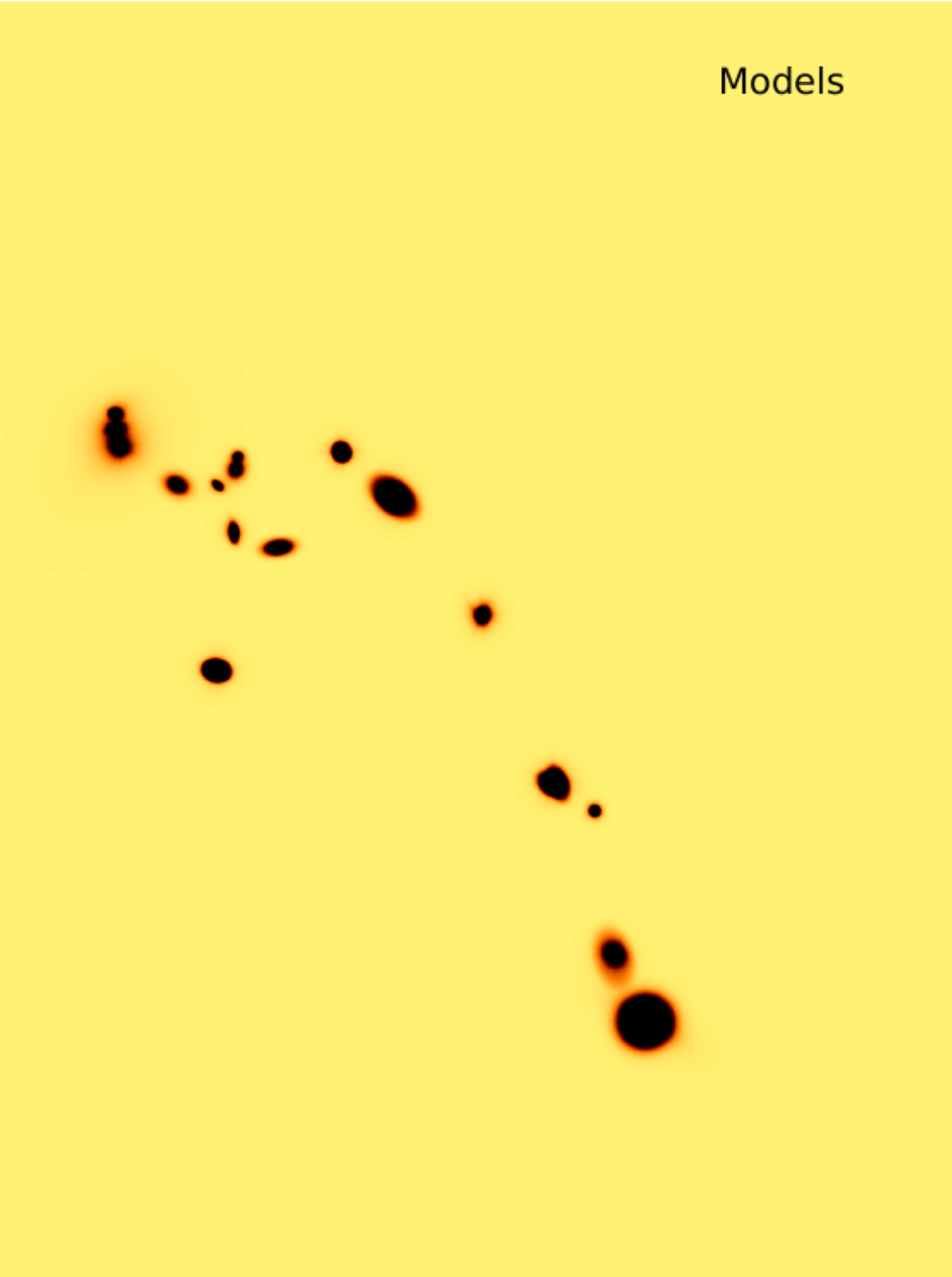}
\end{minipage}
\begin{minipage}{0.33\textwidth}
\centering
\includegraphics[width=1\textwidth, height=0.3\textheight]{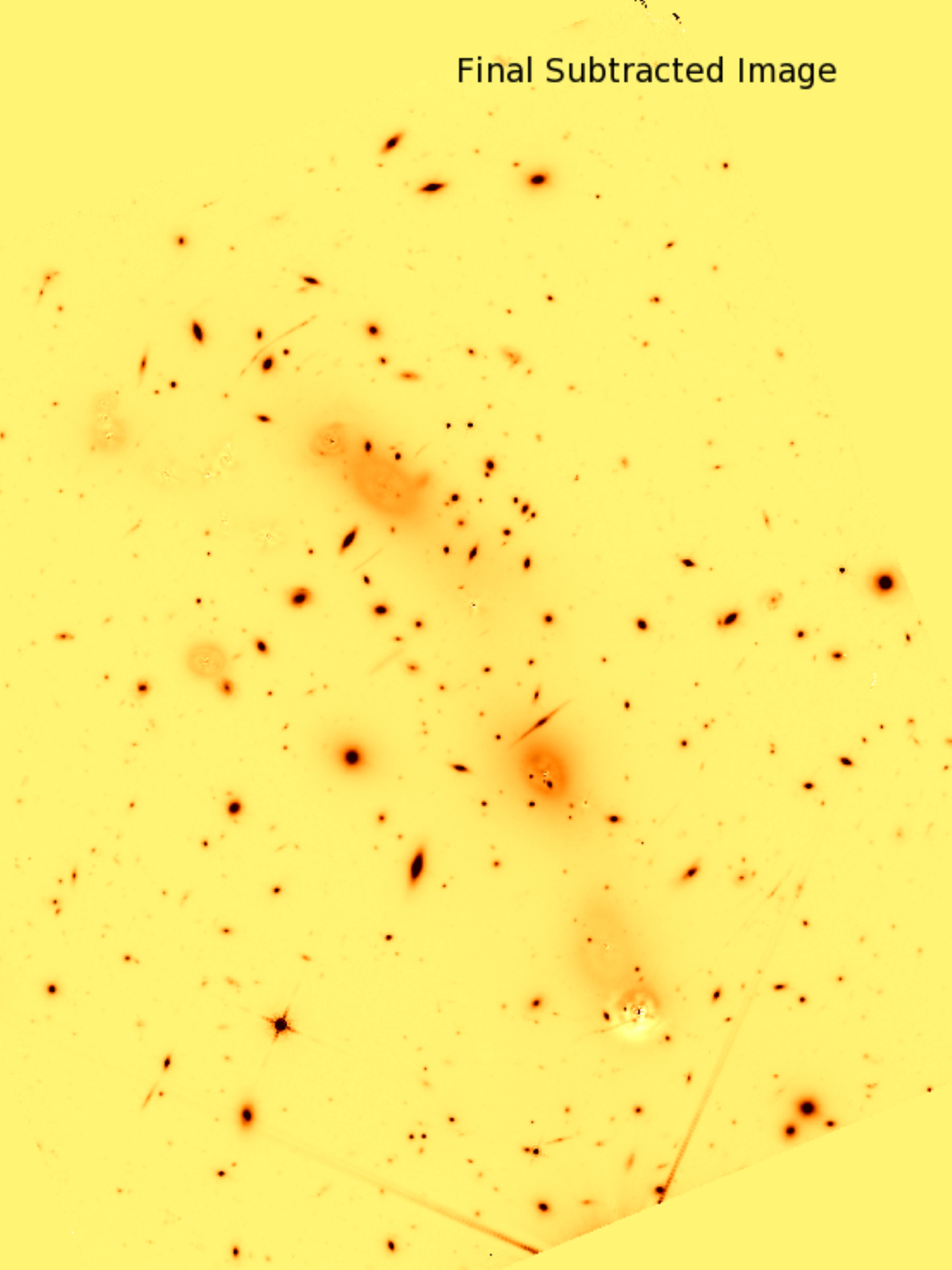}
\end{minipage}
\caption{Results of our subtraction procedure on the MACSJ0416 cluster. Left to right: Original F160W detection image on which we applied our "Divide and Conquer" procedure, models of galaxies, final subtracted image. Some of the faint sources that were obscured by the light of bright galaxies are clearly seen after the subtraction procedure.}
\label{fig:sub_results}
\end{figure*}

After obtaining reasonable models for all the secondary and primary sources in a stamp, we create a new object file such that the stamp pixel coordinates correspond to the pixel coordinates in the original image, along with all the other model parameters. GALFIT is then run on this new object file, resulting in an image with models only, which could be subtracted from the original image. This models-only image is then subtracted out from the original image before we move on to fit the next bright galaxy on the critical line. Steps 1 to 3 are then repeated in iteration until all the models for all the stamps are sequentially subtracted from the original image.   Fig.~\ref{fig:sub_results} shows the final results of our procedure.

\subsection{Multiwavelength photometry}
 \label{sec:photometry}
Having subtracted the massive galaxies on the critical line of the cluster in the H-band, we proceed to construct a multiwavelength photometry catalog from 0.4 to 4.5$\mu$m. In this section we explain how we obtained our photometric measurements for MACS0416 cluster and its parallel field starting from our subtracted H-band.

\subsubsection{HST images}
To measure accurate photometry, we first PSF-match all the other \textit{HST} bands to the PSF of the lowest resolution H-band (F160W) using PSFMATCH task in IRAF. Empirical PSFs were generated by stacking the images of several isolated and unsaturated stars in the field. We then use SExtractor in dual image mode with our subtracted H-band as detection image and use the same detections/apertures to perform photometry on the rest of the bands. SExtractor relies on the GAIN parameter to calculate accurate flux uncertainties. We therefore compute the effective gains for each band separately using EFFECTIVE GAIN $=$ INSTRUMENT GAIN $\times$ EXPOSURE TIME (the instrument gain is different for ACS and WFC3/IR; 2 for ACS and 2.5 for WFC3/IR) and use this value while performing photometry.      

\begin{figure}
\includegraphics[width=\columnwidth]{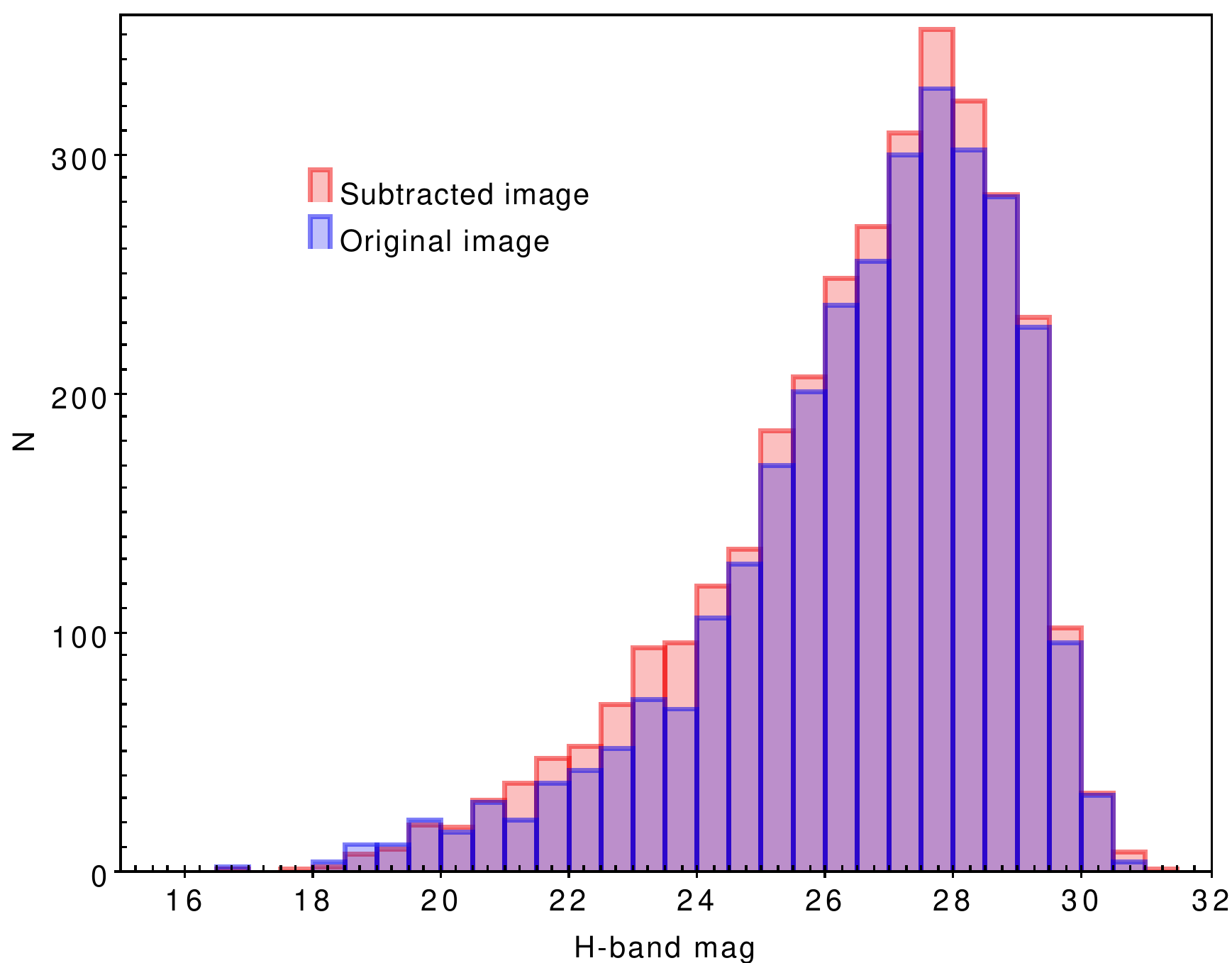}
\caption{Comparison of detections in the original H-band image and the subtracted H-band image for the MACS0416 cluster. The subtraction procedure has enhanced the detection of faint sources in the cluster field by $\sim8$ per cent.}
\label{fig:detect_comp}
\end{figure}

Although MAG\_AUTO (derived from flux in a flexible Kron-like elliptical aperture) generally gives the best estimate of the magnitude irrespective of the settings, it still underestimates the magnitudes for faint objects.  On the other hand, MAG\_ISO computes the magnitude in an isophotal area roughly the same shape as the object, and when using separate detection and photometry images, gives you the most accurate colours \citep{Bertin1996}. Following previous work (e.g., \citealt{Galametz2013, Guo2013}) we therefore adopt MAG\_ISO magnitudes in our analysis. Since our aim is to detect faint galaxies at $z\gtrsim6$, we employ an aggressive detection strategy with following SExtractor parameters: \textit{DETECT{\_}MINAREA} 4 pixels, \textit{DETECT{\_}THRESH} 0.7, \textit{DEBLEND{\_}NTHRESH} 64 and \textit{DEBLEND{\_}MINCONT} 0.0001. We employ the same method on the parallel field, with the exception of subtraction procedure, as there are no bright foreground galaxies in the parallel field. To analyze how much our subtraction procedure has enhanced the detection of faint sources in the cluster field, we run SExtractor with the same parameters on the original H-band image of MACS0416 cluster. We detect 3051 objects in the original H-band and 3293 objects in the subtracted H-band, showing that the subtraction procedure has enhanced the detection of faint sources in the cluster field by $\sim8$ per cent (See Fig.~\ref{fig:detect_comp}).

In order to compute stellar mass from spectral energy distribution (SED) fitting, we need to estimate the total flux in all the bands. For this, we use the same approach as \citet{Guo2013}, and first derive an aperture correction to total flux for each source in H-band as $\mathrm{aprcor = FLUX\_AUTO\_F160W/FLUX\_ISO\_F160W}$. We then convert the SExtractor isophotal fluxes and their uncertainties into $\mu$Jy first and then into total fluxes and uncertainties as:

$\mathrm{FLUX\_TOTAL = aprcor \times FLUX\_ISO}$
$\mathrm{FLUXERR\_TOTAL = aprcor \times FLUXERR\_ISO}$

We then build a catalog of total fluxes and the associated uncertainties that combines the photometry of the MACS0416 cluster and the parallel field.

\subsubsection{$K_{s}$ band and Spitzer images}
\begin{figure*}
\centering
\begin{minipage}{0.45\textwidth}
\centering
\includegraphics[width=1\textwidth, height=0.3\textheight]{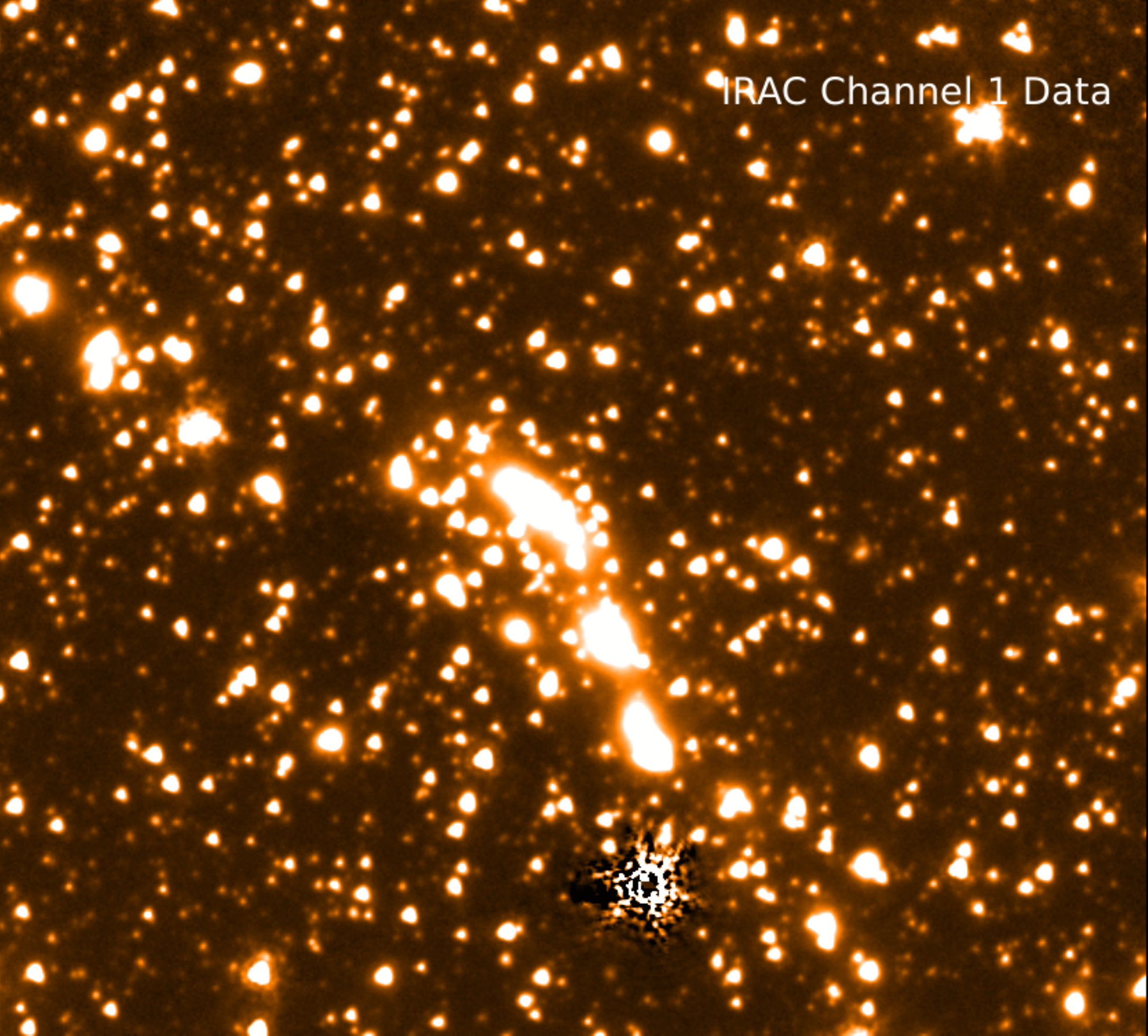}
\end{minipage}
\begin{minipage}{0.45\textwidth}
\centering
\includegraphics[width=1\textwidth, height=0.3\textheight]{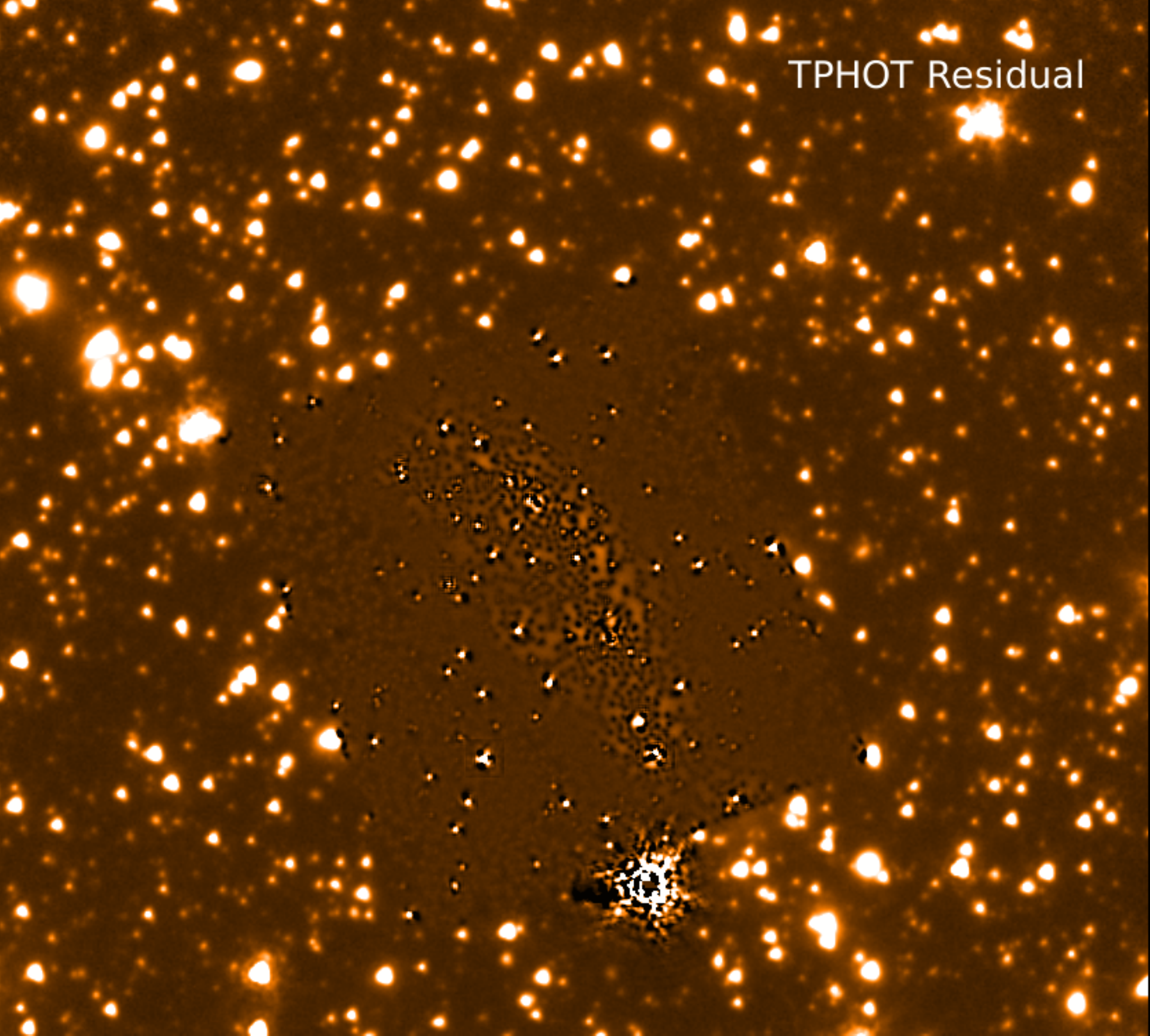}
\end{minipage}
\begin{minipage}{0.45\textwidth}
\centering
\includegraphics[width=1\textwidth, height=0.3\textheight]{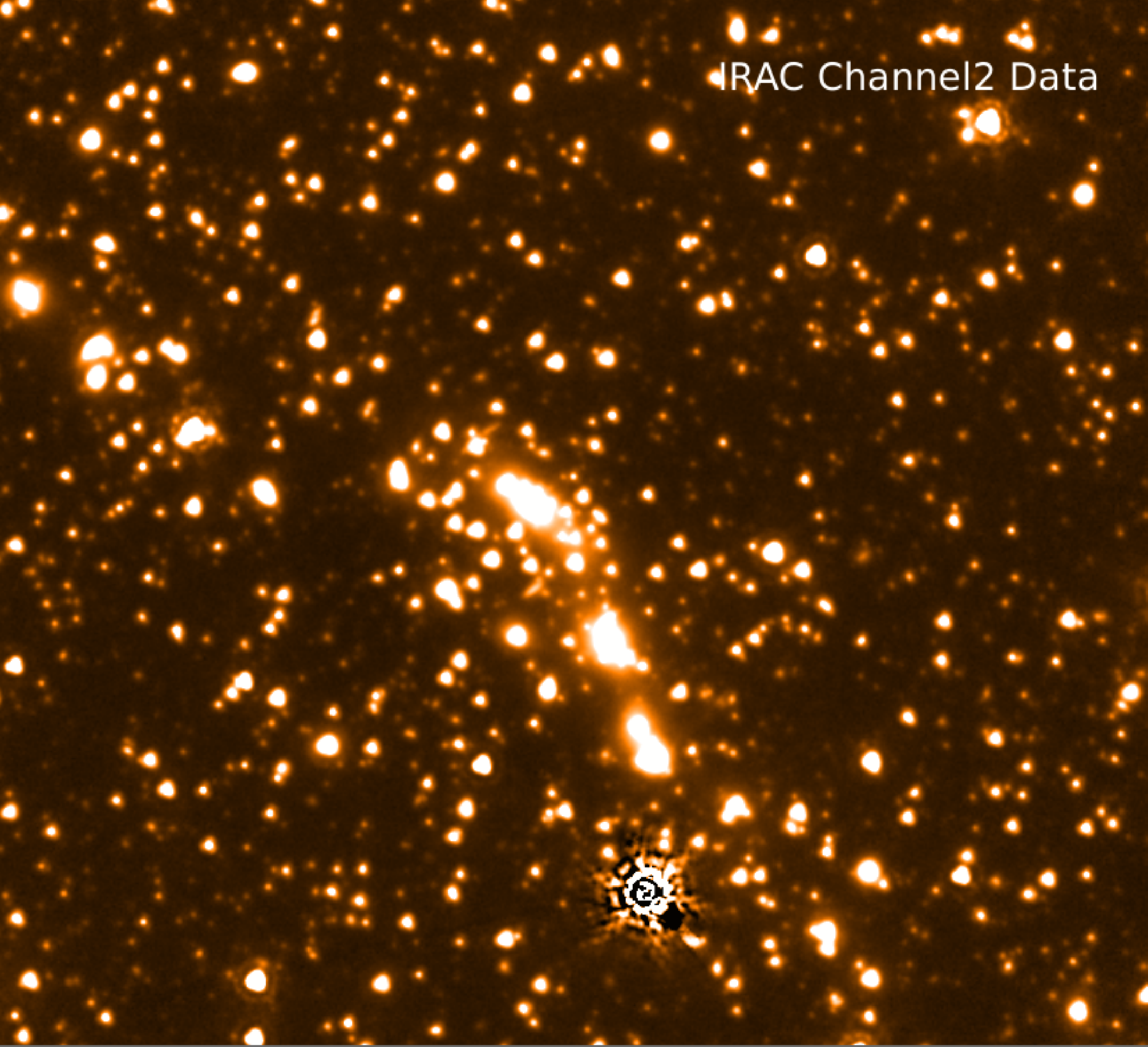}
\end{minipage}
\begin{minipage}{0.45\textwidth}
\centering
\includegraphics[width=1\textwidth, height=0.3\textheight]{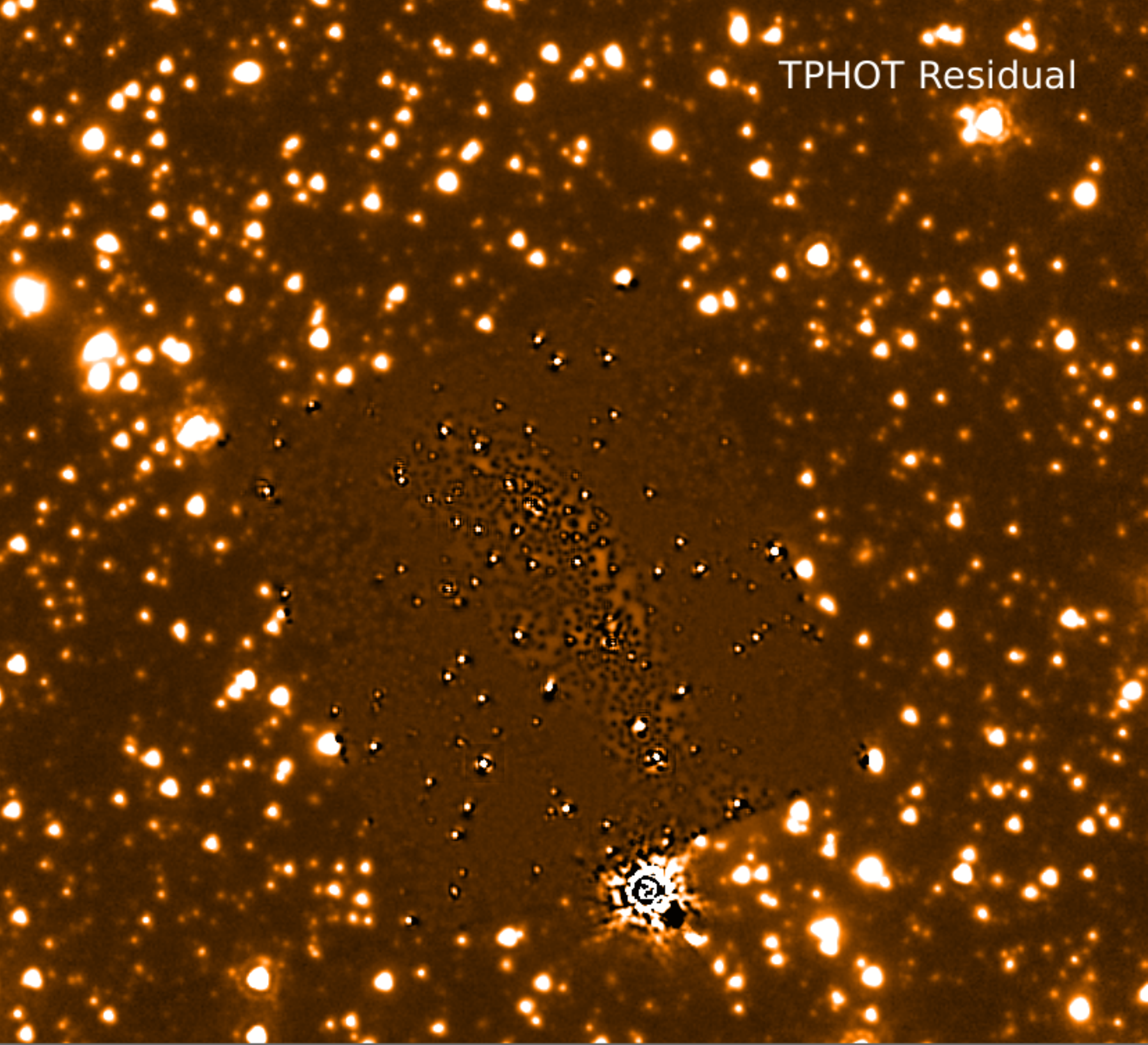}
\end{minipage}
\caption{IRAC 3.4 $\mu$m and 4.5 $\mu$m imaging of the MACS0416 cluster. Top left: Example of severely blended sources in IRAC channel 1 low resolution image. Top right: Residual image from T-PHOT. Bottom left: Original IRAC channel 2 image. Bottom right: Residual image from T-PHOT. }
\label{fig:irac_images}
\end{figure*}

We also include longer wavelength $K_{s}$ band and \textit{Spitzer} data in our multiwavelength photometry catalog for reasons mentioned in Section \ref{sec:vltdata} and Section \ref{sec:spitzerdata}. However, extracting accurate photometry from low resolution images ($\sim1\farcs8$ FWHM PSF of \textit{Spitzer} as opposed to $\sim0\farcs18$ FWHM of WFC3) is challenging because the sources are severely blended (See Fig.~\ref{fig:irac_images}), rendering aperture photometry and PSF fitting photometry unreliable. Also, PSF matching from \textit{HST} to \textit{Spitzer data} is not necessarily the best solution because we are then effectively throwing away all the useful spatial information from \textit{HST} when we smooth it out.

To overcome these issues, we perform photometry on the $K_{s}$ band and \textit{Spitzer} imaging using the T-PHOT code \citep{Merlin2015}, built on TFIT \citep{Laidler2007} and CONVPHOT \citep{DeSantis2007}. T-PHOT is a software designed to use information from the high resolution image (such as source positions and morphologies), and use this information as priors to measure fluxes in the low resolution image. T-PHOT uses a combination of input priors such as: 
\begin{enumerate}
  \item a list of sources to obtain cutouts from the high resolution image.
  \item analytical models from codes such as GALFIT.
  \item unresolved, point-like sources.
\end{enumerate} 

In our study, we use the first method, which uses the true high resolution priors. First, a list of source positions is obtained by SExtractor in order to obtain the high resolution cutouts of the sources as priors. The priors are then convolved with a suitable convolution kernel in order to degrade them to the resolution of the low resolution image. These low resolution (normalized) model templates are then placed at appropriate positions given by the high resolution image source catalogue and scaled by a $\chi^{2}$ minimization technique to give the measured fluxes of sources in the low resolution image (see \citealt{Merlin2015} for more details).

Several steps had to be performed before running T-PHOT as explained below:

\begin{enumerate}
\item The units of IRAC images are in MJy/sr, whereas the units of $K_{s}$ band are in counts/s. We first convert all the low resolution images to the same flux units of $\mu$Jy so that the fluxes from T-PHOT can be directly fed to the photometric redshift fitting code EAZY \citep{Brammer2008}.
\item Since the resolution of IRAC is about 10 times lower than that of F160W, we use the IRAC PSF itself as the kernel to convolve the F160W images. We assume that the actual kernels would 
not be so different from the IRAC PSFs. For $K_{s}$ band images, however, we run IRAF/PSFMATCH to compute the kernel. We derive empirical PSFs for both IRAC and $K_{s}$ band via neighbor-masked median-scaled stacks of isolated stars.
\item T-PHOT requires that the lower resolution image (as well as the rms maps) be at the same orientation as the high resolution image. We therefore resample all IRAC and  $K_{s}$ band images and the rms maps to the F160W pixel scale of $\sim0\farcs06$ and reproject to \textit{HST} astrometry using SWarp \citep{Bertin2002}.
\end{enumerate}
To correct the spatial distortion/misregistration and to obtain more astrometrically precise results, we run T-PHOT in two passes. The first pass cross-correlates the model image and the low resolution image to compute the shifts and build a new set of kernels for regions under consideration. The second pass uses these shifted kernels to reduce the misalignment between the templates and low resolution images to obtain more precise results (see \citealt{Merlin2015}).

\begin{figure}
\includegraphics[scale=0.6]{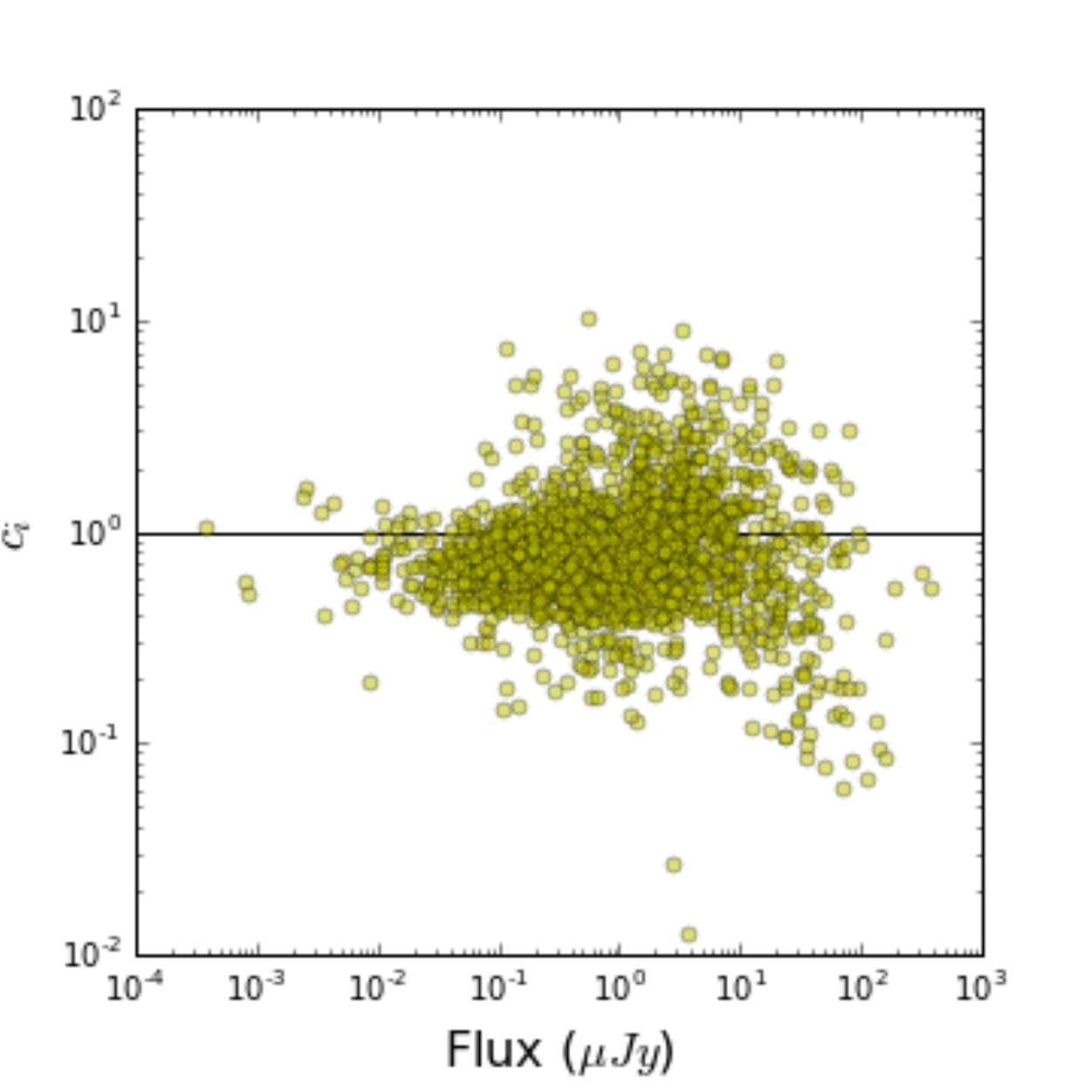}
\centering
\caption{Covariance index $c_{i}$ as a function of measured fluxes from T-PHOT for the MACS0416 cluster. 527 (16 per cent) sources from our catalog lie above $c_{i}>1$ and are treated with caution. }
\label{fig:cov_index}
\end{figure}

The accuracy of this procedure with T-PHOT on IRAC images is illustrated in Fig.~\ref{fig:irac_images}. A comparison between the original image and the residual image demonstrates that T-PHOT does a remarkably good job at fitting sources in the low resolution image, with the exception of the very bright sources in the cluster. We notice that if the bright sources lie close to the faint sources, then their residuals affect the photometry of nearby faint sources significantly. For example, for such faint sources the S/N was significantly high (or negative) in the output catalogue generated by T-PHOT even if those faint sources were physically absent in the low resolution IRAC image (e.g., \citealt{Song2016} have noticed the same effect). We found that such contaminated sources caused some unfortunate high-z solutions. T-PHOT helps to identify unreliable measurements for sources, especially contaminated from neighbours, with the help of a flag designated as "$c_{i}$", the covariance index (which is a ratio between a source's maximum covariance term and its variance in the covariance matrix) flag and suggests users to treat sources with $c_{i}>1$ with caution (see \citealt{Merlin2015}). Fig.~\ref{fig:cov_index} shows the values of covariance index as a function of estimated fluxes in IRAC 3.4 $\mu$m in the MACS0416 cluster. Out of 3293 sources in our catalog, 527 (16 per cent) objects have $c_{i}>1$. We therefore visually inspected each of these sources carefully to ensure that their photometric measurements are reliable.

\subsection{Photometric redshifts and source selection}

\subsubsection{Photometric redshifts}
There are two widely used methods for computing photometric redshifts -- i) The Lyman Break (LB) \citep{Steidel1996} also known as the "drop-out" technique that relies on the large break in the continuum flux from an object in bands blueward of the Lyman break and two-color selection in bands redward of the break, and ii) Template fitting method in which photometric redshifts are derived by fitting synthetic template spectra to the observed photometry. There are several codes, such as HYPERZ \citep{Bolzonella2000}, EAZY \citep{Brammer2008}, BPZ \citep{Benitez2000}, LePHARE \footnote[4]{http://www.cfht.hawaii.edu/~arnouts/LEPHARE/lephare.html} \citep{Arnouts2011}, in place that use their own methods to calculate photometric redshifts. In our work, we determine the photometric redshifts for our multiwavelength catalog using EAZY \citep{Brammer2008}. We use the default reduced template set provided with EAZY, based on the PEGASE stellar population synthesis models of \citet{Fioc1997}, which includes contribution from emission lines, and also an additional template based on the spectrum of \citet{Erb2010} that includes features such as strong optical emission lines and a high Ly$\alpha$ equivalent width; characteristics peculiar to young, unreddened, low-metallicity galaxies at high redshift, similar to \citet{Duncan2014}.

\subsubsection{Sample selection criteria}
\label{sec:selectioncriteria}

After computing photometric redshifts, we construct a sample of galaxies in the redshift range  $5.5\leq z\leq9.5$. To do this, instead of simply relying on the best-fit redshift value, we take advantage of the full redshift probability distribution function (PDF) ($P(z)\propto\exp(-\chi^{2}/2$), using the $\chi^{2}$ distribution from EAZY, similar to previous work (e.g., \citealt{Finkelstein2012, Finkelstein2015, Duncan2014}). We then form galaxy samples in four redshift bins centered at z $\sim$ 6, 7, 8, and 9 with $\triangle z$ =  1, for both the MACS0416 cluster and the parallel field, by applying a set of additional selection criteria following \citet{Duncan2014} as: 

\begin{equation}
\int_{z_{\mathrm{s}}-0.5}^{z_{\mathrm{s}}+0.5}p(z)dz>0.4
\end{equation}

\begin{equation}
\int_{z_{\mathrm{p}}-0.5}^{z_{\mathrm{p}}+0.5}p(z)dz>0.6
\end{equation}

\begin{equation}
(\chi_{\mathrm{min}}^{2}/N_{\mathrm{filters}}-1)<3,
\end{equation}

\noindent where $z_{s}=$ 6, 7, 8 and 9 for the respective bins and $z_{p}$ is the primary redshift peak.

With the first criterion we ensure that a significant area of the PDF occupies the redshift range of our interest. The second criterion ensures that at least 60 per cent of the PDF lies under the peak of the distribution, making sure if the high-redshift solution is picked, then it is the dominant one. With the third criterion we ensure that EAZY provides a reasonable fit. In addition to the above three criteria, we also place a S/N cut such that S/N$(J_{125})>3.5$ and S/N$(H_{160})>5$. This ensures the secure detection of candidates in primary filters with very high significance, excluding spurious detections. 

\begin{figure*}
\centering
\begin{minipage}{0.49\textwidth}
\centering
\includegraphics[width=1\textwidth, height=0.42\textheight]{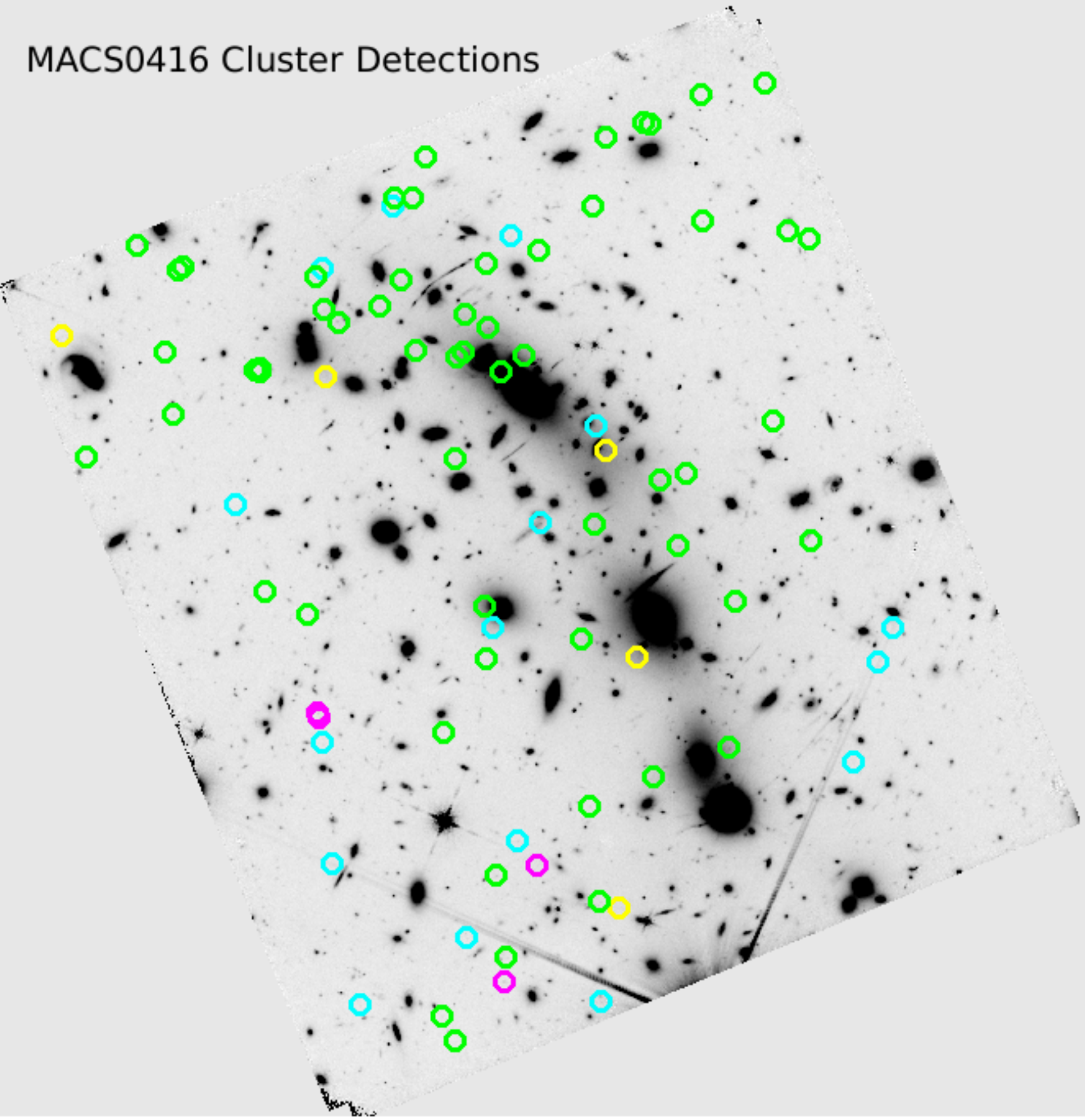}
\end{minipage}
\begin{minipage}{0.49\textwidth}
\centering
\includegraphics[width=1\textwidth, height=0.42\textheight]{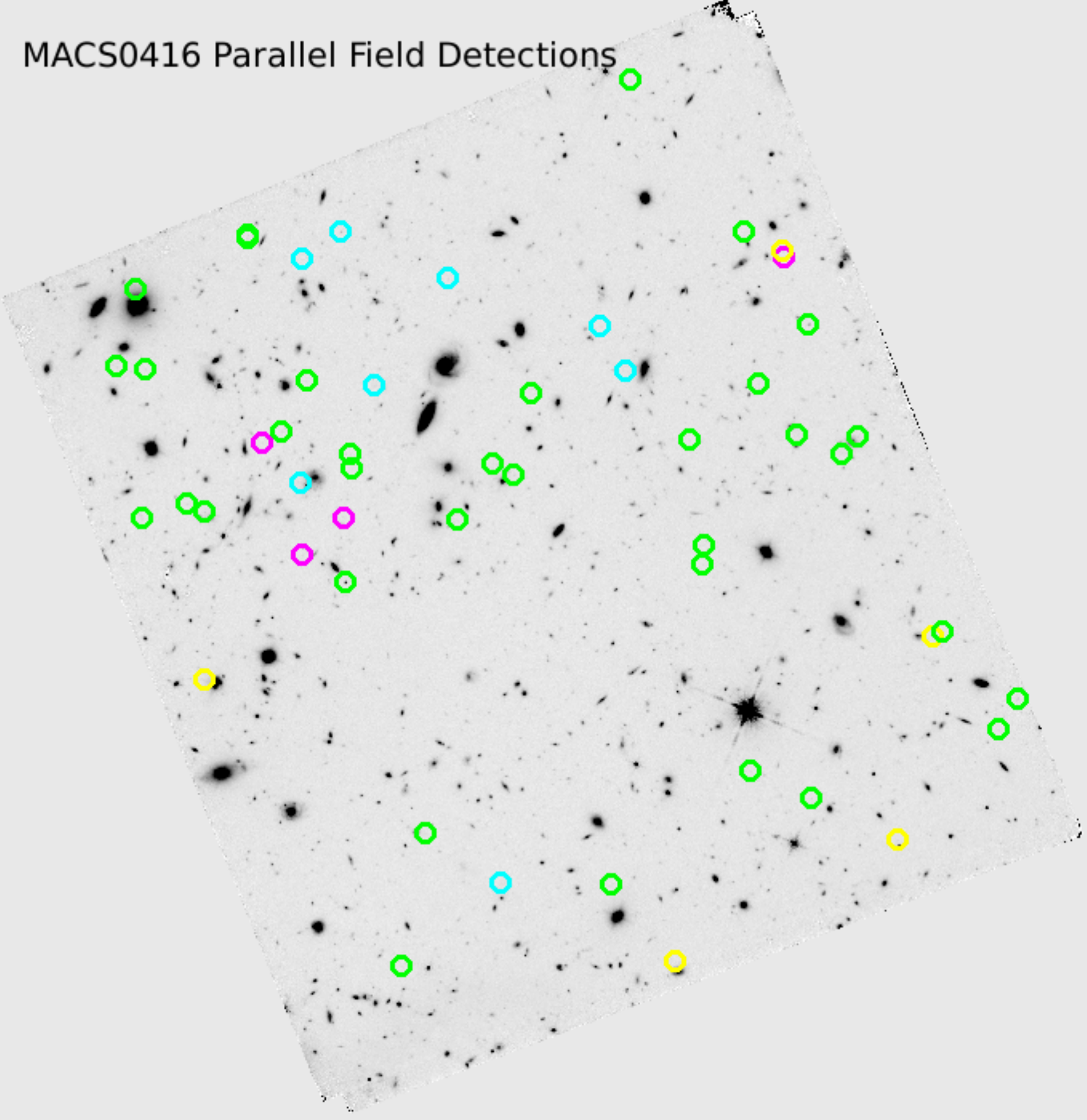}
\end{minipage}
\caption{H-band image of the MACS0416 cluster (left) and the parallel field (right). The positions of our high-z galaxy sample at $z=6,7,8$ and $9$ are shown by green, cyan, yellow, and magenta circles, respectively.}
\label{fig:detections}
\end{figure*}

Once we have the sample of high-redshift candidates in each redshift bin for both the cluster and the parallel field, we inspect each object thoroughly to eliminate the potential contaminants such as stars, stellar diffraction spikes, sources at the edge of the images, sources with flagged photometry etc. Our final samples then contains 134 galaxies: 82 in the MACS0416 cluster, and 52 in the parallel field, out of which, 92 are at $z\sim6$, 24 are at $z\sim7$, 10 are at $z\sim8$ and 8 are at $z\sim9$ (cluster and parallel field combined), as shown in Fig.~\ref{fig:detections}. 

\begin{figure}
\includegraphics[scale=0.45]{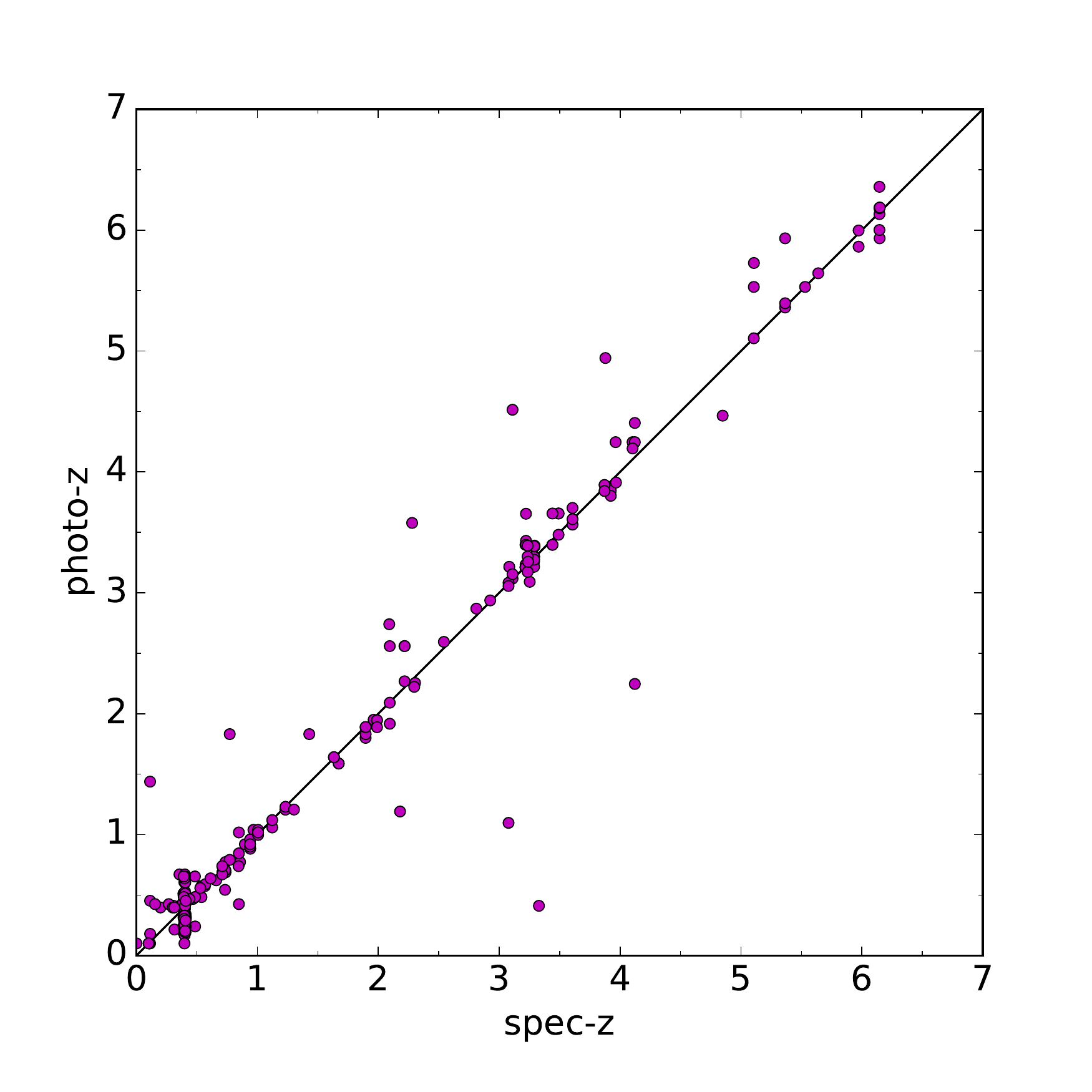}
\centering
\caption{Photometric redshift and spectroscopic redshift comparison for sources in our catalog for the MACS0416 cluster, with spectroscopic redshifts from the published redshift catalog of the MACS0416 cluster from the combination of the VIMOS CLASH-VLT campaign \citep{Balestra2016} and the MUSE spectroscopic study presented in \citet{Caminha2017}.}
\label{fig:photz_specz_comp}
\end{figure}

In order to do comparisons between our estimated photometric redshifts and available spectroscopic redshifts, we use the published redshift catalog of the MACS0416 cluster from the combination of the VIMOS CLASH-VLT campaign \citep{Balestra2016} and the MUSE spectroscopic study presented in \citet{Caminha2017} (VLT programme IDs 186.A-0798,  094.A-0115(B), 094.A-0525(A)). We match our photometric redshifts catalog and the spectroscopic redshifts catalog within 1 arcsec and include only those redshifts with QualityFlag = 3, 4, 5 and 9. Following \citet{Dahlen2013}, we define outliers as $|\Delta z/(1+z_{\mathrm{spec}})|\geq0.15$, where $\Delta z=(z_{\mathrm{spec}-}z_{\mathrm{phot}})$. After excluding the outliers, we compute $\Delta z/(1+z_{\mathrm{spec}})$ and find $\sigma_{\Delta z/(1+z_{\mathrm{spec}})}=0.041$. The comparison between photometric and spectroscopic redshifts for the MACS0416 cluster are shown in Fig.~\ref{fig:photz_specz_comp}.

\subsubsection{Comparison of $8.5\leq z\leq9.5$ sample with previous work}
Following the comparison of our photometric redshifts with available spectroscopic redshifts, we now compare our sample of high redshift candidates with previous studies. While the number counts of our lower redshift sources are in general agreement with previous work such as \citet{Livermore2017}, given the strong interest in the very high redshift candidates, in this section we only directly compare our highest redshift sample at $8.5\leq z\leq9.5$ with previous works of \citet{Laporte2015}, \citet{Mcleod2016} and \citet{Castellano2016} who study MACS0416 and its parallel field using all the bands (\textit{HST}, VLT and IRAC) available for the HFF.

\citet{Laporte2015} report four candidates at $z\sim8$ behind the MACS0416 lensing cluster. Our results are in excellent agreement with theirs, in that we are able to recover all four of their sources in our work (our IDs 126, 1040, 1065, 393, See Appendix~\ref{sec:catalogs}), and the photometric redshift estimates agree within the error bars. \citet{Mcleod2016} furthermore report two sources (their HFF2C-9-3 and HFF2C-9-5) in addition to the three sources (so five in total) found by \citet{Laporte2015} (their Y1, Y2, and Y4) at $z\sim8.4$. However, these two sources are rejected in our defined selection criteria (See Section \ref{sec:selectioncriteria}) and are therefore not included in our sample. Examining the parallel field, \citet{Mcleod2016} report five candidates at $z\sim8.4$ and we recover three of these (Ours IDs 1524, 1957 and 2660) with the correct photometric redshifts within the error bars. However, their HFF2P-9-2 and HFF2P-9-5 sources get rejected in our defined selection criteria.

Finally, we compare our detections with \citet{Castellano2016} who determine the photometric redshifts of their high-z sample from the ASTRODEEP catalog \citep{Merlin2016} constructed by subtracting the massive galaxies in the cluster, similar to this work. They report four sources at $8.5\leq z\leq9.5$ in the MACS0416 cluster and we are able to recover two of them (their ID 99 and 743, our IDs 126 and 1065). Of the other two sources, one source (their ID 2385, which is their HFF2C-9-3 in \citet{Mcleod2016}) is rejected in our sample selection criteria and the other source (their ID 538 reported at $z=8.7_{-3.6}^{+3.6}$) appears to be a possible low-z interloper at $z=0.27_{-0.143}^{+1.581}$ based on our analysis. In addition, \citet{Castellano2016} report eight candidates at $z\geq9.5$ in the MACS0416 cluster, out of which five appear to be possible noisy detections, one appears to be a potential low-z interloper (at $z=0.14_{-0.037}^{+0.001}$ in our analysis), and the other two (their IDs 2362 and 2543) get rejected in our sample selection criteria, and therefore we are not able to recover any of their reported sources. In the parallel field, \citet{Castellano2016} find seven sources at $8.5\leq z\leq9.5$. We recover two of these (their ID 1272 and 1656, our IDs 1524 and 1957), while three of these sources get rejected by our sample selection criteria, and two appear to be potential low-z interlopers at $z=1.51_{-0.70}^{+0.943}$ and $z=0.654_{-0.135}^{+0.044}$ respectively in our analysis. Lastly, they report two candidates at $z\geq9.5$, one of which appears to be a possible noisy detection in our analysis, and the other one again seems to be a potential low-z interloper at $z=2.22_{-0.374}^{+7.437}$. However, because of the large uncertainty on the photometric redshift, it is difficult to ascertain the exact nature of this source.

\subsection{Completeness simulations}
\label{sec:comp_sim}
Completeness correction helps us better constrain the low-mass end of the GSMF as well as the faint-end of UV LF. However, an overcorrection can lead to an artificial steepening of slope. We therefore perform an extensive set of simulations to estimate the completeness of our high-redshift sample by accounting for both the image incompleteness and the sample selection effects.

The gold standard way of doing this is by inserting 1000s of mock galaxies into the imaging data  and perform the same analysis for source detection and recovery, photometric redshift estimation and sample selection as for real data. This can be separated into two parts: a) completeness from the method used to extract the objects -- Here we insert 1000s of fake galaxies to a blank image in the source place, assuming different intrinsic source sizes, positions and magnitudes, and attempt to recover them with the same SExtractor parameters as for building the original catalog, and b) selection efficiency of an object in a given sample of particular redshift because of the employed selection criteria -- This involves calculating the photometric redshifts of a sample of galaxies and passing them through the same sample selection criteria as our real catalog and estimating their recovery fraction, which is then folded into completeness from part a).

\subsubsection{Detection completeness}
\label{sec:detect_comp}
This is determined by inserting thousands of synthetic sources in the detection image (H-band in our case) and attempting to recover them with the same SExtractor parameters as used for building the original catalogue. 

For this, we first make an artificial list of galaxies using the IRAF/GALLIST function. GALLIST produces a list of x and y coordinates, magnitudes, morphological types, half-light radii, axial ratios, and position angles for a sample of galaxies based on user defined input. Various studies have looked into the evolution of sizes of galaxies at high-redshift. For example, \citet{Grazian2012} and \citet{Ono2013} have found that faint galaxies at $z\geqslant7$ have extremely small half-light radii of $0.3-0.5$ kpc. \citet{Kawamata2015} report the detection of a few lensed galaxies at $z\sim6-8$ in Abell 2744 with sizes as small as 0.08kpc. More recently, \citet{Bouwens2017} have shown that the detection of highly magnified galaxies as a function of shear is highly dependent on galaxy sizes. With the help of simulations, they have demonstrated that only the most compact galaxies can be detected in high-shear regions, indicating that extremely faint $z\sim2-8$ galaxies have near point-source profiles, resulting in smaller completeness corrections and hence shallower faint-end slopes than reported in recent HFF studies. 

To quantify this, we perform our completeness simulations with two different methods: i) Prompted by previous work such as \citet{Ferguson2004}, \citet{Grazian2011} and \citet{Oesch2010}, we draw galaxy sizes from a log-normal distribution of mean $=$ 0.15 arcsec and $\sigma=$ 0.075, and ii) With galaxies having near point-source profiles i.e., unresolved point sources. For both the methods, the S\'ersic indices are chosen from a log-normal distribution in the range $0.5\leq n\leq4.0$ with the majority of the galaxies having disc-like morphologies with $n\leq2$ \citep{Ravindranath2006}, whereas the position angle is selected from a uniform distribution between $0^{\circ}$ and $360^{\circ}$. The axial ratio we use is also lognormal with a peak at 0.8 and the H-band magnitudes range from $21<H<35$. In each iteration of the simulation, the redshift is selected from a uniform distribution of $z=5-10$. To avoid confusion due to excessive number of sources, and also due to blending with nearby sources in the field, we insert only 200 fake sources in each iteration.  

The next step is to insert the mock sources in the source plane and incorporate the effect of lensing on them. Selecting the galaxy positions randomly in the source plane and then lensing them back to the image plane caused the sources to fall out of the edge of the image. We therefore choose the galaxy positions randomly in the image plane and then use IRAF/MKOBJECTS function to insert fake sources to a blank image at those positions in the source plane. This image is lensed back to the image plane using LENSTOOL \citep{JulloKneib2009,Jullo2007,Kneib1993} with the help of the latest lens models by the Sharon team (\citealt{Johnson2014}) made available to the Frontier Fields project\footnote[5]{https://archive.stsci.edu/prepds/frontier/lensmodels/}. The image is then convolved with the H-band PSF after its been transformed to the image plane; convolving it before the transformation made the galaxies look much bigger than they should. SExtractor is then run on this convolved image to recover the lensed galaxies and the parameters such as magnitudes, positions, half-light radii etc are stored in an input catalog. These lensed, fake sources are then added to the subtracted H-band image and SExtractor (with the same parameters as used for building the original catalog) is used to recover them and to construct a new catalog. Finally, this catalog is matched to the input catalog with fake sources within a $\sim0\farcs2$ matching radius. 

We repeat the above process 2000 times and ultimately combine the results to calculate the recovery fraction as a function of position, input magnitude and profile type. The same process was repeated for the parallel field, with the exception of the lensing equations. 

\begin{figure}
\includegraphics[width=1.1\columnwidth, height=1\columnwidth]{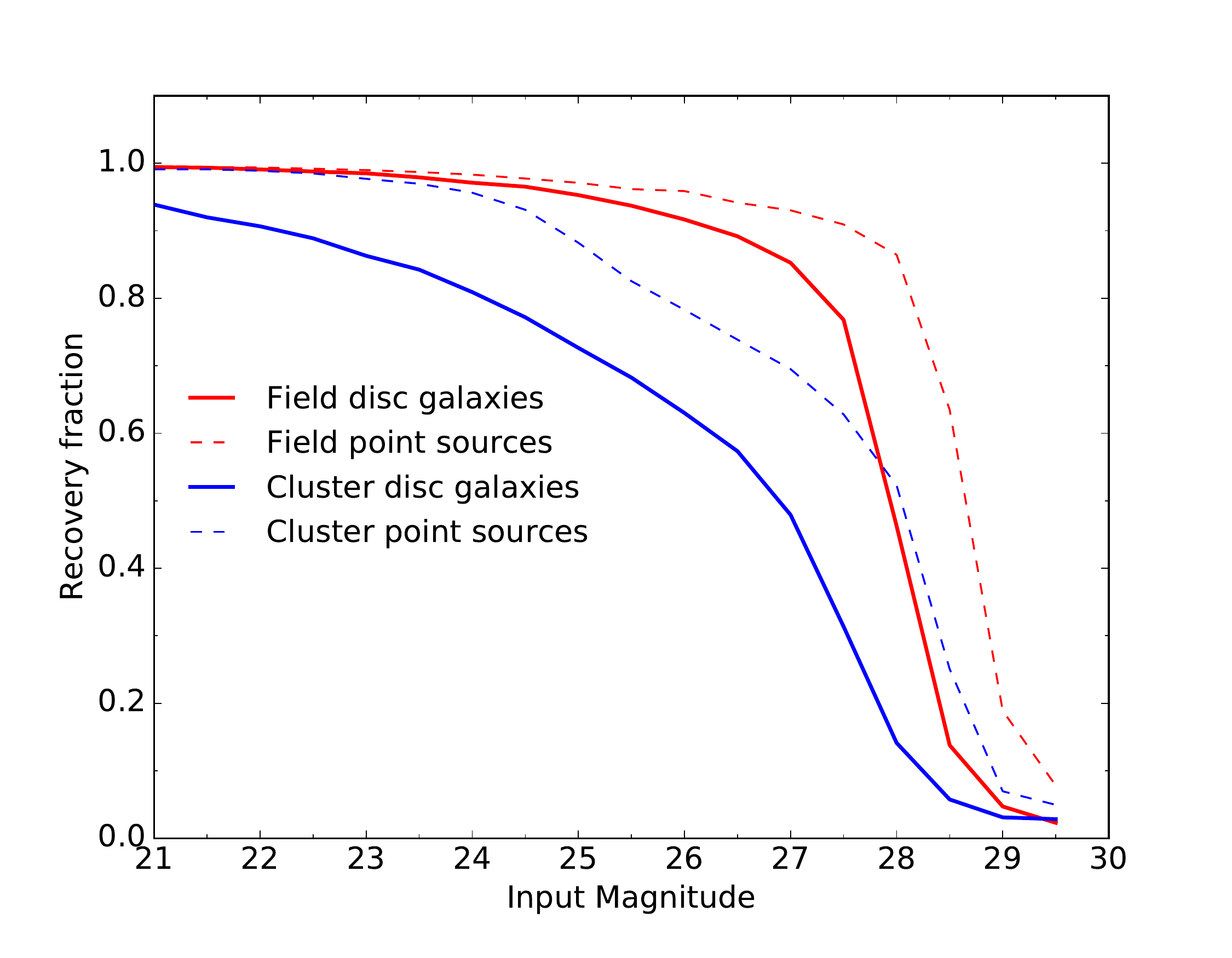}
\centering
\caption{Recovery fraction as a function of H-band input magnitude for unresolved point-like (blue and red dashed lines for cluster and field respectively) and disc-like sources (solid blue and solid red line for cluster and field respectively) in the MACS0416 cluster and the parallel field. }
\label{fig:completeness_curves}
\end{figure}

Fig.~\ref{fig:completeness_curves} shows the recovery fraction as a function of input magnitudes for the MACS0416 cluster and the parallel field both with disc-like profile galaxies and unresolved point sources. As expected, we find that with point sources the completeness goes deeper. On the other hand, not all the sources are not recovered even at brighter magnitudes due to confusion and blending with nearby sources.

\subsubsection{Selection function}
\label{sec:selecfunc}
To determine the selection function efficiency, we build a custom mock catalog of high redshift galaxies using the Theoretical Astrophysical Observatory (TAO) \citep{Bernyk2016}. The TAO is an online cloud-based virtual laboratory that allows web based access to mock observations of extragalactic survey data. To build our mock photometry catalog, we use the existing CANDELS mock cone from redshift $z=0$ to $z=9$ on the TAO and generate SEDs from the single stellar populations of \citet{Bruzual2003}, using the initial mass function (IMF) of \citet{chabrier2003}. Dust is then applied using the dust model of \citet{Calzetti2000} and the final catalog is made with a H-band distribution of magnitudes in the range $21<H_{160}<35$, such that the magnitudes in the rest of the filters are determined based on the range of H band magnitudes.

In order to generate errors for the mock photometry catalog, we bin the fluxes of sources from our real catalog and calculate the standard deviation and mean of the flux errors in those bins, giving us a Gaussian distribution of the errors. Photometric errors for each filter in the mock catalog were then simulated by picking random errors each time from the Gaussian distribution in the corresponding bins. Furthermore, we ensure that the colors of the mock galaxies are a sufficiently close representation of the colors of real galaxies in our sample. For eg., at $z=9$, our catalog had brighter galaxies as compared to the mock catalog. Therefore, as there were not enough bright galaxies intrinsic to the catalog, we rescaled the faint sources of the mock catalog to the magnitudes we need so that the mock data samples our real data very well (See also \citet{Duncan2014} who use CANDELS mock cones similar to us and compare the color distributions of the sources). Following these steps, we run EAZY on this mock catalog and compute the photometric redshifts.

\begin{figure*}
\begin{minipage}{1\textwidth}
\centering
\includegraphics[width=1.1\textwidth, height=0.3\textwidth]{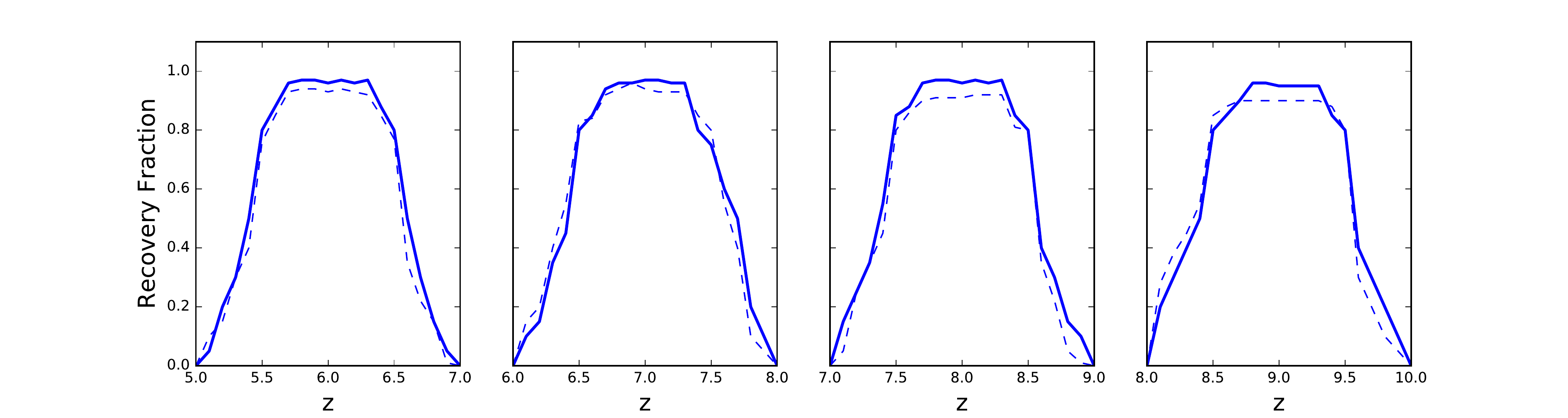}
\end{minipage}
\caption{Selection efficiencies for the MACS0416 cluster, showing recovery fraction as a function of redshift and input magnitude. The solid line is for objects with $H_{160}=25$ and the dashed line is for $H_{160}=27$, finding very similar results. }
\label{fig:select_func}
\end{figure*}

Finally, to determine the selection efficiency for each of the redshift bins, we pass the high redshift sources from the mock catalog through the same sample selection criteria (See Section~\ref{sec:selectioncriteria}) as our real sample. From this, we measure the fraction of simulated galaxies that pass the selection criteria. This is then folded into the detection completeness computed in Section~\ref{sec:detect_comp}. Fig.~\ref{fig:select_func} shows the selection efficiencies for the MACS0416 cluster.

\subsection{Stellar masses, rest-frame magnitudes and star formation rates}
\subsubsection{Stellar masses and rest-frame magnitudes}
\label{sec:sedfitting}
To examine the evolution of the GSMF and UV LF from $z=6-9$, we first estimate the stellar masses and rest-frame magnitudes for our sample in different redshift bins for both the cluster and the parallel field using a stellar population fitting technique described in detail in \citet{Duncan2014}. 

Briefly, we compute the stellar masses and rest-frame magnitudes using a custom template-fitting routine \textbf{SMpy}\footnote[6]{https://github.com/dunkenj/smpy} \citep{Duncan2014}, which matches the observed spectral energy distribution (SED) of each source to one of the models it generates. First, single stellar population models of \citet{Bruzual2003} are used to generate synthetic SEDs for a user-defined combination of parameters such as metallicity, age and star formation history (SFH), assuming a Chabrier IMF \citep{chabrier2003}. The ages are allowed to vary from 5 Myr up to the age of the Universe at the redshift step being fit, dust attenuation is varied in the range $0\leq A_{v}\leq2$, and metallicities of 0.02, 0.2 and 1 Z$_{\odot}$ are used. The widely adopted parametrization of the SFH (SFR $\propto e^{-t/\tau}$) is used with $\tau=0.05,0.25,0.5,1,2.5,5,10,-0.25,-0.5,-1,-2.5,-5,-10$ and $1000$ (constant SFR) Gyr, (similar to \citealt{Duncan2014}), where negative $\tau$ values represent exponentially increasing histories.

Galaxy properties such as age, mass and star formation rate deduced from SED fitting are affected to different degrees depending on the inclusion (or not) of nebular emission. \citet{Schaerer2012} discussed the importance of accounting for nebular emission in the SEDs of high redshift galaxies while analysing a large sample of Lyman break galaxies from $z\sim3-6$ and found that the majority of objects were better fit with SEDs that include nebular emission, irrespective of star formation history. Several other studies have also shown that significantly younger ages and lower masses could be obtained when nebular emission lines are included in SED fitting (e.g., \citealt{Schaerer2009,Schaerer2010,Ono2010,McLure2011,Duncan2014}). It has therefore become evident that nebular emission must be taken into account while understanding the photometric measurements of the SEDs of star-forming galaxies at high redshift.

Hence, we apply nebular emission lines on the model SEDs in this work, assuming an escape fraction of $f_{\mathrm{esc}}=0$ (See \citealt{Duncan2014} for a detailed description of the method for including nebular emission lines). Dust extinction is then applied using the law described by \citet{Calzetti2000}. We then redshift each model SED in the range $0\leq z\leq11$ in steps of $\triangle z=0.02$ and apply the attenuation by neutral hydrogen according to \citet{Madau1995}. Finally, each model spectrum is convolved through the photometric filters and the resulting SED grid is fit to the observed photometry. For each model, we compute the absolute magnitude at 1500 angstrom  by fitting a 100 \AA -wide top-hat filter centered on 1500 \AA. The model SEDs are fitted to the observed photometry using a Bayesian-like approach, resulting in a likelihood distribution of stellar mass and rest-frame magnitudes. The stellar masses and rest-frame magnitudes thus are computed by summing the likelihoods. 

The uncertainties of the stellar masses and rest-frame magnitudes are computed via a Monte Carlo analysis. For this, the observed flux of each source was perturbed by randomly choosing a point from a Gaussian distribution with standard deviation equal to the 1$\sigma$ uncertainty on the flux in a given filter. The stellar mass and rest-frame magnitude for each source is then estimated with the simulated photometry. This process is repeated 500 times and the final uncertainty is the standard deviation of the distribution of these 500 values. We employ the same method to account for errors on photometric redshifts on the estimation of stellar masses. The uncertainty due to photometric redshifts is then folded into the uncertainty from flux measurement by summing both errors in quadrature. This is then taken as the final uncertainty on stellar mass, which are found to be $\sim0.2-0.3$ dex.

\subsubsection{Star formation rates}
To calculate the UV SFR, we use the rest-frame absolute magnitudes ($M_{1500}$) obtained from the SED fitting in Section~\ref{sec:sedfitting}. We first convert the absolute magnitudes to UV luminosities, and then use the \citet{Kennicutt1998} relation that converts the UV luminosities to star formation rates assuming a Chabrier IMF:

\begin{equation}
\mathrm{SFR_{UV}(M_{\odot}yr^{-1})=1.4\times10^{-28}L_{UV}}(\mathrm{erg}\mathrm{s^{-1}Hz^{-1})}.
\end{equation}

\noindent This, however, does not take into account the dimming of light caused by dust, which can significantly affect the measurements of SFRs. In order to study the effects of dust on the UV continuum, \citet{Meurer1999} fitted the observed UV continuum of a sample of local starburst galaxies to a power law expressed by,

\begin{equation}
f(\lambda)\propto\lambda^{\beta},
\end{equation}

\noindent where $f(\lambda)$ is the observed flux density and $\beta$ is the power-law index. Likewise, we measure the UV slope ($\beta$) by fitting a power-law to each model spectrum (See Section~\ref{sec:sedfitting}) using the 10 windows defined by \citet{Calzetti1994} and calculate the dust extinction using the \citet{Meurer1999} relation as:

\begin{equation}
A_{1600}=4.43+1.99\beta,
\end{equation}

\noindent which relates the UV slope $\beta$ and the dust extinction at 1600 \AA. We then use this to calculate the final dust-corrected SFRs.

\subsection{Lensing magnification}
\label{sec:lensing}
Reliable lensing models are required in order to interpret many of the properties of background lensed galaxies accurately. For the HFF, the lens models were produced by seven independent teams for all six clusters using different methods, assumptions and software. The primary difference between the lensing models is that some assume that the cluster mass substructure is traced by the luminous cluster galaxies, while others make no assumption about light tracing mass, and their models are instead solely constrained by lensing observables, and thus probe a broader range of possible mass distributions. We refer the reader to the MAST website \footnote[7]{https://archive.stsci.edu/prepds/frontier/lensmodels/} for more details of the lensing models. Five models that assume that light-traces-mass are: CATS \citep{Jauzac2014}, Sharon \citep{Johnson2014}, GLAFIC \citep{Oguri2010,Ishigaki2015}, Zitrin-NFW and Zitrin-LTM \citep{Zitrin2013}. The rest of the models that work without this assumption are provided by Bradac \citep{Bradac2009}, Williams \citep{Grillo2015} and Merten \citep{Merten2011}. 

In order to demagnify the masses and the rest-frame magnitudes, we first need to compute the per pixel magnification factor. Each lens modelling team has provided maps of mass surface density (kappa), and weak lensing shear (gamma) from which magnifications at any redshift can be derived. As a first step, we regrid the gamma and kappa maps to match the \textit{HST} pixel scale and then use these maps to calculate the magnification values at each pixel using, 

\begin{equation}
\mu=\frac{1}{(1-\kappa)^{2}-\gamma^{2}},
\end{equation}

\noindent where $\kappa$ and $\gamma$ both scale with the distance ratio $D_{LS}/D_{S}$, $D_{LS}$ being the angular diameter distance between the source and the cluster, and $D_{S}$ the angular diameter distance between the observer and the source. Finally, we calculate the median magnification value, $\mu_{\mathrm{med}}$, for all eight different lensing models in order to exclude possible outliers. This median magnification value was used to demagnify the masses and the rest-frame magnitudes for our sample in each redshift bin for the MACS0416 cluster (we do not demagnify the masses for parallel field). 

We find that our subtraction procedure and the lensing effect allows us to probe stellar masses as low as $10^{6.8}M_{\odot}$; lower than previous studies (e.g., \citealt{Gonzalez2011, Duncan2014, Grazian2015, Song2016}) at $z\sim6$ (See Fig.~\ref{fig:gsmf}).  Similarly, we are probing masses as low as $10^{7.4}M_{\odot}$, $10^{7.8}M_{\odot}$, and $10^{8.3}M_{\odot}$ at $z\sim7$, $z\sim8$, and $z\sim9$ respectively. With these methods we also find that we are able to probe magnitudes as faint as $M\mathrm{_{UV}=-13.5}$, $M\mathrm{_{UV}=-15.5}$, $M\mathrm{_{UV}=-15.6}$ and $M\mathrm{_{UV}=-18.8}$ at $z\sim6$, $z\sim7$, $z\sim8$ and $z\sim9$ respectively.

\subsection{Number densities}
\label{sec:number_densities}
\subsubsection{Volume estimation}
To estimate the volume for our luminosity and mass functions we use an enhancement of the $1/V_{\mathrm{max}}$ method \citep{Schmidt1968} introduced by \citet{Avni1980} and treat our high redshift galaxies as a ``coherent' sample (nomenclature of \citet{Avni1980}) including the cluster and the parallel field with their corresponding depths, following \citet{Avni1980}, \citet{Eales1993}, \citet{Ilbert2005} and \citet{Duncan2014}.

Strong gravitational lensing with massive clusters enable the detection of faint galaxies by providing a magnified boost to the light from background sources. However, the disadvantage is that a high magnification region will essentially reduce the survey volume. To account for this,  we calculate the maximum observable comoving volume in which a galaxy can remain detectable as:

\begin{equation}
V_{\mathrm{obs,i}}=\sum_{k}^{N_{\mathrm{fields}}}\int_{z1,k}^{z2,k}\frac{\mathrm{d\mathit{V}}}{\mathrm{d\mathit{z}}}f(z,m,\mu)\mathrm{d\mathit{A_{k}(\mu,z)\mathrm{d\mathit{z}}}}
\end{equation}

\noindent where the summation, $k$, is over the cluster and the parallel field with their corresponding survey area, $\mathrm{d\mathit{A_{k}}}$, $z_{1,k}$, $z_{2,k}$ are the integration limits, $f$ is the completeness function, and $\mathrm{d}V/\mathrm{d}z$ is the comoving volume. For the cluster, the survey area $\mathrm{d\mathit{A}}$ is the delensed survey area in the source plane, which is a function of the magnification computed in Section~\ref{sec:lensing} and redshift, whereas for the parallel field the survey area is just the survey area without the effects of lensing. Similarly, the completeness function $f$, which includes the corrections for incompleteness and the selection function calculated in Section~\ref{sec:comp_sim}, is a function of redshift, magnitude $m$ and magnification factor $\mu$ for the cluster, but does not incorporate the effects of lensing for the parallel field. The integration limits are given by $z_{1,k}=z_{\mathrm{min}}$ and $z_{2,k}=\mathrm{min\{\mathit{z_{\mathrm{max}},}\mathit{z\mathrm{(}z_{\mathrm{j}},}\mathit{m_{\mathrm{j}},m_{\mathrm{max,k}}\mathrm{)}\mathrm{\}}}}$
, where $z_{\mathrm{min}}$ and $z_{\mathrm{max}}$ are the minimum and maximum redshift to which the source could be pushed and still be included in the sample (e.g. \citealt{Avni1980, Eales1993, Ilbert2005, Duncan2014}). The function, $z(z_{\mathrm{j}},m_{\mathrm{j}},m_{\mathrm{max,k}})$, gives the maximum redshift at which a source of apparent magnitude $m_{j}$, observed redshift $z_{j}$, could be observed and included in the sample given the depth $m_{\mathrm{max,k}}$ of the field/cluster. So for e.g. for the $z=9$ sample ($8.5<z<9.5$), $z_{\mathrm{min}}=8.5$ and $z_{\mathrm{max}}=9.5$ in the case of the object being still sufficiently bright to be detected beyond this redshift.

Once we have the estimated volume, the  number density in each magnitude (or mass) bin is calculated as:

\begin{equation}
\phi(M)\mathrm{d\mathit{M=\sum_{i}^{N\mathrm{gal}}\frac{\mathrm{\mathit{N_{i}}}}{V_{\mathrm{obs,i}}}}}
\end{equation}

\subsubsection{Errors on number densities}
To calculate the errors on number densities, we perform a Monte Carlo analysis in which the stellar mass (or rest-frame UV magnitude) and redshift of each galaxy is varied along a Gaussian distribution with standard deviation equal to the 1$\sigma$ uncertainty of the stellar masses (or rest-frame UV magnitudes) and redshifts. We then repeat the number density calculation explained in this section on 500 simulated redshifts and stellar masses (or rest-frame UV magnitudes) to obtain simulated number densities. The standard deviation of these values then is the uncertainty  due to measurement error. We then calculate the Poisson uncertainty for the number counts and add these in quadrature to the uncertainty from the simulations. 

Furthermore, the choice of lensing maps impacts the magnification factor, the survey area and hence the effective volume. This can introduce large uncertainties, in particular at the very faint end of the UV LF/GSMF where magnification factors exceed 10x. These uncertainties can reach a factor >2x on the selection volume as shown in \citet{Atek2018} and therefore needs to be accounted for. For this, we calculate the effective volume and hence the number densities explained in this section using all eight lensing models used in this study. The standard deviation of these values is then the uncertainty due to lensing maps and is added in quadrature to the uncertainties from simulations and the Poisson uncertainty calculated above, which is then taken as the final uncertainty on number densities.

\section{Results}
\label{sec:results}
\subsection{Ultra-Violet luminosity function}

One of the primary and popular methods for comparing and understanding galaxy evolution is to use the rest-frame Ultraviolet (UV) luminosity function (LF) and to measure its evolution (e.g., \citealt{McLure2013, Finkelstein2015, Bouwens2015}). This is largely because the UV luminosity function is the most straightforward way of measuring a galaxy property and determining how it changes with time. The UV LF, which gives the number densities of galaxies at different UV luminosities, is the simplest measure of galaxies at these redshifts, which requires the fewest assumptions.  Furthermore, the rest-frame ultraviolet luminosity function (as well as the stellar mass function) allows us to answer fundamental questions about the way the first galaxies and stars formed. This is becoming even more true with the advent of specific predictions for the shape and normalization of the $z>8$ UV LF (and MF).

There are many different physical processes that are responsible for the formation of the first galaxies, and thus the creation of UV light, including the metallicity of the gas and stars, the dust content and form of extinction laws, various forms of feedback, the density of gas, as well as perhaps magnetic fields. The predictions for the formation of these first galaxies can vary significantly depending on the assumptions, and the luminosity function is a powerful approach for understanding this issue, as models of its distribution are degenerate at lower redshifts, but differ significantly at higher redshifts.

Previous studies of luminosity functions at a wide range of redshifts have shown that the number densities of galaxies as a function of luminosity can be characterized by a Schechter function \citep{Schechter1976} of the form:

\begin{equation}
\phi(L)=\Phi^{\ast}(\frac{L}{L^{\ast}})^{\alpha}\exp(-\frac{L}{L^{\ast}}).
\end{equation}

The Schechter function can also be expressed in log space as, 

\begin{equation}
\phi(M)=0.4ln(10)\phi^{*}10^{-0.4(M-M^{*})(\alpha+1)}e^{-10^{-0.4(M-M^{*})}},
\end{equation}

\noindent where $\phi^{\ast}$ provides the normalization, $M^{\ast}$ (referred to as the knee of the Schechter function) corresponds to the characteristic magnitude at which the function turns over from a power law into an exponential form and $\alpha$ is the faint-end slope.

\subsubsection{Best fit Schechter parameters and their uncertainties for UV LF}
\label{sec:mcmc}
Once we have the number densities (See Section \ref{sec:number_densities}) in each luminosity bin, we proceed to determine the rest-frame UV luminosity function at $z=6, 7, 8$ and $9$ by fitting a Schechter function as defined above to the number densities in each redshift bin. 

To determine the best fit Schechter parameters and their uncertainties for UV LF, we use a pure-Python implementation of Goodman \& Weare's Affine Invariant Markov chain Monte Carlo (MCMC) Ensemble sampler \citep{Goodman2010,Foreman2013} that examines the three-dimensional parameter space of Schechter parameters. For each redshift, we use $10^{2}$ MCMC chains of $10^{4}$ steps each to explore the full parameter space, building a distribution of $M_{\mathrm{UV}}^{*}$, $\Phi^{*}$ and $\alpha$. For the priors, we limit the parameter space to $-24<M_{\mathrm{UV}}^{*}<-17$, log$(\phi^{*}/\mathrm{Mpc^{-3})>-8}$ and $\alpha>-4$. However at $z=9$, where the sample size is small, we fix $M_{\mathrm{UV}}^{*}$ to the value estimated at $z=8$. 

As a first step, we compute the likelihood function, which is simply a Gaussian, and numerically optimize it. The starting position of the chain(s) is then a small Gaussian ball around the maximum likelihood result. The chains start in small distributions around the maximum likelihood values and then they quickly diverge and start exploring the full posterior distribution. To minimize the dependence of the posterior distribution on the starting point, we discard the first 10 per cent of steps in the burn-in phase before running each chain. For our final result, we join the chains together giving a distribution of $10^{6}$ values of Schechter function parameters at each redshift. The best-fit values for each Schechter function parameter are the median of this distribution, with the uncertainties covering the central 68 per cent of the distribution. 

\begin{figure*}
\centering
\begin{minipage}{0.49\textwidth}
\centering
\includegraphics[width=1\textwidth, height=0.3\textheight]{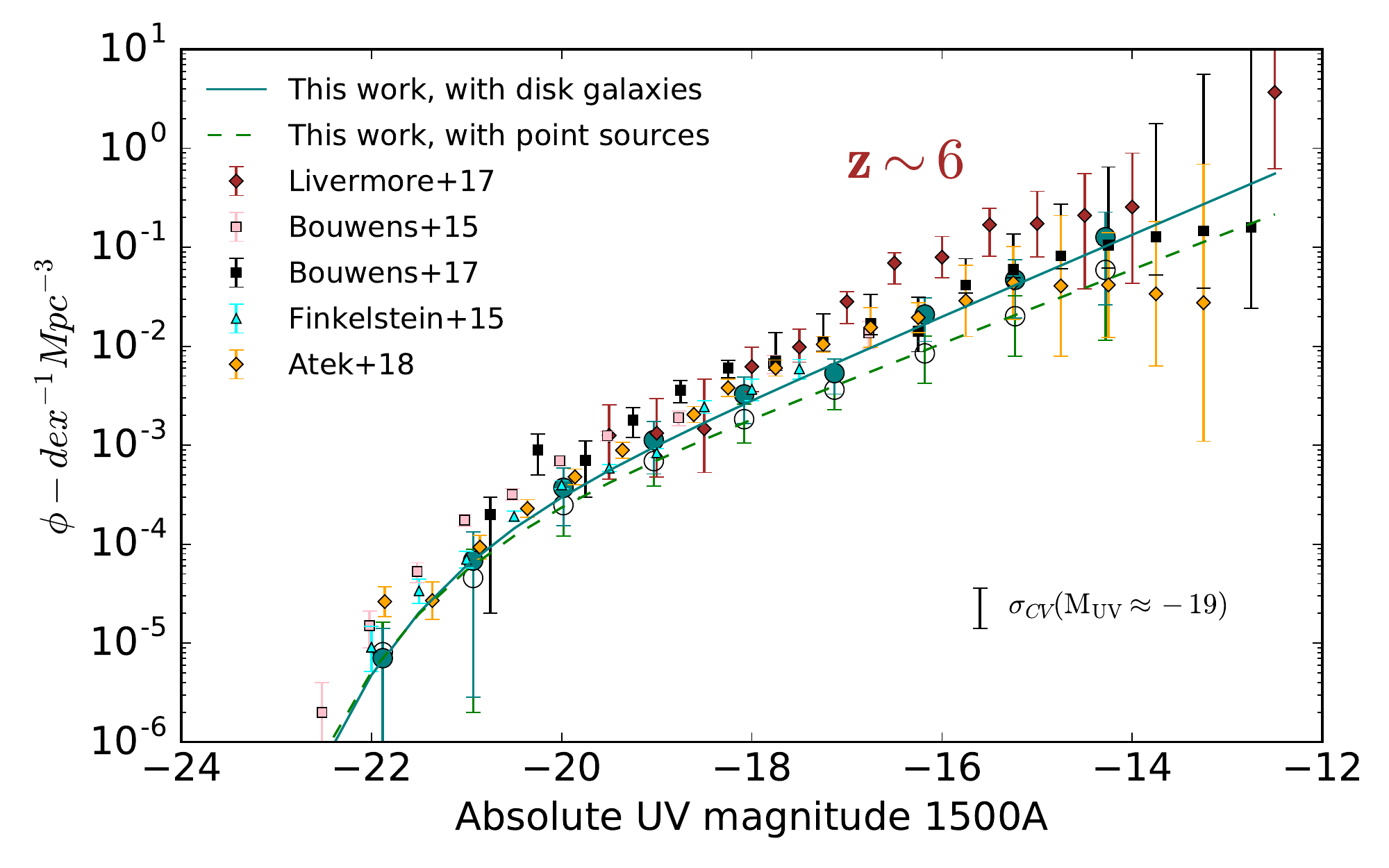}
\end{minipage}
\begin{minipage}{0.49\textwidth}
\centering
\includegraphics[width=1\textwidth, height=0.3\textheight]{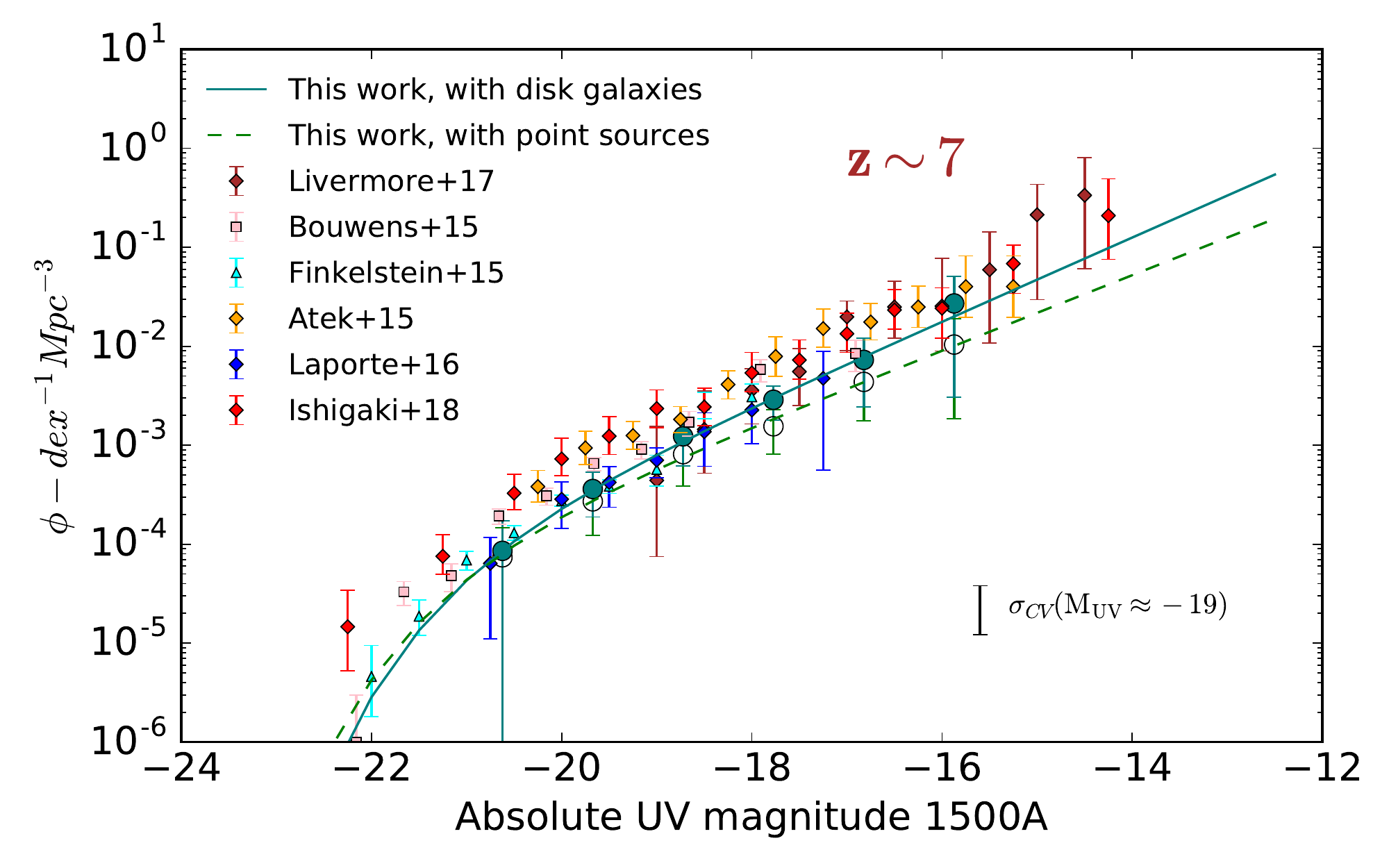}
\end{minipage}
\begin{minipage}{0.49\textwidth}
\centering
\includegraphics[width=1\textwidth, height=0.3\textheight]{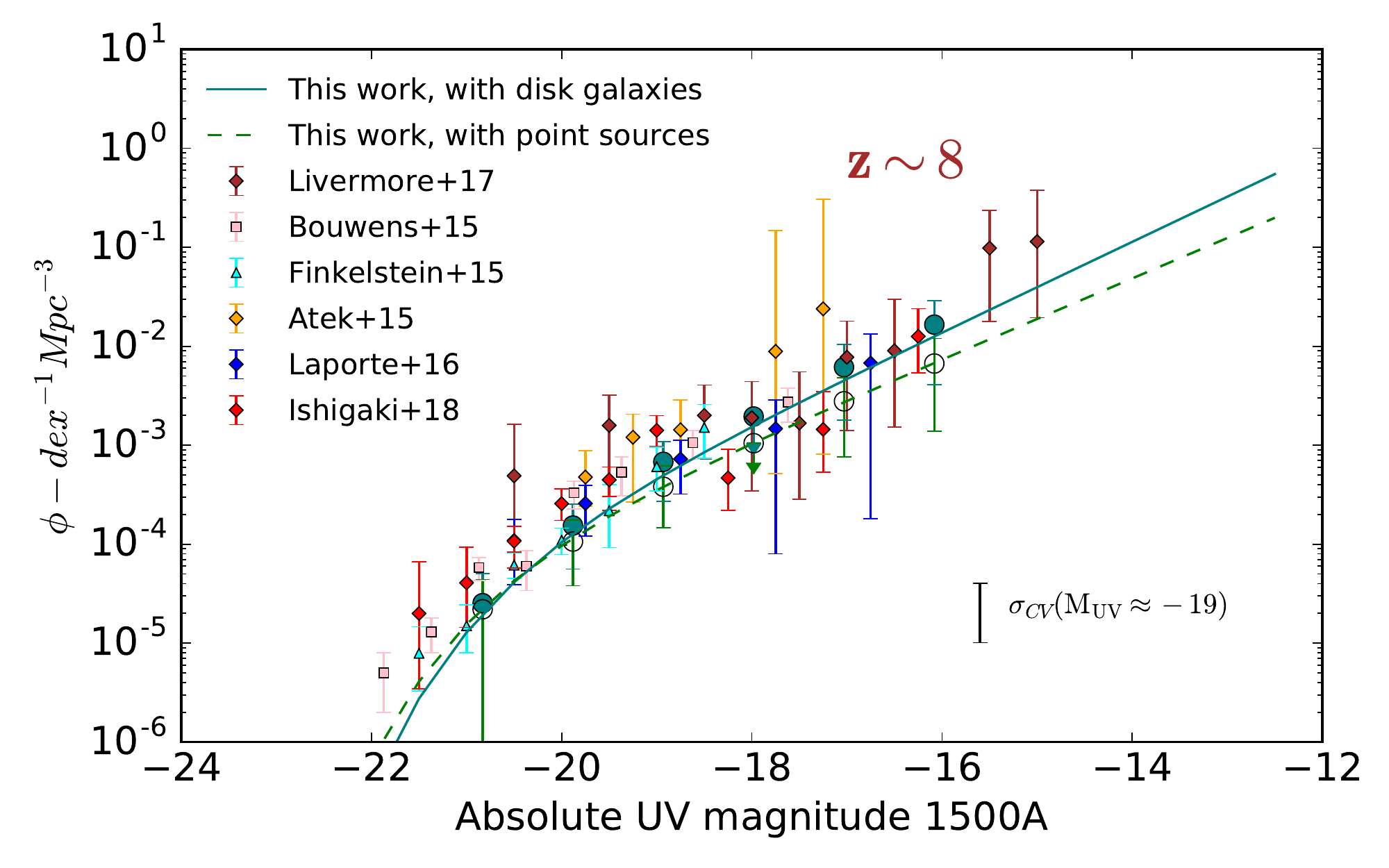}
\end{minipage}
\begin{minipage}{0.49\textwidth}
\centering
\includegraphics[width=1.01\textwidth, height=0.3\textheight]{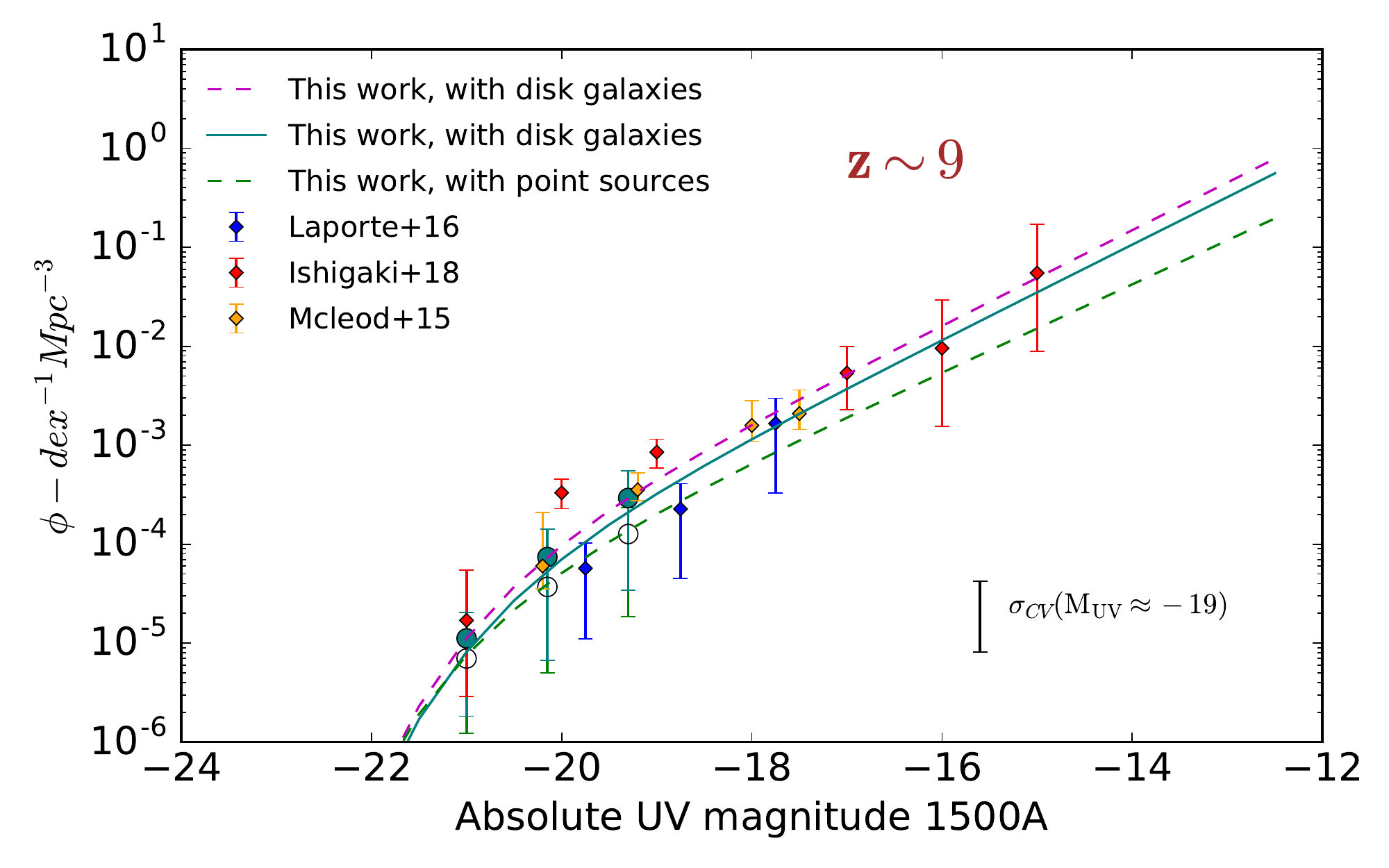}
\end{minipage}
\caption{The UV LF at $z=6-9$ in the MACS0416 cluster and its parallel field. The solid green line is our best fit Schechter function derived from the MCMC analysis to the green circles, using disc-like galaxies in completeness simulations. The dashed green line is our best fit Schechter function from the MCMC analysis to the open green circles, using point sources. At $z=9$, the dashed magenta line shows the best fit Schechter function derived using a simple chi squared minimization technique, using disk-like galaxies. The error bars on our data points take into account the errors on photometric redshifts, errors on magnitudes, Poisson errors and uncertainties due to lensing maps, but does not include the errors due to cosmic variance. The representative error bar in the bottom right corner of each plot shows the measured error on number densities for galaxies of $\mathrm{M_{UV}\approx-19}$ due to cosmic variance. }
\label{fig:uvlf}
\end{figure*}

In order to investigate the effect of galaxy sizes on completeness corrections and hence on the faint-end slopes, we derive the best-fit Schechter parameters and their uncertainties for our UV luminosity functions in this way using the completeness curves described in Section \ref{sec:comp_sim}, with both disc-like galaxies and point sources. Our results, along with the values from the literature, are listed in Table~\ref{tab:uvlftable} and the resulting UV luminosity functions plotted along with previous work are shown Fig.~\ref{fig:uvlf}. At $z=9$, we also show the best-fit Schechter function derived using a simple chi squared minimization technique using disk-like galaxies. The error bars on our data points in Fig.~\ref{fig:uvlf} take into account the errors on photometric redshifts, errors on magnitudes, Poisson errors and uncertainties due to lensing maps, but does not include the errors due to cosmic variance. 

To estimate the fractional uncertainty in the number densities due to cosmic variance, we use the cosmic variance calculator tool by \citet{Trenti2008} using the estimated completeness and survey area as inputs. The average excess uncertainty in the moderately lensed area (e.g., $\sim1-10$x magnification) is typically only $10-30$ per cent higher than an equivalent blank field (Brant Robertson, private discussion). We therefore estimate the cosmic variance of our sample by computing the blank field cosmic variance and then scaling the rms cosmic variance by $\sim1.2$. For the highest magnification objects, we compute the effective area and volume for their magnification and then estimate the cosmic variance as if the sources were in a blank field of that (much smaller) size. Finally, since we are observing two fields, we estimate the joint uncertainty on the combined sample by adding the uncertainties in quadrature. We find that the cosmic variance errors span from $30-90$ per cent from the low luminosity/mass end to the high luminosity/mass end in all redshift bins. In Fig.~\ref{fig:uvlf} we show the measured error on number densities for galaxies of $\mathrm{M_{UV}\approx-19}$.

We find through these Schechter fits that the faint-end slope $\alpha$ of the UV LF becomes steeper at higher redshifts, such that the ratio of lower to higher luminosity galaxies increases back to where we can measure these systems. The faint-end slope $\alpha$ changes from $-2.03_{-0.10}^{+0.12}$ at $z=6$ to $-2.20_{-0.47}^{+0.51}$ at $z=9$ using disc-like galaxies in our completeness simulations. These measurements are consistent with a continuation of the trends seen from lower redshift, however more data is needed to confirm the trend at these redshifts. \citet{Bouwens2017} suggest shallower faint end slopes than what has been derived in some recent studies (by $\Delta\alpha\gtrsim0.1-0.3$) when simulated galaxies for completeness have near point-source profiles. When point sources are considered in our analysis, we find that the faint-end slopes are indeed shallower ($\Delta\alpha\sim0.09-0.12$, See Table~\ref{tab:uvlftable}) in the redshift range probed.  We note that these systematic offsets in slope are less than those estimated by \citet{Bouwens2017} but we can attribute this to the difference in assumed profiles for input galaxies with our broader range of assumed S\'ersic indices resulting in a higher completeness for the same half-light radii (i.e. $0.5\leq n\leq4.0$, vs $n = 1$ for \citet{Bouwens2017}). We also find a decrease in $\Phi^{*}$ with increasing redshift at $z=6-9$, and a slight evolution in $M_{\mathrm{UV}}^{*}$ is also observed, with it decreasing with redshift from $z=6-8$.

  \begin{table}
	 \centering
	 \renewcommand{\arraystretch}{1.2}
	 \setlength{\tabcolsep}{3pt}
	  \caption{Best fit Schechter function parameters and their uncertainties for our UV LFs. The quoted best fit values and 1$\sigma$ errors of the Schechter parameters constitute the median and the central 68 per cent of posterior distribution of each parameter from our MCMC analysis. \textbf{Note}: We only show the Schechter function parameters and their uncertainties derived using a classical Schechter function fit for \citet{Atek2018} and \citet{Bouwens2017b} and do not show the values they derive with a modified Schechter function that allows for a curvature at very faint magnitudes.}
	 \label{tab:uvlftable}
	 \begin{tabular}{lccr} 
		 \hline
		 Redshift & $M_{\mathrm{UV}}^{*}$  & $\alpha$ & $\phi^{*}(10^{-4} \mathrm{Mpc^{-3})}$\\
		 \hline
		 $z\sim6$ & & &\\
		 This work (disc galaxies) & $-20.94_{-0.26}^{+0.31}$ & $-2.03_{-0.10}^{+0.12}$& $2.01_{-0.60}^{+0.85}$\\
		 This work (point sources) & $-21.00_{-0.30}^{+0.36}$ & $-1.93_{-0.11}^{+0.11}$& $1.61_{-0.62}^{+0.88}$\\
		 Livermore et al. (2017) & $-20.83_{-0.04}^{+0.05}$ & $-2.10_{-0.03}^{+0.04}$& $2.23_{-0.100}^{+0.273}$\\
		 Finkelstein et al. (2015) & $-21.13_{-0.31}^{+0.25}$ & $-2.02_{-0.10}^{+0.10}$& $1.86_{-0.80}^{+0.94}$\\
		 Bouwens et al. (2015) & $-20.94_{-0.20}^{+0.20}$ & $-1.87_{-0.10}^{+0.10}$& $5.00_{-1.6}^{+2.2}$\\
		 Bouwens et al. (2017) $^\textrm{a}$ $^\textrm{b}$ & $-20.94$ (fixed) & $-1.91 _{-0.02}^{+0.02}$& $6.6_{-0.4}^{+0.4}$\\
		 Atek et al. (2018) $^\textrm{a}$ &$-20.74_{-0.20}^{+0.21}$ & $-1.98_{-0.09}^{+0.11}$& $3.72_{-1.43}^{+2.31}$\\
		\hline
		 $z\sim7$ & & &\\
		This work (disc galaxies) & $-20.85_{-0.38}^{+0.40}$ & $-2.06_{-0.15}^{+0.17}$& $1.70_{-0.70}^{+1.20}$\\
		 This work (point sources) & $-21.04_{-0.40}^{+0.45}$ & $-1.95_{-0.18}^{+0.20}$& $1.20_{-0.51}^{+1.00}$\\
		 Livermore et al. (2017) & $-20.80_{-0.05}^{+0.06}$ & $-2.06_{-0.05}^{+0.05}$& $2.13_{-0.188}^{+0.260}$\\
		 Finkelstein et al. (2015) & $-21.03_{-0.50}^{+0.37}$ & $-2.03_{-0.20}^{+0.21}$& $1.57_{-0.95}^{+1.49}$\\
		 Bouwens et al. (2015) & $-20.87_{-0.26}^{+0.26}$ & $-2.06_{-0.13}^{+0.13}$& $2.9_{-1.2}^{+2.1}$\\
		 Atek et al. (2015) & $-20.89_{-0.72}^{+0.60}$ & $-2.04_{-0.13}^{+0.17}$& $2.88_{-1.86}^{+5.82}$\\
		 Laporte et al. (2016) & $-20.33_{-0.47}^{+0.37}$ & $-1.91_{-0.27}^{+0.26}$& $3.7_{-1.1}^{+1.2}$\\
		 Ishigaki et al. (2018) & $-20.89_{-0.13}^{+0.17}$ & $-2.15_{-0.06}^{+0.08}$& $1.65_{-0.484}^{+0.684}$\\
		 
		 \hline
		 $z\sim8$ & & &\\
		 This work (disc galaxies) & $-20.40_{-0.54}^{+0.52}$ & $-2.14_{-0.32}^{+0.37}$& $1.49_{-0.83}^{+1.83}$\\
		 This work (point sources) & $-20.58_{-0.58}^{+0.57}$ & $-2.02_{-0.36}^{+0.40}$& $1.09_{-0.62}^{+1.58}$\\
		 Livermore et al. (2017) & $-20.72_{-0.14}^{+0.18}$ & $-2.01_{-0.08}^{+0.08}$& $1.69_{-0.439}^{+0.592}$\\
		 Finkelstein et al. (2015) & $-20.89_{-1.08}^{+0.74}$ & $-2.36_{-0.40}^{+0.54}$& $0.72_{-0.65}^{+2.52}$\\
		 Bouwens et al. (2015) & $-20.63_{-0.36}^{+0.36}$ & $-2.02_{-0.23}^{+0.23}$& $2.08_{-1.1}^{+2.3}$\\
		 Laporte et al. (2016) & $-20.32_{-0.26}^{+0.49}$ & $-1.95_{-0.40}^{+0.43}$& $3.01_{-1.9}^{+8.5}$\\
		 Ishigaki et al. (2018) & $-20.35_{-0.30}^{+0.20}$ & $-1.96_{-0.15}^{+0.18}$& $2.5_{-1.25}^{+1.03}$\\
		 \hline
		 $z\sim9$ & & &\\
		 This work (disc galaxies) & $-20.40$ (fixed) & $-2.20_{-0.47}^{+0.51}$& $0.98_{-0.60}^{+1.65}$\\
		 This work (point sources) & $-20.58$ (fixed) & $-2.11_{-0.47}^{+0.56}$& $0.55_{-0.35}^{+1.10}$\\
		  Laporte et al. (2016) & $-20.45$ (fixed) & $-2.17_{-0.43}^{+0.41}$& $0.70_{-0.30}^{+0.30}$\\
		 Ishigaki et al. (2018) & $-20.35$ (fixed) & $-1.96$ (fixed) & $1.31_{-0.318}^{+0.266}$\\
		   Mcleod et al. (2015) & $-20.1$ (fixed) & $-2.02$ (fixed) & $2.51_{-1.39}^{+1.46}$\\
		 \hline
		 \hline
		 \multicolumn{4}{l}{ $^\textrm{a}$ Using classical Schechter function} \\
		 \multicolumn{4}{l}{ $^\textrm{b}$ Using the CATS model} \\
	 \end{tabular}
 \end{table}

\subsubsection{Comparison with previous work}

As can be seen from Fig.~\ref{fig:uvlf}, there are some differences and similarities between our work and previous work\footnote[8]{Note: We restrict the comparison of our results with previous work to the best-fit Schechter parameters derived using disc-galaxies in our completeness simulations, and do not compare our results derived using point sources.}. First, at $z\sim6$ we observe a disagreement in the binned luminosity function values at various luminosities for the number densities across the entire luminosity range. We, for example, do not see the upturn in the luminosity function near $M_{\mathrm{UV}}=-18$ as seen by \citet{Livermore2017}.  We, however, also do not find the possible downturn that \citet{Atek2018} and \citet{Bouwens2017b} find in the UV luminosity function near $M_{\mathrm{UV}}=-15$. Instead, we find that at $z\sim6$ the luminosity function is well fit by a Schechter function, as well as a single power-law, up to the faintest limits we go, $M_{\mathrm{UV}}=-13.5$. Similarly, looking at the faint-end of the LF, we find a difference in the estimated error bars between our work and \citet{Atek2018} and \citet{Bouwens2017b}. Our errors bars are likely underestimated due to the fact that we have only used the median magnification factors to demagnify the magnitudes for individual sources and also have not included the errors due to cosmic variance in our analysis.

At this redshift, we generally find that our binned luminosity function values are lower than previous studies. This can be attributed to the significantly smaller volume that we are probing as compared to \citet{Finkelstein2015}, \citet{Bouwens2015}, \citet{Bouwens2017b} and \citet{Atek2018} (\citet{Bouwens2015} utilize data sets from CANDELS, HUDF09, HUDF12, ERS, and the BoRG/HIPPIES programs \citep{Trenti2011}, \citet{Bouwens2017b} combine the data from the first four HFF clusters; Abell 2744, MACS 0416, MACS 0717, and MACS 1149, \citet{Finkelstein2015} use data sets from CANDELS/GOODS, HUDF, the parallel fields near MACS0416 and Abell 2744 from the HFF program, whereas \citet{Atek2018} combine the data sets of all 6 clusters from the HFF program). The difference in binned luminosity function values with \citet{Livermore2017} is, however, unclear since they are probing a similar volume as our study (They combine data from the Abell 2744 and the MACS0416 clusters). Further discrepancy is observed in the Schechter function parameters. \citet{Livermore2017} are able to probe magnitudes as faint as $M_{\mathrm{UV}}=-12.5$, with higher number densities resulting in a steeper faint-end slope than ours. Comparing our detections with \citet{Livermore2017}, we find that their faintest source at $M_{\mathrm{UV}}=-12.5$ lie in the Abell 2744 cluster, which we have not included in our analysis in this study, and the next faintest source at $M_{\mathrm{UV}}=-14$ is in the MACS0416 cluster, but is not recovered in our analysis (it appears to be a possible noisy detection). In the case of $M_{\mathrm{UV}}^{*}$ values, we are unable to put robust constraints on $M_{\mathrm{UV}}^{*}$ as there are very few galaxies in the brightest bins ($\sim1-3$), given the smaller volume we are probing in this study.

At $z\sim7$, our binned luminosity function values are in excellent agreement with \citet{Laporte2016} who combine data from the HFF MACSJ0717 cluster and its parallel field. However, our binned luminosity function values are lower as compared to other studies, once again possibly because of the smaller volume that we are probing. We also find that our faint-end slope is generally in agreement with previous studies, except \citet{Laporte2016} and \citet{Ishigaki2017} who are finding significantly shallower and steeper faint-end slopes respectively. This is because we are going fainter than \citet{Laporte2016} at this redshift, whereas \citet{Ishigaki2017} are probing $\sim2$ magnitudes fainter than us and finding higher number densities with the complete HFF data, resulting in observed differences in the measured values of the faint-end slopes.

Comparing to previous studies at $z\sim8$, we find large discrepancies due to smaller number statistics in all the studies. For example, there are inconsistencies in the binned luminosity function values at various luminosities. Our number densities are in agreement with \citet{Finkelstein2015} in brighter bins upto $M_{\mathrm{UV}}=-19$, after which they find higher number densities in their faintest bin at $M_{\mathrm{UV}}=-18.5$, possibly leading to a steeper faint-end slope than ours. Similarly, the number densities of \citet{Ishigaki2017} are higher than ours, except that they find a drastic drop in their number densities at $M_{\mathrm{UV}}=-18.25$. Similarly, there is disagreement in the measured values of faint-end slopes at this redshift, in that our faint-end slope is steeper, except \citet{Finkelstein2015}.  

Finally, at $z\sim9$, we find that our faint-end slope is in agreement with \citet{Laporte2016}. \citet{Mcleod2016} and \citet{Ishigaki2017} are finding shallower faint-end slopes than ours, however, we note that they held the value of faint-end slope $\alpha$ to be fixed while fitting their data points.

In general, at all redshifts, the difference in errors bars on the Schechter parameters between our work and previous work (e.g., \citealt{Finkelstein2015, Bouwens2015, Atek2018}) can be attributed to the fact that we do not include the errors due to cosmic variance on our number densities.

\subsection{Galaxy stellar mass function}
The galaxy stellar mass function (GSMF) is a valuable mechanism to probe the growth of stellar mass, as it includes all processes such as star formation and mergers, which contribute to building up the mass of a galaxy. Cosmological simulations produce a distribution of dark matter halos by hierarchical assembly following the initial tiny perturbations in the early young Universe. The mass distribution of these halos follows the Schechter form and for this reason a Schechter function for the galaxy mass distribution is also expected. We therefore fit our number densities in each redshift bin derived in Section \ref{sec:number_densities} with a Schechter function of the form:

\begin{equation}
\Phi(M)=ln(10)\varPhi^{\ast}10^{(M-M^{\ast})(\alpha+1)}e^{-10^{(M-M^{\ast})}},
\end{equation}

\noindent where $\phi^{\ast}$ provides the normalization, $M^{\ast}$ corresponds to the characteristic stellar mass at which the function turns over from a power law into an exponential form, and $\alpha$ is the slope of the low-mass end.
$\alpha$ is usually negative, implying large numbers of galaxies with low masses.

\subsubsection{Best fit Schechter parameters and their uncertainties for GSMF}
\label{sec:mcmc_gsmf}

\begin{figure}
\includegraphics[width=1\columnwidth, height=1\columnwidth]{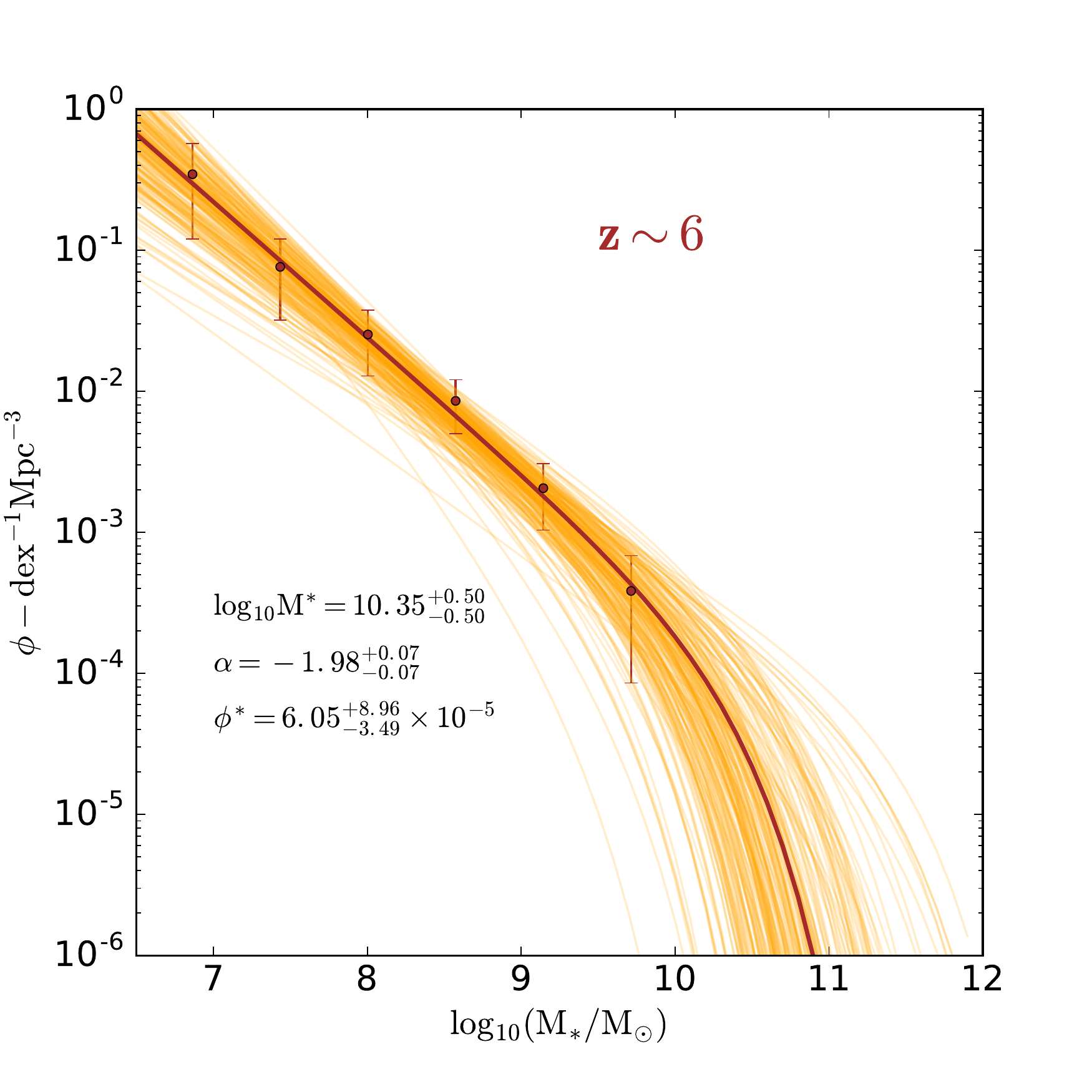}
\centering
\caption{Distribution of 100 random samples drawn from the chain at $z\sim6$ from our MCMC analysis and projected into the space of the observed data.}
\label{fig:mcmcfit}
\end{figure}

To determine the best fit Schechter parameters and their uncertainties for GSMF, we perform a MCMC analysis described in \ref{sec:mcmc} that examines the three-dimensional parameter space of Schechter parameters. Just as for the UV LF, for each redshift, we use $10^{2}$ MCMC chains of $10^{4}$ steps each to explore fully the parameter space, building a distribution of $M^{*}$, $\Phi^{*}$ and $\alpha$. For the priors, we limit the parameter space to $8<\mathrm{log(\mathit{M^{*}/M_{\odot}})<13}$, log$(\phi^{*}/\mathrm{Mpc^{-3})>-8}$ and $\alpha>-4$. However at $z=9$, where the sample size is small, we fix $M^{*}$ to the value estimated at $z=8$. Fig.~\ref{fig:mcmcfit} shows 100 random samples from the chain plotted on the top of our data points for our GSMF at $z\sim6$.

\begin{figure*}
\begin{minipage}{1\textwidth}
\centering
\includegraphics[scale=0.6]{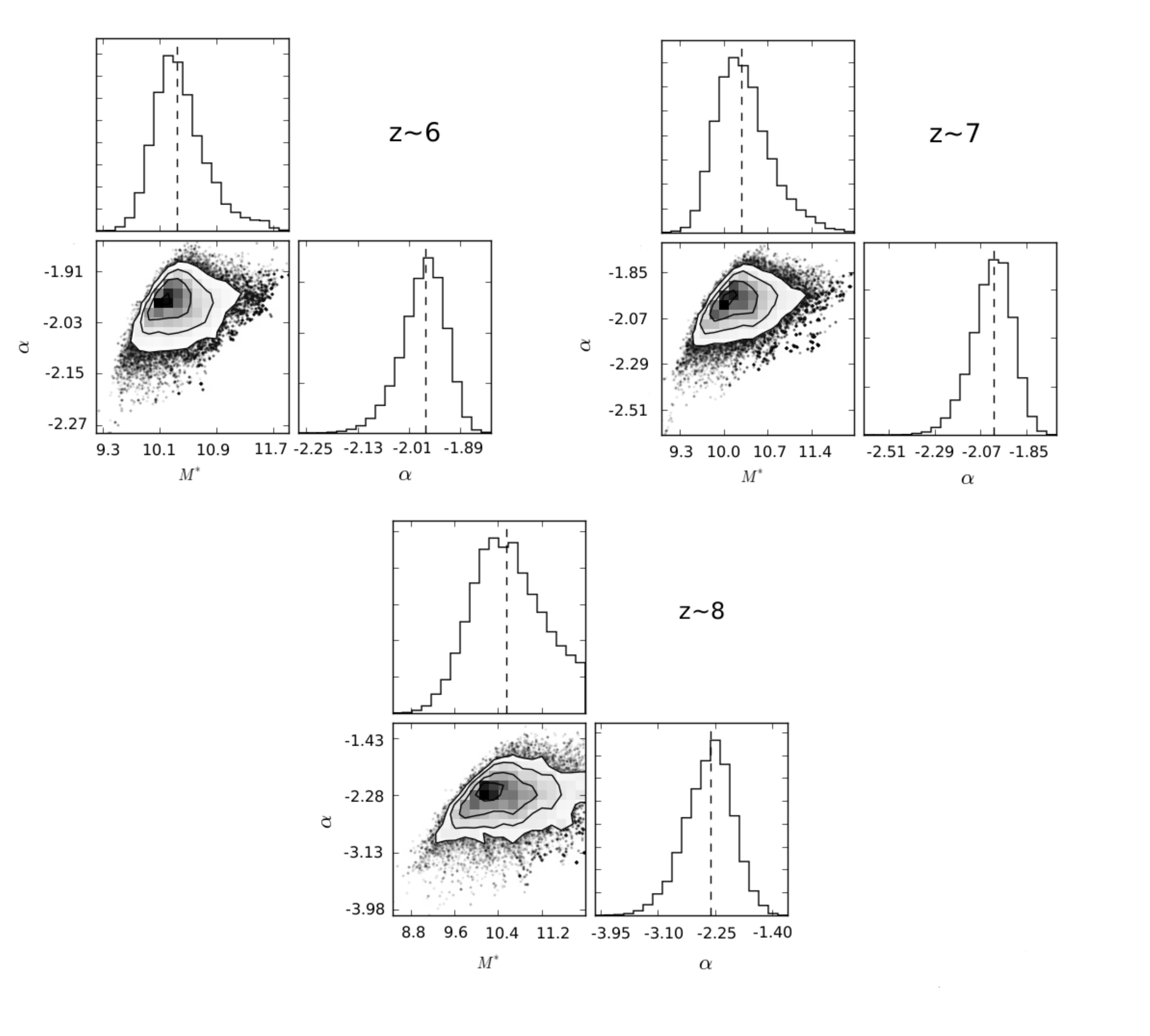}
\end{minipage}
\caption{Corner plots showing the two dimensional posterior probability distributions of characteristic stellar mass $M^{*}$ and faint end slope $\alpha$ at $z=6,7$ and $8$ with contours shown at 0.5, 1, 1.5, and 2 sigma. Top left: Posterior probability distribution at $z\sim6$. Top right: Posterior probability distribution at $z\sim7$. Bottom: Posterior probability distribution at $z\sim8$. The marginalized distribution for each parameter is shown independently in the histograms, with the dashed line being the median value of the distribution.}
\label{fig:corner_plot}
\end{figure*}

For our final result, just as for the UV LF, we join the chains together giving a distribution of $10^{6}$ values of Schechter function parameters at each redshift. The best-fit values for each Schechter function parameter are the median of this distribution, with the uncertainties covering the central 68 per cent of the distribution. Fig.~\ref{fig:corner_plot} shows the two dimensional posterior probability distributions of characteristic stellar mass $M^{*}$ and faint end slope $\alpha$ at $z=6-8$. 

\begin{figure*}
\centering
\begin{minipage}{0.49\textwidth}
\centering
\includegraphics[width=1\textwidth, height=0.3\textheight]{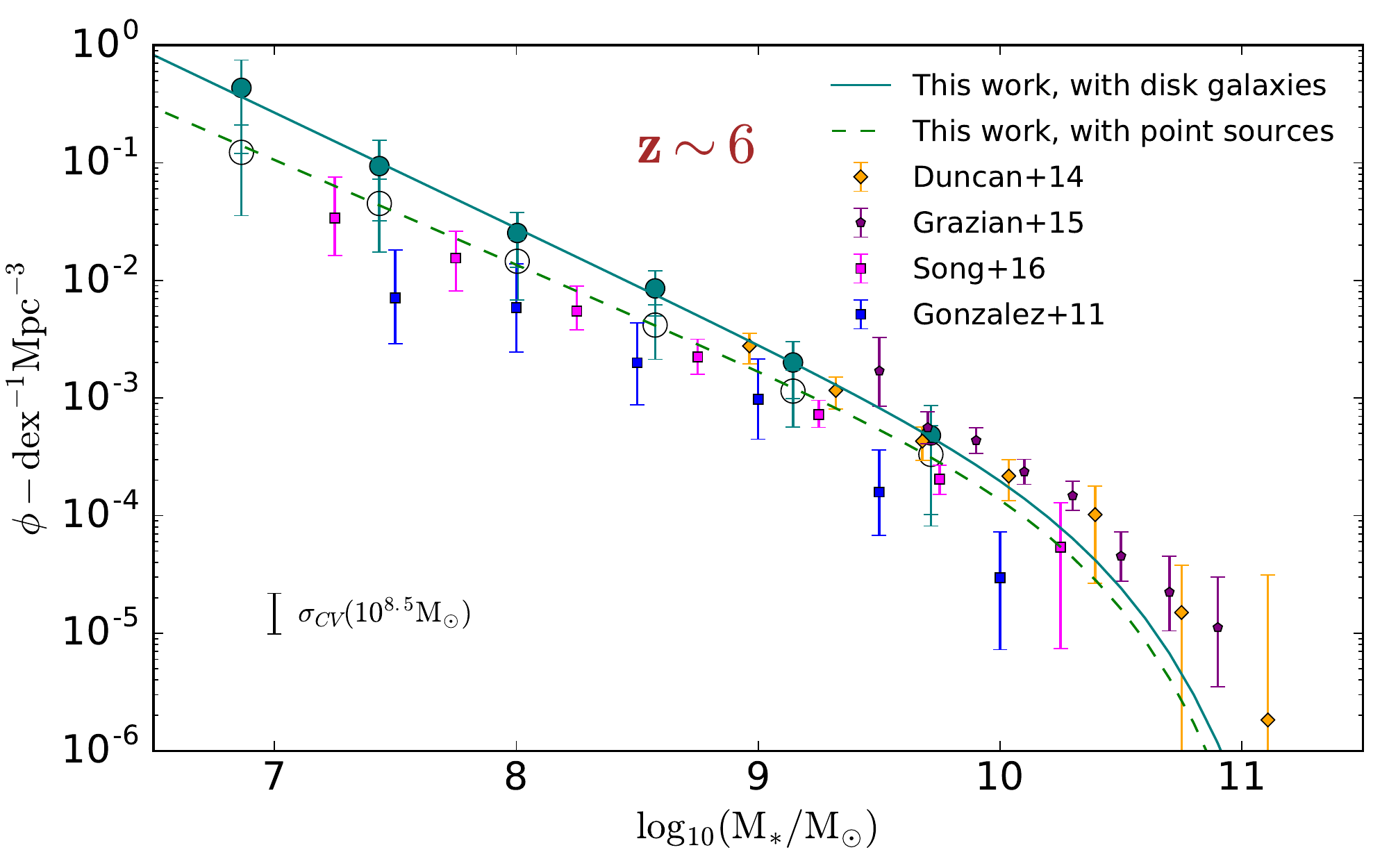}
\end{minipage}
\begin{minipage}{0.49\textwidth}
\centering
\includegraphics[width=1\textwidth, height=0.3\textheight]{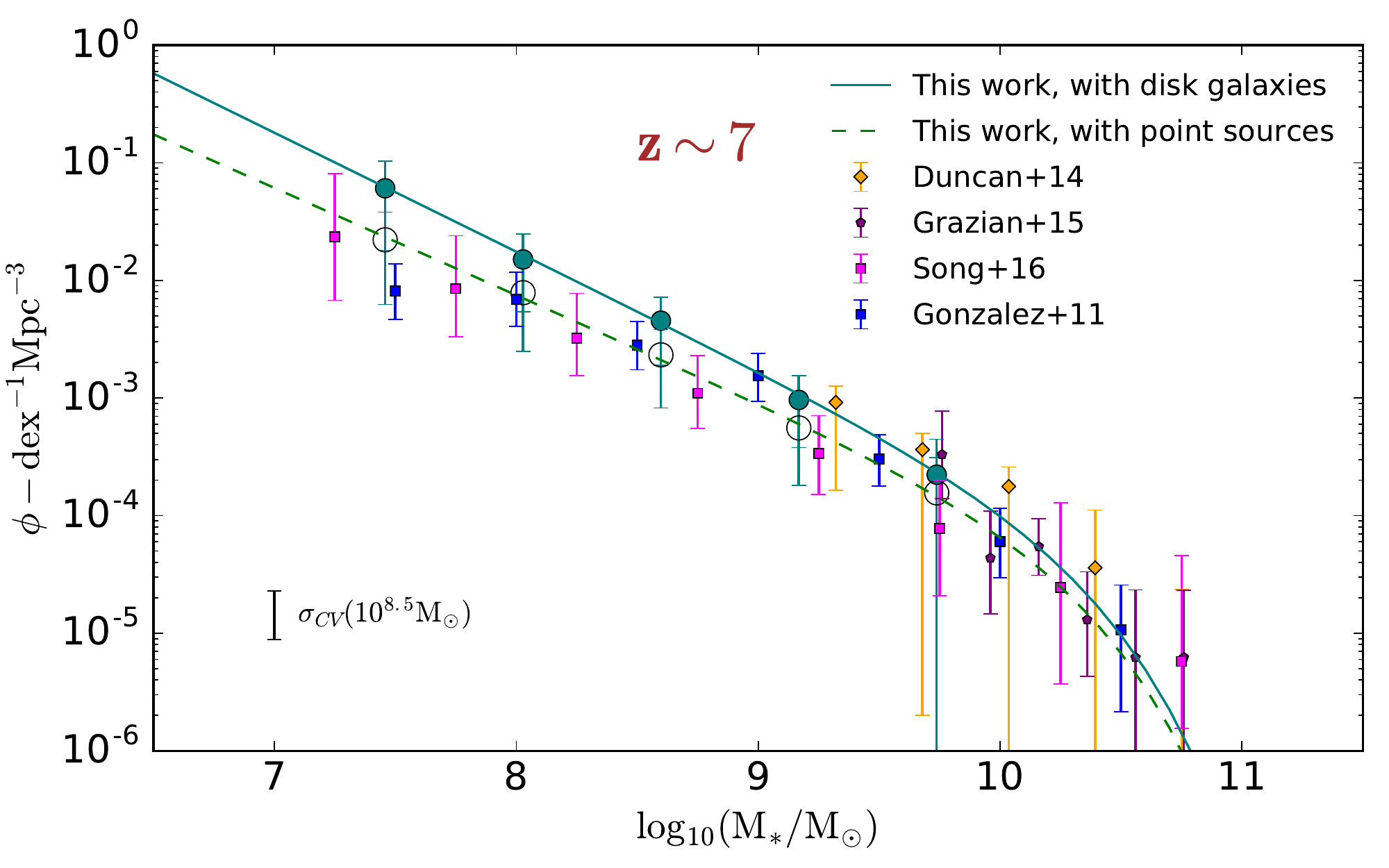}
\end{minipage}
\begin{minipage}{0.49\textwidth}
\centering
\includegraphics[width=1\textwidth, height=0.3\textheight]{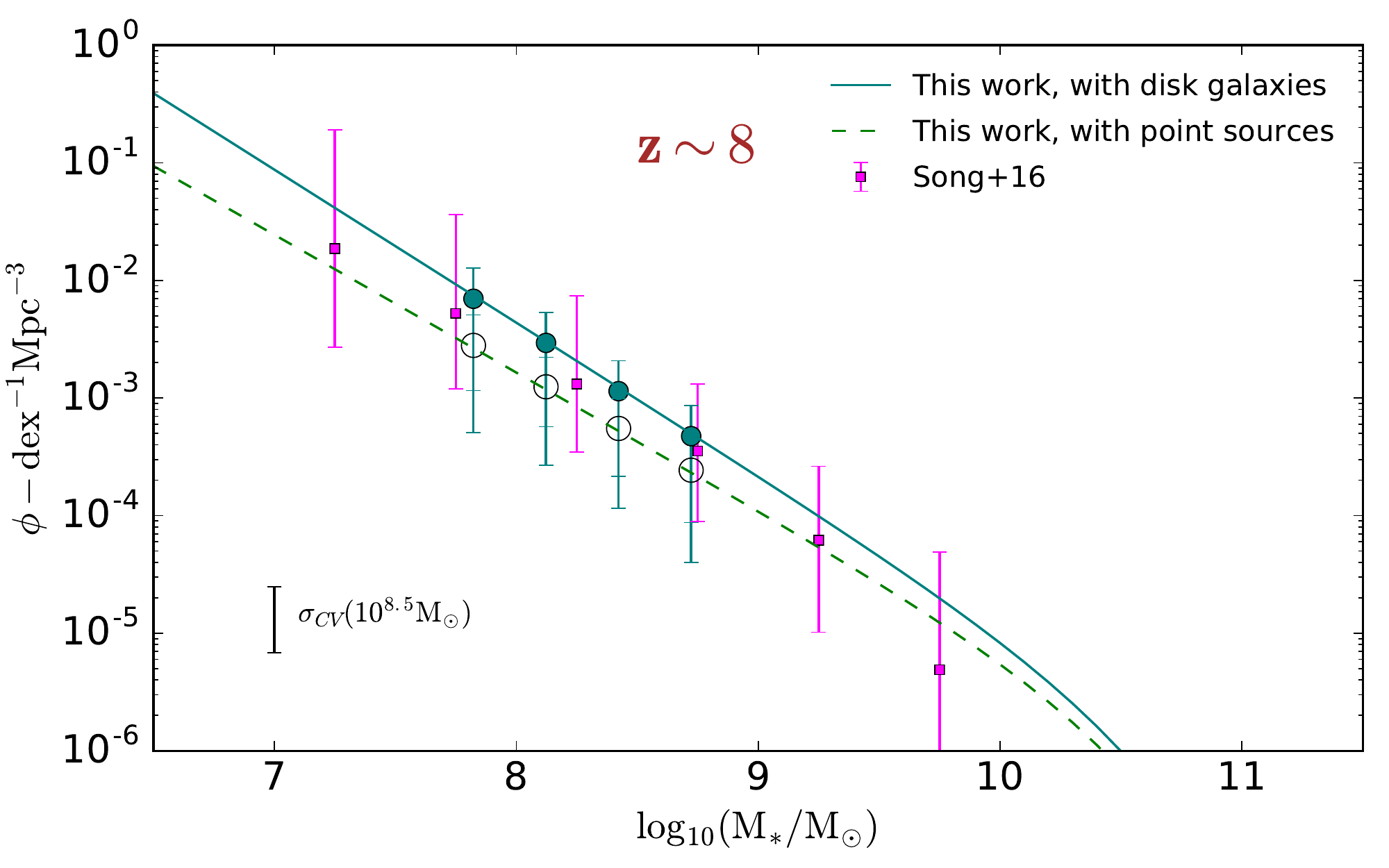}
\end{minipage}
\begin{minipage}{0.49\textwidth}
\centering
\includegraphics[width=1\textwidth, height=0.3\textheight]{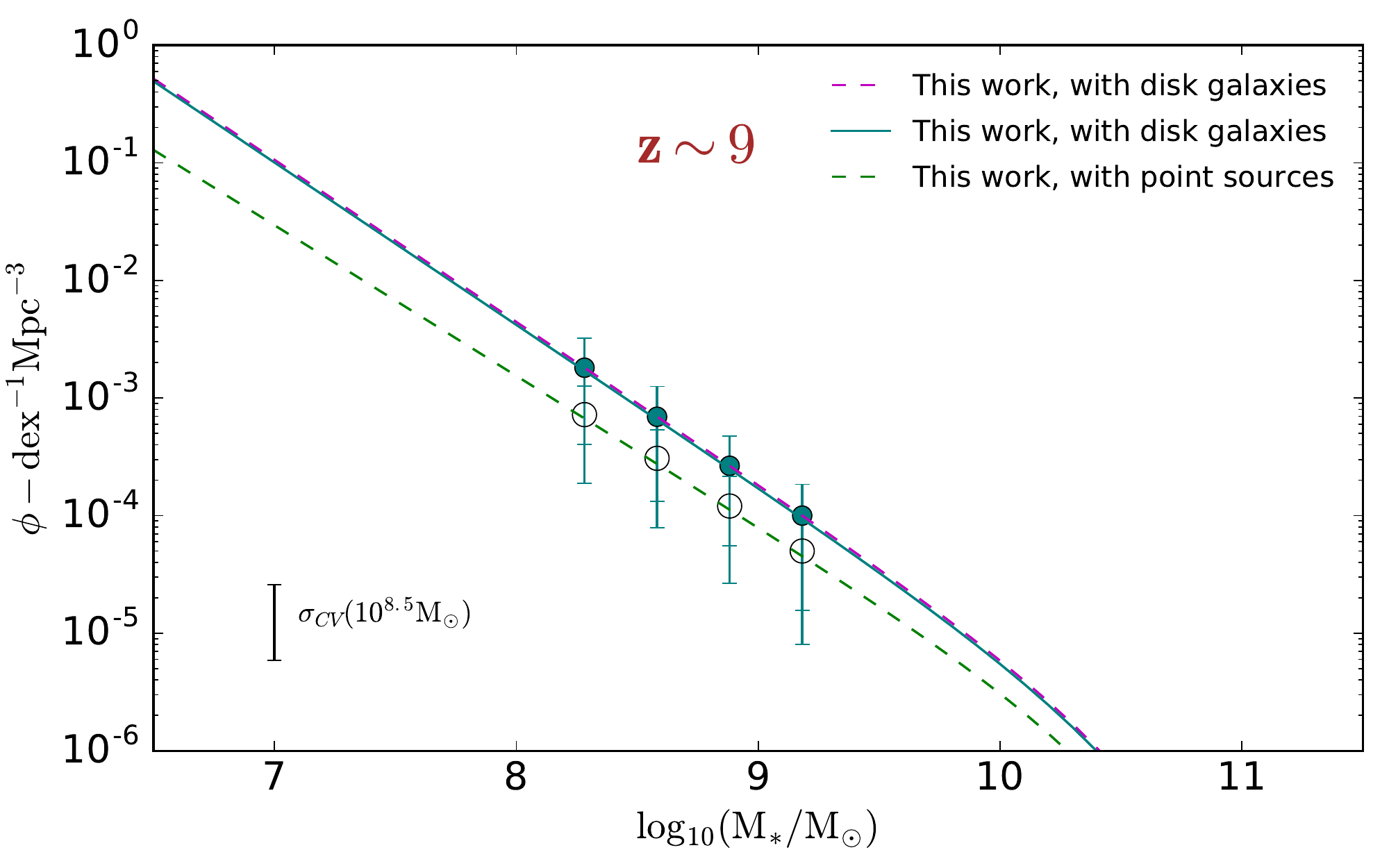}
\end{minipage}
\caption{The GSMF at $z=6-9$ in the MACS0416 cluster and its parallel field. The solid green line is our best fit Schechter function from our MCMC analysis to the green circles, considering disc-like galaxies in completeness simulations. The dashed green line is our best fit Schechter function from MCMC analysis to the open green circles, using point sources. At $z=9$, the dashed magenta line shows the best fit Schechter function derived using a simple chi squared minimization technique, using disk-like galaxies. The error bars on our data points take into account errors on photometric redshifts, errors on stellar mass, Poisson errors and uncertainties due to lensing maps, but does not include errors due to cosmic variance. The representative error bar in the bottom left corner of each plot shows the measured error on number densities for galaxies of $10^{8.5}\mathrm{M_{\odot}}$ due to cosmic variance.}
\label{fig:gsmf}
\end{figure*}

Similar to the UV LF, we derive the best-fit Schechter parameters and their uncertainties for our GSMF using the completeness curves described in \ref{sec:comp_sim}, with both disc-like galaxies and point sources, in order to investigate the effect of galaxy sizes on completeness corrections, and hence on the faint-end slopes. Our results, along with the values from the literature, are listed in Table~\ref{tab:gsmf_table} and the resulting GSMFs plotted alongside previous work are shown in Fig.~\ref{fig:gsmf}. Similar to UV LF, at $z=9$ we also show the best-fit Schechter function derived using a simple chi squared minimization technique using disk-like galaxies. The error bars on our data points in Fig.~\ref{fig:gsmf} take into account the errors on photometric redshifts, errors on magnitudes, Poisson errors and uncertainties due to lensing maps, but does not include the errors due to cosmic variance. We estimate the fractional uncertainty in number densities due to cosmic variance using the method described in Section \ref{sec:mcmc} and in Fig.~\ref{fig:gsmf} we show the measured error on number densities for galaxies of $10^{8.5}\mathrm{M_{\odot}}$.

\begin{table}
	 \centering
     \renewcommand{\arraystretch}{1.2}
	  \caption{Best fit Schechter function parameters and their uncertainties for our GSMFs. The quoted best fit values and 1$\sigma$ errors of the Schechter parameters constitute the median and the central 68 per cent of posterior distribution of each parameter from our MCMC analysis.}
	  \label{tab:gsmf_table}
	 \begin{tabular}{lccr} 
		 \hline
		 Redshift & log$_{10}M^{*}$ & $\alpha$ & $\phi^{*}(10^{-5} \mathrm{Mpc^{-3})}$\\
		 \hline
		 $z\sim6$ & & &\\
		 This work (disc galaxies) & $10.35_{-0.50}^{+0.50}$ & $-1.98_{-0.07}^{+0.07}$& $6.05_{-3.49}^{+8.96}$\\
		 This work (point sources) & $10.29_{-0.67}^{+0.65}$ & $-1.89_{-0.10}^{+0.09}$& $5.43_{-3.29}^{+8.16}$\\
		 Song et al. (2016) & $10.72_{-0.30}^{+0.29}$ & $-1.91_{-0.09}^{+0.09}$&$1.35_{-0.75}^{+1.66}$\\
		 Duncan et al. (2014) & $10.87_{-0.54}^{+1.13}$ & $-2.00_{-0.40}^{+0.57}$& $1.40_{-1.4}^{+41.1}$\\
		 Grazian et al. (2015) & $10.49_{-0.32}^{+0.32}$ & $-1.55_{-0.19}^{+0.19}$& $6.91_{-4.57}^{+13.5}$\\
		
		 \hline
		 $z\sim7$ & & &\\
		This work (disc galaxies) & $10.27_{-0.67}^{+0.60}$ & $-2.01_{-0.13}^{+0.17}$& $3.90_{-2.85}^{+9.20}$\\
		 This work (point sources) & $10.25_{-0.71}^{+0.67}$ & $-1.91_{-0.14}^{+0.18}$& $2.93_{-1.99}^{+6.29}$\\
		 Song et al. (2016) & $10.78_{-0.28}^{+0.29}$ & $-1.95_{-0.18}^{+0.18}$& $0.53_{-0.38}^{+1.10}$\\
		 Duncan et al. (2014) & $10.51_{-0.32}^{+0.36}$ & $-1.89_{-0.61}^{+1.39}$& $3.6_{-0.35}^{+3.01}$\\
		 Grazian et al. (2015) & $10.69_{-1.58}^{+1.58}$ & $-1.88_{-0.36}^{+0.36}$& $0.57_{-0.56}^{+59.68}$\\
		 \hline
		 $z\sim8$ & & &\\
		 This work (disc galaxies) & $10.54_{-0.94}^{+1.00}$ & $-2.30_{-0.46}^{+0.51}$& $0.095_{-0.080}^{+0.56}$\\
		 This work (point sources) & $10.48_{-0.92}^{+1.19}$ & $-2.17_{-0.53}^{+0.55}$& $0.090_{-0.078}^{+0.51}$\\\
		 Song et al. (2016) & $10.72_{-0.29}^{+0.29}$ & $-2.25_{-0.35}^{+0.72}$& $0.035_{-0.030}^{+0.246}$\\
		 \hline
		 $z\sim9$ & & &\\
		 This work (disc galaxies) & $10.54$ (fixed) & $-2.38_{-0.88}^{+0.72}$& $0.057_{-0.050}^{+0.65}$\\
		 This work (point sources) & $10.48$ (fixed) & $-2.28_{-0.98}^{+0.83}$& $0.045_{-0.040}^{+0.63}$\\
		 \hline
		 \hline
	 \end{tabular}
 \end{table}

\begin{figure*}
\centering
\begin{minipage}{0.3\textwidth}
\centering
\includegraphics[scale=0.43]{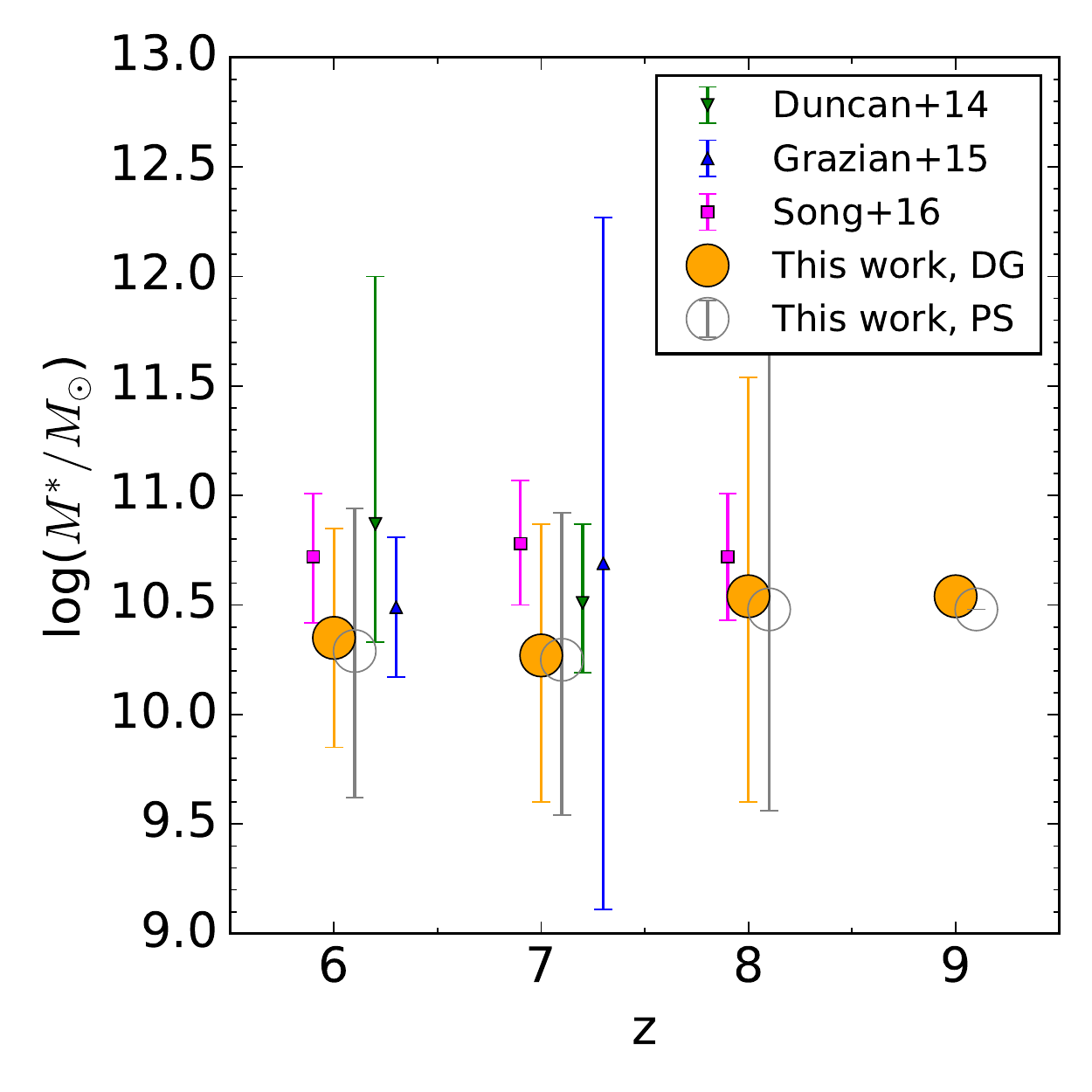}
\end{minipage}
\begin{minipage}{0.3\textwidth}
\centering
\includegraphics[scale=0.43]{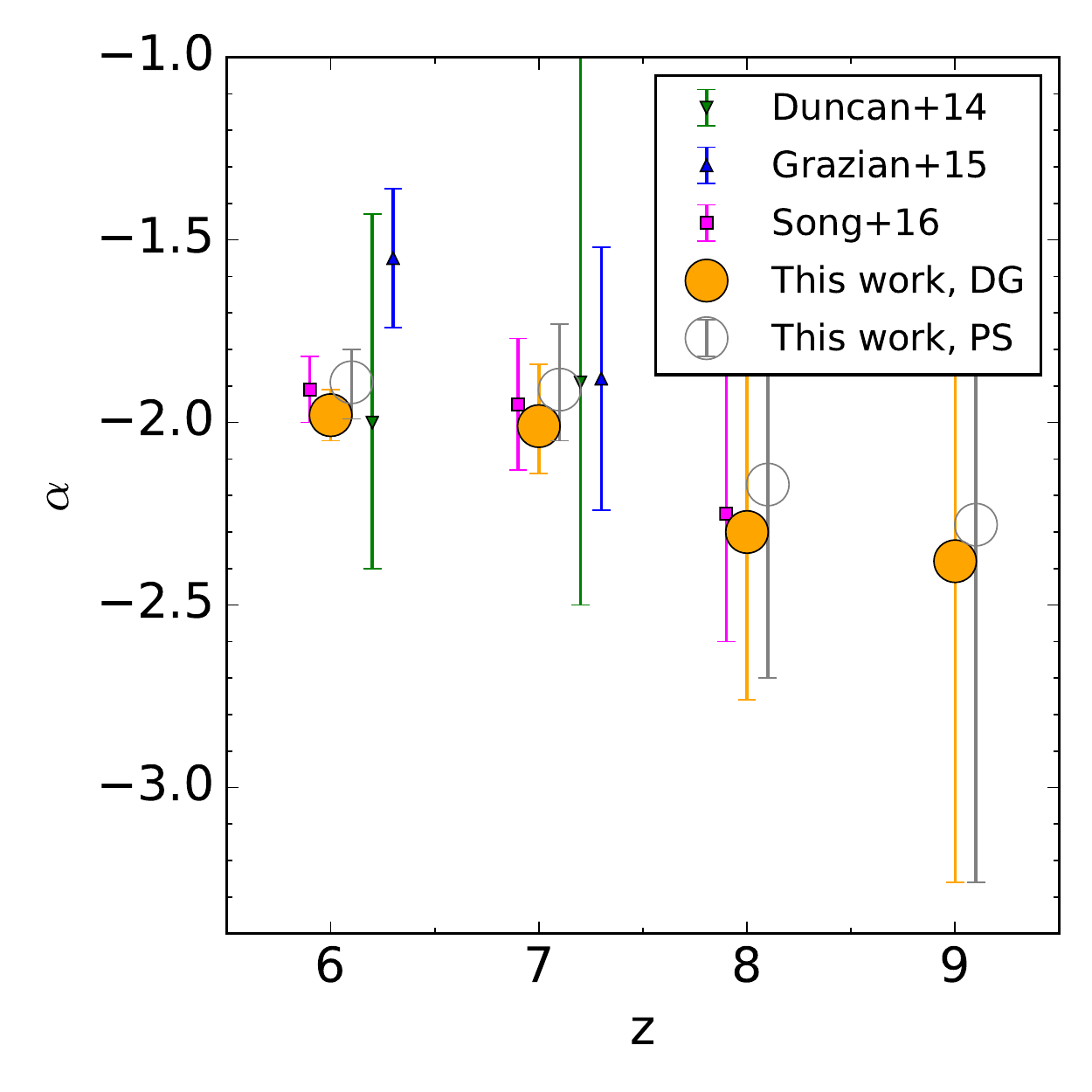}
\end{minipage}
\begin{minipage}{0.3\textwidth}
\centering
\includegraphics[scale=0.43]{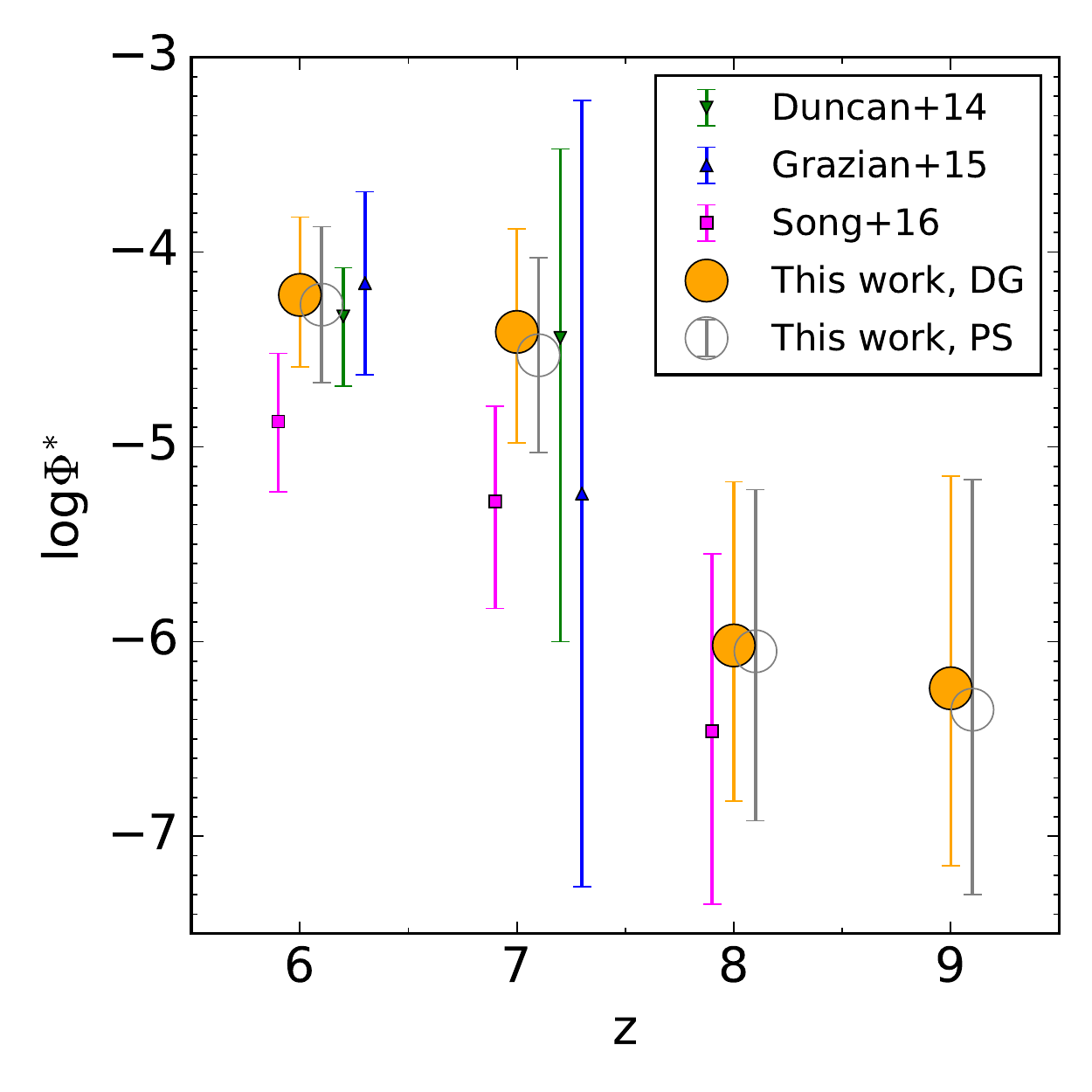}
\end{minipage}
\caption{Redshift evolution of best fit Schechter function parameters for our stellar mass functions. The solid orange circles represent the best fit Schechter parameters considering disc-like galaxies in our completeness simulations, whereas the open circles are our results considering point sources. The low mass end slope $\alpha$ steepens with increasing redshift, number density $\Phi^{*}$ decreases with increasing redshift, whereas no evolution in the characteristic mass $M^{*}$ is observed. \textbf{Note:} The error bars on characteristic mass $M^{*}$ at $z\sim9$ are not shown since we keep the value of $M^{*}$ fix to the value estimated at $z\sim8$.}
\label{fig:sche_par_evolution}
\end{figure*}

Examining our best fit Schechter parameters, our results show a steepening of the low mass end slope $\alpha$ with increasing redshift ($-1.98_{-0.07}^{+0.07}$, $-2.01_{-0.13}^{+0.17}$, $-2.30_{-0.46}^{+0.51}$ and $-2.38_{-0.88}^{+0.72}$ at $z=6,7,8$ and $9$ respectively). However we cannot rule out a constant $\alpha$ between $z=6-9$ and more data is needed to confirm the trend at these redshifts. There is also a decrease in $\Phi^{*}$ with increasing redshift, but no evolution in $M^{*}$ is observed. We notice a similar trend for the low mass end slope $\alpha$ (steepening with increasing redshift) when point sources are considered in our completeness simulations, albeit with a shallower slope as compared to disc galaxies ($\Delta\alpha\sim0.09-0.13$). A decrease in $\Phi^{*}$ with increasing redshift is also observed in the case of point sources, but once again no evolution in $M^{*}$ is observed.  Fig.~\ref{fig:sche_par_evolution} shows the redshift evolution of the Schechter parameters for the stellar mass functions along with the values from the literature.

\subsubsection{Comparison with previous work}
\label{sec:gsmf_comp}

Fig.~\ref{fig:gsmf} shows the comparison of our GSMF results at $z=6-9$ with previous literature work \footnote[9]{Note: We restrict the comparison of our results with previous work to the best-fit Schechter parameters derived using disc-galaxies in our completeness simulations, and do not compare our results derived using point sources.}. The first thing that we notice here is the dissimilarity of normalization values between our work and the literature values at all redshifts. We are finding higher normalization values than \citet{Gonzalez2011}, \citet{Duncan2014}, \citet{Grazian2015} and \citet{Song2016} at all redshifts, except at $z\sim6$ where the normalization values of \citet{Grazian2015} are higher than ours. The reason for this discrepancy is unclear, however, it could be attributed to the shallower slopes of our best fitted $\mathrm{log_{10}(M_{*}/M\odot)-M_{UV}}$ relation that we are finding (See Section~\ref{sec:mstar_muv}) as compared to previous studies (See Table~\ref{tab:muvmstar_table}). A shallower slope of $\mathrm{M_{*}-M_{UV}}$ results into higher normalization values and steeper low mass end slope \citep{Song2016}. Although, we do not use our $\mathrm{M_{*}-M_{UV}}$ relation to calculate our GSMF, we point out that our results of higher normalization values and steep low mass end slope do strengthen the argument.

Inspecting the high mass end of our GSMFs, we find a deficit of bright/massive galaxies in comparison to the literature, making it difficult for us to put any robust constraints on the high mass end of the GSMF. The lack of brighter galaxies can be attributed to the smaller survey area that we are probing in our work. Interestingly, we notice that the massive galaxies that we find in our study all come from the parallel field alone (considering the relatively larger volume that it allows us to probe as compared to the cluster) i.e., the cluster area is devoid of massive galaxies. At $z\sim 6$, the number densities of our highest mass bin (log$(M_{*}/M_{\odot})=9.7-10.3$) are in agreement with \citet{Duncan2014} and \citet{Grazian2015}, but higher than \citet{Song2016} and \citet{Gonzalez2011}. At $z \sim 7$, the number densities in our highest mass bin are in agreement with \citet{Gonzalez2011}, are lower than \citet{Duncan2014}, but once again higher than \citet{Song2016}. 

At the low mass end, even though our survey volume is significantly smaller than those of other studies, the strong gravitational lensing effect allows us to probe masses as low as $10^{6.8}M_{\odot}$, enabling us to put robust constraints on the low mass end slope $\alpha$. Since the lowest mass bin of \citet{Grazian2015} is log$(M_{*}/M_{\odot})\sim9$, we compare our low mass end of the GSMF with \citet{Gonzalez2011}, \citet{Duncan2014} and \citet{Song2016} who are able to reach lower in masses. At $z \sim 6$, our number densities are higher than these studies. We also find a steep faint end slope $\alpha$ of $-1.98_{-0.07}^{+0.07}$, which is steeper than all the literature values except \citet{Duncan2014}, who are finding a slope of $-2.00_{-0.40}^{+0.57}$, albeit with large error bars. At $z \sim 7$, we are again finding higher number densities and a steeper slope than previous studies. This could be attributed to the shallower slopes of our best fitted $\mathrm{log_{10}(M_{*}/M\odot)-M_{UV}}$ relation that we are finding as mentioned earlier in this section. Finally, at $z \sim 8$, we are finding higher number densities and steeper slope than \citet{Song2016}.

\subsection{$\mathrm{M_{*}-M_{UV}}$ relation}
\label{sec:mstar_muv}

Scaling relations that describe the connection between the physical properties of galaxies such as metallicity, star formation rate, stellar mass, luminosity, etc. provide meaningful insights on galaxy formation and evolution. Using the stellar mass and the rest-frame absolute UV magnitudes estimated in Section \ref{sec:sedfitting}, we now examine the mass-to-light ratios of our sample for each of our redshift bins to determine the scaling relation between these properties and how it extends to the lowest masses.

\begin{figure*}
\centering
\begin{minipage}{0.49\textwidth}
\centering
\includegraphics[width=1.1\textwidth, height=0.35\textheight]{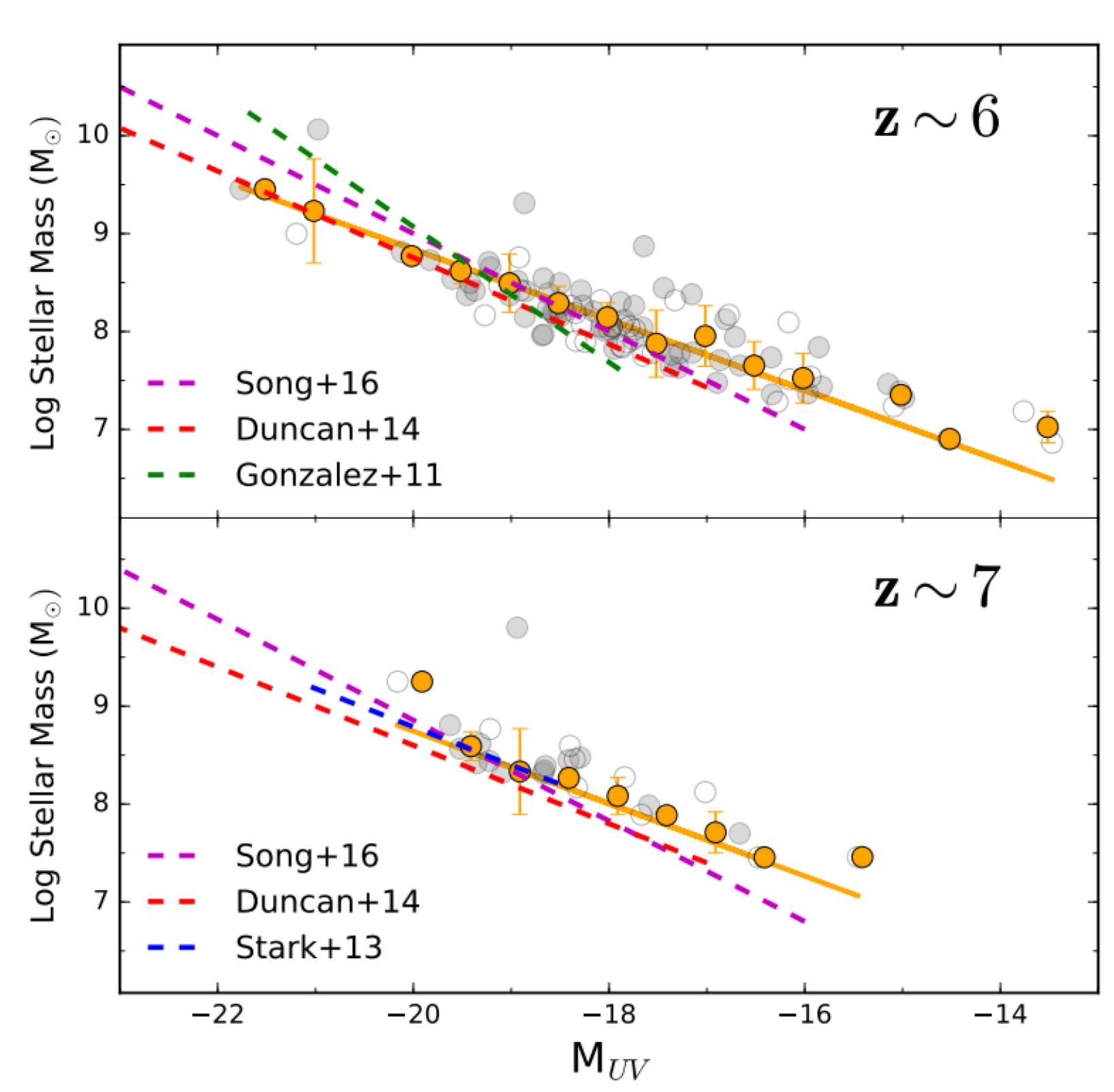}
\end{minipage}
\begin{minipage}{0.49\textwidth}
\centering
\includegraphics[width=0.75\textwidth, height=0.35\textheight]{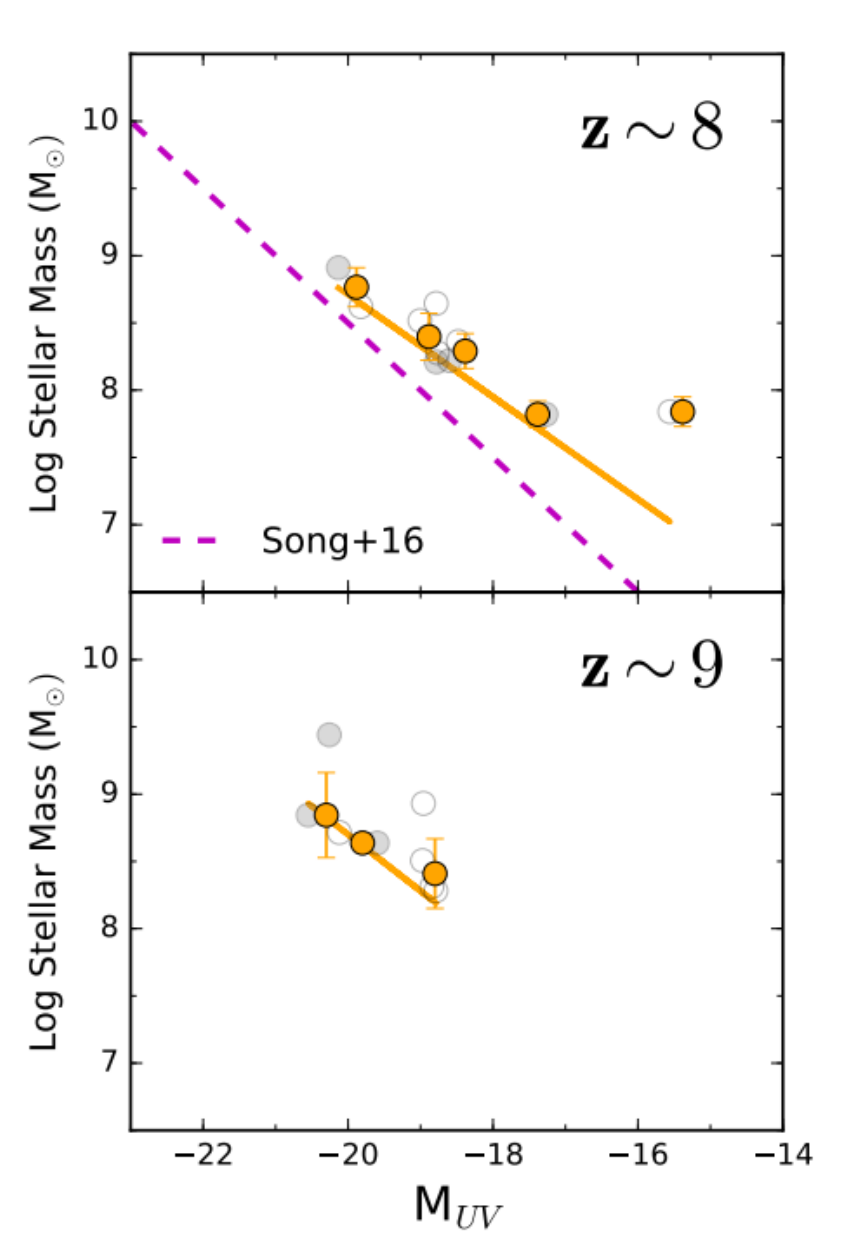}
\end{minipage}
\caption{Stellar mass as a function of rest-frame absolute UV magnitude at 1500 angstrom at $z=6-9$. Grey filled circles represent sources with a S/N ratio $>3\sigma$ at $3.6\mu m$ in IRAC, whereas grey open circles are those with a S/N ratio $\lesssim3\sigma$. Filled yellow circles represent the median stellar masses in each rest-frame absolute UV magnitude bin of 0.5 mag, and the yellow error bars represent the standard deviation in stellar mass in each UV magnitude bin. The best-fit $\mathrm{log_{10}(M_{*}/M\odot)-M_{UV}}$ relation is shown by solid yellow line in each redshift bin. The best-fit $\mathrm{log_{10}(M_{*}/M\odot)-M_{UV}}$ lines of \citet{Gonzalez2011}, \citet{Stark2013}, \citet{Duncan2014} and \citet{Song2016} are also shown by green, blue, red and magenta dashed lines, respectively, for comparison.
}
\label{fig:muvmstar}
\end{figure*}

For this, we derive the best-fit $\mathrm{log_{10}(M_{*}/M\odot)-M_{UV}}$ relation by fitting a linear function to the median masses of the sample in each rest-frame absolute UV magnitude bin of 0.5 mag. Fig.~\ref{fig:muvmstar} shows the best-fit relation and Table~\ref{tab:muvmstar_table} lists the best fitting values for our sample in each redshift bin. Whilst we notice a few high mass galaxies with faint UV magnitudes, there is a clear positive linear trend of increasing stellar mass with increasing rest-frame absolute magnitude (with a large scatter in all the redshift bins). 

At $z\sim6$, our results are in agreement with \citet{Duncan2014} at the bright end ($\mathrm{M_{UV}\leq-20}$) when they include nebular emission lines while estimating their masses, but their stellar masses at the faint-end ($\mathrm{M_{UV}\geq-19}$) are lower than ours, resulting in a steeper slope. Comparing with \citet{Song2016}, their estimated masses with the inclusion of nebular emission lines at the bright end ($\mathrm{M_{UV}\leq-20}$) are higher than ours, but they are finding lower measurements of stellar masses at the faint-end (($\mathrm{M_{UV}\geq19}$), resulting in a steeper slope than ours. \citet{Gonzalez2011} estimate their masses without the inclusion of nebular emission lines and find higher estimates of stellar masses at $\mathrm{M_{UV}\leq-20}$, but obtain lower masses at $\mathrm{M_{UV}\geq-19}$, once again resulting in a steeper slope than we find in this study. We discover a similar trend when comparing our results to previous studies at $z\sim7$ and $z\sim8$, in that previous studies (e.g., \citealt{Duncan2014, Song2016}) are finding lower masses in faint UV bins at $\mathrm{M_{UV}\geq-19}$, resulting in steeper slopes than ours. 

In general, we find that at fainter magnitudes, our estimated masses are $\sim0.2$ dex higher than previous studies. It is not clear why this is the case but this could be due to a couple of reasons: The discrepancy is likely a result of different fitting methods for stellar masses giving results which can differ by $\sim0.2$ dex between otherwise similar methods (e.g., \citealt{Mobasher2015}). The inconsistency could also be a result of IRAC deblending. Although we rule out the possibility of IRAC deblending by carefully inspecting our sample, we do notice that in spite of our deep IRAC data, the S/N of galaxies in faint UV bins is usually low ($\lesssim3\sigma$). A detailed stacking analysis by \citet{Song2016} reveals that at faint UV luminosities, the stacked points are usually lower than the median values, reflecting that the stellar masses of low S/N galaxies in the faint UV bins are on average biased towards higher masses. However, see also \citet{Behroozi2018} for a further discussion on systematic differences in the UV mass-to-light ratios, where they recalculate the median UV-stellar mass relations from the \citet{Song2016} SED stacks for $z=4-8$ galaxies, and argue that the masses found by them are too low, possibly a result of implausibly young ages being fit while estimating the stellar masses.

In order to see the effect of $\lesssim3\sigma$ sources in faint UV bins on the best-fit $\mathrm{log_{10}(M_{*}/M\odot)-M_{UV}}$ relation, we implement two fits to our data. By performing a linear fit to the full luminosity range probed in our study, we find that we are observing shallower slopes in all the redshift bins than previous studies of \citet{Gonzalez2011}, \citet{Duncan2014} and \citet{Song2016}. This could be the reason for the higher observed normalization values in our derived GSMFs at all redshifts as mentioned previously in Section \ref{sec:gsmf_comp}. However, by restricting our analysis to brighter magnitudes ($\mathrm{M_{UV}\leq-16}$) at $z\sim6$, $z\sim7$ and $z\sim8$ and excluding the $\lesssim3\sigma$ points from our sample, we observe a steepening of slopes in comparison (See Table~\ref{tab:muvmstar_table}), but these are still shallower than previous studies. This furthermore highlights the importance of deeper imaging with JWST to obtain accurate photometry of galaxies at $>2\mu m$. Nevertheless, in both the cases, we find that the slopes of our best fitted $\mathrm{log_{10}(M_{*}/M\odot)-M_{UV}}$ relation are close to a constant mass-to-light ratio of $-0.40$, suggesting no strong evolution of mass-to-light ratio with luminosity. We notice that normalization, on the other hand, evolves very weakly from $z\sim6$ to $z\sim9$, with a decrease in normalization with increasing redshift.

\begin{table}
	 \centering
	  \caption{Best-fit $\mathrm{log_{10}(M_{*}/M\odot)-M_{UV}}$ relation. The values in parentheses are the best-fit values when we restrict our analysis to brighter magnitudes ($\mathrm{M_{UV}\leq-16}$). The quoted errors represent the $1\sigma$ uncertainties.}
	 \label{tab:muvmstar_table}
	 \begin{tabular}{lcccr} 
		 \hline
		 \hline
		 $z$ & log $M_{*(\mathrm{M_{UV}=-19.5)}}$  & slope\\
		 \hline
		 6 & $8.66\pm0.05$ & $-0.38\pm0.07$ \\
		   & ($8.66\pm0.07$) & ($-0.41\pm0.06$)\\
		 7 & $8.56\pm0.08$ & $-0.37\pm0.09$ \\
		   & ($8.58\pm0.08$) & ($-0.40\pm0.05$)\\
		 8 & $8.52\pm0.18$ & $-0.38\pm0.14$ \\
		   & ($8.50\pm0.16$) & ($-0.40\pm0.15$)\\
		 9 & $8.49\pm0.28$ & $-0.42\pm0.21$ \\
		 \hline
	 \end{tabular}
 \end{table}

\subsection{Stellar mass density}
\begin{figure}
\includegraphics[width=1\columnwidth, height=1\columnwidth]{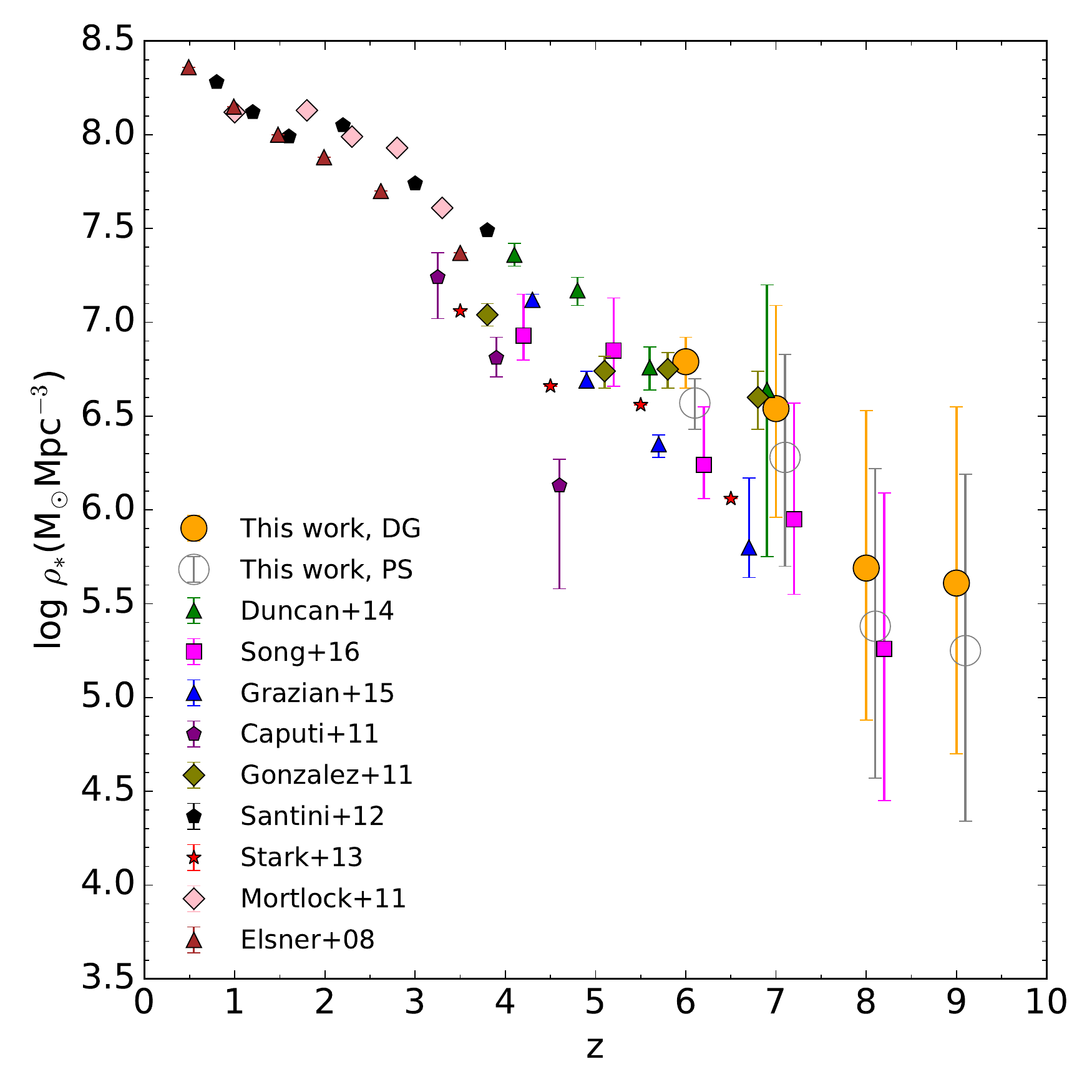}
\centering
\caption{Evolution of total SMD as a function of redshift. We calculate the total SMD by integrating the best fit Schechter function from $M_{*}=10^{8}$ to $10^{13}M_{\odot}$ for our GSMFs. The quoted 1$\sigma$ error bars represent the minimum and maximum range of stellar mass densities within the 1$\sigma$ contours of Schechter parameters obtained from our MCMC analysis. Our results are shown as solid orange points when disc-like galaxies are considered in our completeness simulations, and as open grey circles considering point sources. Shown also are the values from the literature converted to a Chabrier IMF, where necessary. }
\label{fig:smd}
\end{figure}

To estimate the total stellar mass density (SMD) at $z=6-9$, we integrate the best-fit Schechter function in each redshift bin from $M_{*}=10^{8}$ to $10^{13}M_{\odot}$. These limits were chosen so as to allow us to compare with the SMD values in the literature. We estimate the 1$\sigma$ uncertainties as the minimum and maximum range of stellar mass densities within the 1$\sigma$ contours of Schechter parameters obtained from our MCMC analysis in Section~\ref{sec:mcmc_gsmf}. Table~\ref{tab:smd_table} lists our estimates of SMD along with their 1$\sigma$ uncertainties and Fig.~\ref{fig:smd} shows the evolution of the SMD. Our results are shown as solid orange points when disc-like galaxies are considered in our completeness simulations, and as open grey circles considering point sources. In Fig.~\ref{fig:smd}, we also show the results from the literature for comparison, converted to Chabrier IMF where required.

We find that at  $z\sim6$ and $z\sim7$, our results are in agreement with \citet{Gonzalez2011} and \citet{Duncan2014}, but higher than \citet{Stark2013}, \citet{Grazian2015} and \citet{Song2016}. At $z\sim6$, our SMD estimates are higher by $\sim$0.5 dex, $\sim$0.6 dex and $\sim$0.2 dex compared to \citet{Grazian2015}, \citet{Song2016} and \citet{Stark2013} respectively. Comparing at $z\sim7$, the estimates are higher by $\sim$0.7 dex, $\sim$0.6 dex and $\sim$0.5 dex by \citet{Grazian2015}, \citet{Song2016} and \citet{Stark2013} respectively. Finally, at $z\sim8$, the only previous SMD estimates are from \citet{Song2016}, which are $\sim$0.4 dex lower than ours.

Our measurements reveal that the integrated stellar mass density has decreased by a factor of $\sim15_{-6}^{+21}$ from $z\sim6$ to $z\sim9$, confirming that the process of ongoing star formation and merging increases the total stellar mass in the Universe with time. We also find that there is a surprisingly high stellar mass density for galaxies in the early universe up to $z \sim 9$. This is an indication that galaxies of masses around 10$^{8}$ M$_{\odot}$ have already formed a significant density by this time. This further indicates that the star formation and assembly history for galaxies is significant in the epochs $z > 9$, which we cannot probe in detail until the launch of the JWST. 
 
 \begin{table}
	 \centering
	 \renewcommand{\arraystretch}{1.2}
	  \caption{Total stellar mass density estimates by integrating the best fit Schechter function from $M_{*}=10^{8}$ to $10^{13}M_{\odot}$ for our GSMFs. The quoted 1$\sigma$ error bars represent the minimum and maximum range of stellar mass densities within the 1$\sigma$ contours of Schechter parameters obtained from our MCMC analysis. Also shown in the table are values of $\rho_{\textrm{UV}}$ and $\rho_{\mathrm{SFR}}$. The $\rho_{\textrm{UV}}$ values are integrated down to $M\mathrm{_{UV}=-13.5}$, the magnitude of the faintest galaxy in our sample. The SFR densities are calculated using the \citep{Kennicutt1998} relation, assuming a Salpeter IMF and a constant SFH over $\geq$100 Myr. The values in the parentheses are computed using point sources in our completeness simulations.}
	 \label{tab:smd_table}
	 \begin{tabular}{lcccr} 
		 \hline
		 \hline
		 $z$ & log $\rho_{*}$ & log $\rho_{\textrm{UV}}$ & log SFR density\\
		     &(M$_{\odot}$Mpc$^{-3}$) & (erg s$^{-1}$Hz$^{-1}$Mpc$^{-3}$) & (M$_{\odot}$yr$^{-1}$Mpc$^{-3}$) \\
		 \hline
		 6 & $6.79_{-0.12}^{+0.13}$ & $26.16_{-0.07}^{+0.08}$ & $-1.69_{-0.07}^{+0.08}$\\
		   & ($6.57_{-0.14}^{+0.13}$) & ($25.94_{-0.07}^{+0.08}$) & ($-1.91_{-0.07}^{+0.08}$)\\
		 7 & $6.54_{-0.55}^{+0.52}$ & $26.09_{-0.08}^{+0.08}$ & $-1.76_{-0.08}^{+0.08}$ \\
		   & ($6.28_{-0.58}^{+0.55}$)& ($25.86_{-0.09}^{+0.09}$) & ($-1.99_{-0.09}^{+0.09}$) \\
		 8 & $5.69_{-0.81}^{+0.83}$ & $25.95_{-0.09}^{+0.10}$ & $-1.90_{-0.09}^{+0.10}$ \\
		   & ($5.38_{-0.81}^{+0.84}$) & ($25.71_{-0.10}^{+0.12}$) & ($-2.14_{-0.10}^{+0.12}$) \\
		 9 & $5.61_{-0.90}^{+0.92}$ & $25.88_{-0.13}^{+0.14}$ & $-1.97_{-0.13}^{+0.14}$ \\
		   & ($5.25_{-0.91}^{+0.94}$)& ($25.55_{-0.14}^{+0.15}$) & ($-2.30_{-0.14}^{+0.15}$) \\
		 \hline
	 \end{tabular}
 \end{table}

\subsection{Specific star formation rates}
Previous studies have shown that the inclusion of emission lines in the estimation of stellar masses results in higher values of specific star formation rates (sSFR) at high redshift (e.g, \citealt{Stark2013, Gonzalez2014, Duncan2014, Schaerer2009, Schaerer2010}). This is in contrast with the results from other studies such as \cite{Stark2009}, \citet{Gonzalez2010} and \citet{Bouwens2012} who show that the sSFR remains constant at $\sim$2 Gyr$^{-1}$ with increasing redshift, and also with theoretical expectations of sSFR evolution (e.g., \citealt{Dave2008,Weinmann2011}).  We therefore use our stellar mass and SFR estimates and investigate this further by calculating the sSFR (sSFR = SFR/M$_{*}$) of our sample in different redshift bins.

\begin{figure}
\includegraphics[scale=0.43]{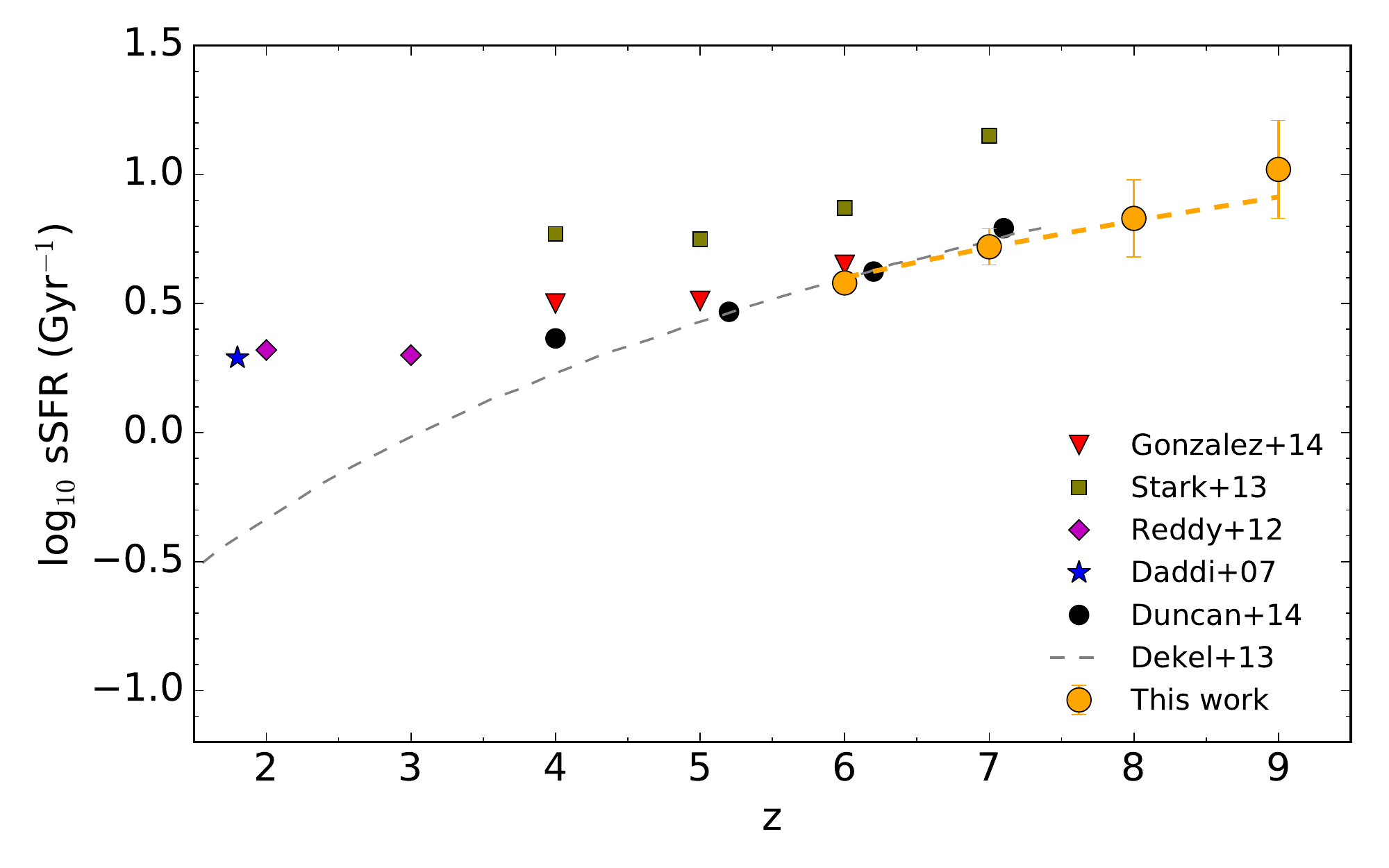}
\centering
\caption{Biweight mean sSFR and error on the mean for galaxies in a fixed stellar mass bin of log$_{10}(M/M_{\odot})=9.7\pm0.3$ as a function of redshift. The dashed yellow line is the best-fitting power law to the observed values of sSFR from $z\sim6-9$. The theoretical model of \citet{Dekel2013} in which the specific accretion rate follows the form sSFR $\propto(1+z)^{2.5}$ is shown by dashed grey line. }
\label{fig:ssfr_z}
\end{figure}

In order to compare with previous studies and to avoid incompleteness, we calculate our sSFR in a fixed stellar mass bin of log$_{10}(M/M_{\odot})=9.7\pm0.3$. Fig.~\ref{fig:ssfr_z} shows our results along with the values of sSFR estimated in the same mass bin from the literature and show a clear trend of increasing sSFR with redshift. We find that our sSFRs are in good agreement with \citet{Duncan2014} and \cite{Gonzalez2014} at $z\sim6$ and $z\sim7$, but are lower than \citet{Stark2013}. However, it turns out that the scatter in the $\mathrm{log_{10}(M_{*}/M\odot)-M_{UV}}$ relation plays a crucial role in the estimated values of sSFR, in that the sSFR will be
overestimated if only the $\mathrm{log_{10}(M_{*}/M\odot)-M_{UV}}$ relation is used without considering the numerous population of lower SFR sources with large mass-to-light ratios. \citet{Stark2013}  do not take this into account and report their results considering zero scatter. Nevertheless, they discuss this and report that if the scatter of 0.5 dex described by \citet{Gonzalez2012} is included, then this would lower the average sSFR by $2.8\times$ at $z\sim4$.  

To understand the evolution of sSFR with redshift, we fit a power-law to the observed values of sSFR, giving us sSFR $\propto(1+z)^{2.01\pm0.16}$. This is still a weaker trend than the theoretical expectations of \citet{Dekel2013}, in which the specific accretion rate follows the form sSFR $\propto(1+z)^{2.5}$, but is closer to that observed by \citet{Duncan2014} (sSFR $\propto(1+z)^{2.06\pm0.25}$).  

In general, we find an increase in the sSFR when we probe galaxies at higher redshifts.  This implies that there is a decline in the relative formation rate for galaxies at $z < 9$, and that the formation rate, relative to mass, increases at higher look back times. 

\subsection{UV luminosity density}
To calculate the UV luminosity density ($\rho_{\mathrm{UV}}$) in each redshift bin, we integrate the best-fit Schechter parameters for our UV LFs down to a faint-end magnitude limit of $M\mathrm{_{UV}=-13.5}$, which is the magnitude of the faintest galaxy in our sample. Since dust extinction decreases with both redshift and UV luminosity (\citealt{Bouwens2014, Finkelstein2015}), and hence has a negligible effect on estimated $\rho_{\textrm{UV}}$ values, we do not dust correct the data while calculating the UV luminosity density. We then use the \citet{Kennicutt1998} relation to convert the UV luminosity density values to cosmic star formation rate density (SFRD), assuming a Salpeter IMF and constant SFH over $\geq$100 Myr . 

\begin{figure}
\includegraphics[scale=0.45]{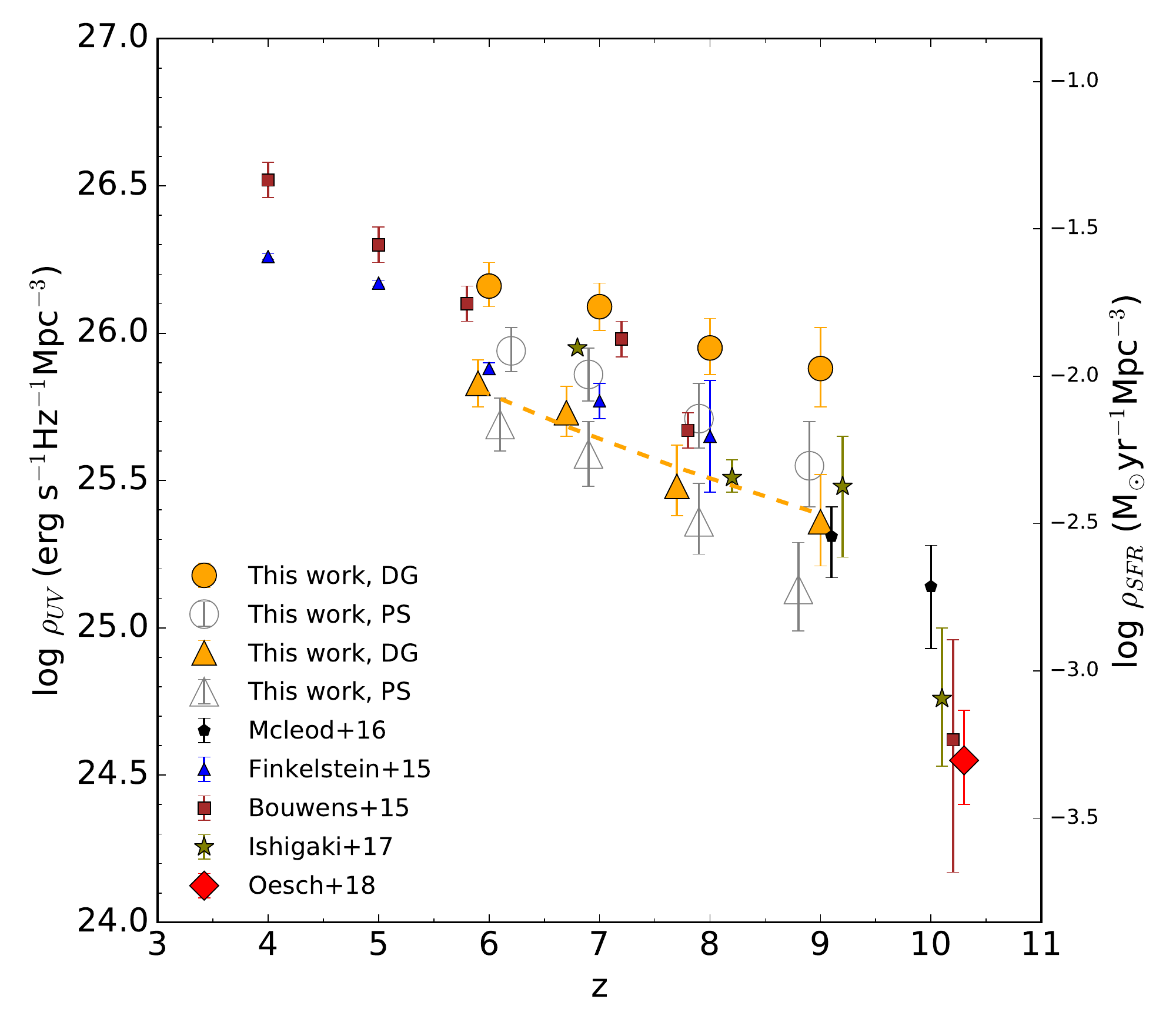}
\centering
\caption{Evolution of UV luminosity density as a function of redshift. $\rho_{\textrm{UV}}$ values calculated with a $M\mathrm{_{UV}=-13.5}$ limit are shown in orange filled circles and $\rho_{\textrm{UV}}$ calculated using a $M\mathrm{_{UV}=-17}$ limit are shown in orange filled triangles. Also shown are the $\rho_{\textrm{UV}}$ estimates considering point sources in our completeness simulations with open grey circles and open grey triangles for $M\mathrm{_{UV}=-13.5}$ and $M\mathrm{_{UV}=-17}$ respectively. The literature points shown are estimated with a $M\mathrm{_{UV}=-17}$ limit. The dashed yellow line is the power-law fit to the data points at $z\sim6-9$, such that $\rho_{\textrm{UV}}$ $\propto(1+z)^{-2.63}$. Fitting a power-law at $z\geq8$ values results in $\rho_{\textrm{UV}}$ $\propto(1+z)^{-4.61}$ (not shown).}
\label{fig:rho_uv}
\end{figure}

In order to facilitate comparison with other studies such as \citet{Bouwens2015}, \citet{Finkelstein2015}, \citet{Mcleod2016}, \citet{Ishigaki2017} and \citet{Oesch2018}, we also estimate $\rho_{\textrm{UV}}$ down to $M\mathrm{_{UV}=-17}$. Fig.~\ref{fig:rho_uv} shows the evolution of UV luminosity density as well as SFR density from $z=6-9$ and Table~\ref{tab:smd_table} shows the derived values of $\rho_{\textrm{UV}}$ and $\rho_{\textrm{SFR}}$ in each redshift bin. In both the cases ($M\mathrm{_{UV}=-13.5}$ and $M\mathrm{_{UV}=-17}$) our results support a smooth decline of $\rho_{\textrm{UV}}$ towards high redshifts when disc-like galaxies are considered in our completeness simulations. However, there appears to be a slight accelerated decline of $\rho_{\textrm{UV}}$ and $\rho_{\textrm{SFR}}$ when using completeness results with point sources (See Fig.~\ref{fig:rho_uv}). The values estimated with point sources are shown in parentheses in Table~\ref{tab:smd_table}. The quoted 1$\sigma$ uncertainties represent the minimum and maximum range of $\rho_{\textrm{UV}}$ and $\rho_{\textrm{SFR}}$ within the 1$\sigma$ contours of Schechter parameters obtained from our MCMC analysis in Section~\ref{sec:mcmc}. 

Performing a simple power-law fit to the data points estimated with disc-like galaxies at $z\sim6-9$, we find that $\rho_{\textrm{UV}}$ $\propto(1+z)^{-2.63}$ (shown by dotted yellow line in Fig.~\ref{fig:rho_uv}), whereas fitting a power-law at $z\geq8$ values results in $\rho_{\textrm{UV}}$ $\propto(1+z)^{-4.61}$. This is even shallower than \citep{Mcleod2016} who find that $\rho_{\textrm{UV}}$ $\propto(1+z)^{-5.8}$ beyond $z\simeq8$, supporting a smooth decline in $\rho_{\textrm{UV}}$. This is in contrast with \citet{Oesch2014}, \citet{Bouwens2015}, \citet{Ishigaki2017} and \citet{Oesch2018} who find an accelerated decline in $\rho_{\textrm{UV}}$ at $z\geq8$ when they integrate it down to $M\mathrm{_{UV}=-17}$.
 
\section{SUMMARY}
\label{sec:summary}
In this paper we have exploited the power of gravitational lensing of massive clusters and combined the \textit{HST}, VLT and \textit{Spitzer} imaging of MACS0416 cluster and its parallel field to probe galaxy evolution with galaxy stellar mass functions and UV luminosity functions out to z$\sim$9. We have developed a novel method to subtract the massive foreground galaxies that lie close to the critical line from the MACS0416 cluster, allowing for a deeper and cleaner detection of the faintest systems at $z\geqslant6$.

We have constructed a multi-wavelength catalog (from 0.4 to 4.5$\mu$m) using all the bands (\textit{HST}, K and IRAC) available for the HFFs, allowing us to put better constraints on redshift estimates as well as obtain robust stellar mass estimates. From this, we have estimated the stellar masses of our high-z sample through SED fitting with the inclusion of nebular emission lines and have derived, for the first time, the GSMF at $z=6-9$ for the HFF program. 

Using the same sample, we have also derived the UV LF at $z=6-9$. For our high-z sample, we have estimated the dust-corrected star formation rates from UV luminosities and UV continuum slopes. From this, we have calculated the sSFR and UV luminosity density in the redshift range probed. Our key conclusions are as follows:

\begin{enumerate}
  \item Our new measurements of the GSMF show an apparent steepening of the low mass end slope $\alpha$ with increasing redshift ($-1.98_{-0.07}^{+0.07}$, $-2.01_{-0.13}^{+0.17}$, $-2.30_{-0.46}^{+0.51}$ and $-2.38_{-0.88}^{+0.72}$ at $z=6,7,8$ and $9$ respectively) within the error bars, statistically steeper than previously observed mass functions at slightly lower redshifts, and we find no evidence of a turnover in the mass range probed. We also find a decrease in normalization $\Phi^{*}$ with increasing redshift, but no evolution in characteristic mass $M^{*}$ is observed.
  \item The faint-end slope of the UV LF also exhibit an apparent steepening with increasing redshift (from $-2.03_{-0.10}^{+0.12}$ at $z=6$ to $-2.20_{-0.47}^{+0.51}$ at $z=9$), without any evidence of a turnover. These measurements are consistent with a continuation of the trends seen from lower redshift, however more data is needed to confirm the trend at these redshifts. The normalization $\Phi^{*}$ of UV LF decreases with increasing redshift and a weak evolution in $M_{\mathrm{UV}}^{*}$ is also observed, with it decreasing with redshift, implying that the evolution of the UV luminosity function with redshift appears to be more consistent with an evolution of density.
  \item The slopes of our best fitted $\mathrm{log_{10}(M_{*}/M\odot)-M_{UV}}$ relation are close to a constant mass-to-light ratio of $-0.40$, suggesting no strong evolution of mass-to-light ratio with luminosity. We notice that normalization, on the other hand, evolves very weakly from $z\sim6$ to $z\sim9$, with a decrease in normalization with increasing redshift.
  \item From our new measurements of the GSMF, we estimate the stellar mass density and find that the SMD increases by a factor of $\sim15_{-6}^{+21}$, from log$_{10}\rho_{*}=5.61_{-0.90}^{+0.92}$ at $z=9$ to log$_{10}\rho_{*}=6.79_{-0.12}^{+0.13}$ at $z=6$.
  \item We estimate the specific star formation rates ($\mathrm{sSFR}=\mathrm{SFR/M_{*}}$) of our sample, and find that for a fixed stellar mass of $5\times10^{9}M_{\odot}$, sSFR $\propto(1+z)^{2.01\pm0.16}$.
  \item From our new measurements, we estimate the UV luminosity density ($\rho_{\textrm{UV}}$) and the cosmic star formation rate density ($\rho_{\textrm{SFR}}$), and find that our results support a smooth decline of $\rho_{\textrm{SFR}}$ towards high redshifts.
\end{enumerate}

Our study exhibits the power of gravitational lensing to probe the faintest and earliest galaxies in the early Universe. While the results from this study are very intriguing, the uncertainties are still large due to a small sample size. The analysis of the complete HFF dataset comprising six clusters and six associated parallel fields will increase the signal to noise of these results, providing robust constraints in the future. This will, however, require a significant amount of effort as subtracting the foreground galaxies takes up the bulk of the effort when studying nearby massive lensing clusters. On the other hand, to probe even higher redshifts requires the advantage of JWST. The advent of JWST will allow us to extend this study all the way back to redshift of $z=12$, if not even earlier. Ultimately, it will also allow us to probe far deeper down the luminosity and mass functions to determine if and where there is a turnover in the number counts as a function of mass and luminosity. Until then, any future studies of lensing clusters must take into account the careful subtraction of the foreground cluster galaxies such as we have done here, otherwise the results will be biased and galaxies at the faintest limits will be missed.
   
\section{ACKNOWLEDGEMENTS}
The authors would like to thank the referee for their comments, which have greatly improved this paper. We would also like to thank Rachael Livermore, Rychard Bouwens and Derek Mcleod for their suggestions and acknowledge the discussions that led to improvements in this paper. In addition, we would like to thank Emiliano Merlin for his support and guidance with the T-PHOT software. This work is based on the observations made with the NASA/ESA Hubble Space Telescope, obtained from the Mikulski Archive for Space Telescopes (MAST) at the Space Telescope Science Institute (STScI), which is operated by the Association of Universities for Research in Astronomy, Inc., under NASA contract NAS 5-26555. This work is also based on observations made with the Spitzer Space Telescope, which is operated by the Jet Propulsion Laboratory (JPL), California Institute of Technology under a contract with NASA and uses the data taken with the Hawk-I instrument on the European Southern Observatory (ESO) Very Large telescope (VLT) from ESO programme 092.A-0472. This work utilizes gravitational lensing models produced by PIs Brada\u{c}, Ebeling, Merten \& Zitrin, Sharon, and Williams funded as part of the \textit{HST} Frontier Fields program conducted by STScI. The lens models were obtained from the MAST. R.A.B acknowledges funding from the Science and Technology Facilities Council (STFC). K.J.D acknowledges support from the ERC Advanced Investigator programme NewClusters 321271.




\bibliographystyle{mnras}
\bibliography{hff_massfunctions_withoutboldtext.bib} 

\begin{thebibliography}{}
\makeatletter
\relax
\def\mn@urlcharsother{\let\do\@makeother \do\$\do\&\do\#\do\^\do\_\do\%\do\~}
\def\mn@doi{\begingroup\mn@urlcharsother \@ifnextchar [ {\mn@doi@}
  {\mn@doi@[]}}
\def\mn@doi@[#1]#2{\def\@tempa{#1}\ifx\@tempa\@empty \href
  {http://dx.doi.org/#2} {doi:#2}\else \href {http://dx.doi.org/#2} {#1}\fi
  \endgroup}
\def\mn@eprint#1#2{\mn@eprint@#1:#2::\@nil}
\def\mn@eprint@arXiv#1{\href {http://arxiv.org/abs/#1} {{\tt arXiv:#1}}}
\def\mn@eprint@dblp#1{\href {http://dblp.uni-trier.de/rec/bibtex/#1.xml}
  {dblp:#1}}
\def\mn@eprint@#1:#2:#3:#4\@nil{\def\@tempa {#1}\def\@tempb {#2}\def\@tempc
  {#3}\ifx \@tempc \@empty \let \@tempc \@tempb \let \@tempb \@tempa \fi \ifx
  \@tempb \@empty \def\@tempb {arXiv}\fi \@ifundefined
  {mn@eprint@\@tempb}{\@tempb:\@tempc}{\expandafter \expandafter \csname
  mn@eprint@\@tempb\endcsname \expandafter{\@tempc}}}

\bibitem[\protect\citeauthoryear{{Arnouts} \& {Ilbert}}{{Arnouts} \&
  {Ilbert}}{2011}]{Arnouts2011}
{Arnouts} S.,  {Ilbert} O.,  2011, {LePHARE: Photometric Analysis for Redshift
  Estimate}, Astrophysics Source Code Library (\mn@eprint {ascl} {1108.009})

\bibitem[\protect\citeauthoryear{{Atek} et~al.,}{{Atek}
  et~al.}{2014}]{Atek2014}
{Atek} H.,  et~al., 2014, \mn@doi [\apj] {10.1088/0004-637X/786/1/60}, \href
  {http://adsabs.harvard.edu/abs/2014ApJ...786...60A} {786, 60}

\bibitem[\protect\citeauthoryear{{Atek}, {Richard}, {Kneib}  \&
  {Schaerer}}{{Atek} et~al.}{2018}]{Atek2018}
{Atek} H.,  {Richard} J.,  {Kneib} J.-P.,   {Schaerer} D.,  2018, preprint,
  \href {http://adsabs.harvard.edu/abs/2018arXiv180309747A} {} (\mn@eprint
  {arXiv} {1803.09747})

\bibitem[\protect\citeauthoryear{{Avni} \& {Bahcall}}{{Avni} \&
  {Bahcall}}{1980}]{Avni1980}
{Avni} Y.,  {Bahcall} J.~N.,  1980, \mn@doi [\apj] {10.1086/157673}, \href
  {http://adsabs.harvard.edu/abs/1980ApJ...235..694A} {235, 694}

\bibitem[\protect\citeauthoryear{{Balestra} et~al.,}{{Balestra}
  et~al.}{2016}]{Balestra2016}
{Balestra} I.,  et~al., 2016, \mn@doi [\apjs] {10.3847/0067-0049/224/2/33},
  \href {http://adsabs.harvard.edu/abs/2016ApJS..224...33B} {224, 33}

\bibitem[\protect\citeauthoryear{{Barden}, {H{\"a}u{\ss}ler}, {Peng},
  {McIntosh}  \& {Guo}}{{Barden} et~al.}{2012}]{Barden2012}
{Barden} M.,  {H{\"a}u{\ss}ler} B.,  {Peng} C.~Y.,  {McIntosh} D.~H.,   {Guo}
  Y.,  2012, \mn@doi [\mnras] {10.1111/j.1365-2966.2012.20619.x}, \href
  {http://adsabs.harvard.edu/abs/2012MNRAS.422..449B} {422, 449}

\bibitem[\protect\citeauthoryear{{Behroozi}, {Wechsler}, {Hearin}  \&
  {Conroy}}{{Behroozi} et~al.}{2018}]{Behroozi2018}
{Behroozi} P.,  {Wechsler} R.,  {Hearin} A.,   {Conroy} C.,  2018, preprint,
  \href {http://adsabs.harvard.edu/abs/2018arXiv180607893B} {} (\mn@eprint
  {arXiv} {1806.07893})

\bibitem[\protect\citeauthoryear{{Benitez}}{{Benitez}}{2000}]{Benitez2000}
{Benitez} N.,  2000, \mn@doi [\apj] {10.1086/308947}, \href
  {http://adsabs.harvard.edu/abs/2000ApJ...536..571B} {536, 571}

\bibitem[\protect\citeauthoryear{{Bernyk} et~al.,}{{Bernyk}
  et~al.}{2016}]{Bernyk2016}
{Bernyk} M.,  et~al., 2016, \mn@doi [\apjs] {10.3847/0067-0049/223/1/9}, \href
  {http://adsabs.harvard.edu/abs/2016ApJS..223....9B} {223, 9}

\bibitem[\protect\citeauthoryear{{Bertin} \& {Arnouts}}{{Bertin} \&
  {Arnouts}}{1996}]{Bertin1996}
{Bertin} E.,  {Arnouts} S.,  1996, \mn@doi [\aaps] {10.1051/aas:1996164}, \href
  {http://adsabs.harvard.edu/abs/1996A%26AS..117..393B} {117, 393}

\bibitem[\protect\citeauthoryear{{Bertin}, {Mellier}, {Radovich}, {Missonnier},
  {Didelon}  \& {Morin}}{{Bertin} et~al.}{2002}]{Bertin2002}
{Bertin} E.,  {Mellier} Y.,  {Radovich} M.,  {Missonnier} G.,  {Didelon} P.,
  {Morin} B.,  2002, in {Bohlender} D.~A.,  {Durand} D.,   {Handley} T.~H.,
  eds,  Astronomical Society of the Pacific Conference Series Vol. 281,
  Astronomical Data Analysis Software and Systems XI. p.~228

\bibitem[\protect\citeauthoryear{{Bolzonella}, {Miralles}  \&
  {Pell{\'o}}}{{Bolzonella} et~al.}{2000}]{Bolzonella2000}
{Bolzonella} M.,  {Miralles} J.-M.,   {Pell{\'o}} R.,  2000, \aap, \href
  {http://adsabs.harvard.edu/abs/2000A%26A...363..476B} {363, 476}

\bibitem[\protect\citeauthoryear{{Bouwens}, {Illingworth}, {Franx}  \&
  {Ford}}{{Bouwens} et~al.}{2007}]{Bouwens2007}
{Bouwens} R.~J.,  {Illingworth} G.~D.,  {Franx} M.,   {Ford} H.,  2007, \mn@doi
  [\apj] {10.1086/521811}, \href
  {http://adsabs.harvard.edu/abs/2007ApJ...670..928B} {670, 928}

\bibitem[\protect\citeauthoryear{{Bouwens} et~al.,}{{Bouwens}
  et~al.}{2009}]{Bouwens2009}
{Bouwens} R.~J.,  et~al., 2009, \mn@doi [\apj] {10.1088/0004-637X/690/2/1764},
  \href {http://adsabs.harvard.edu/abs/2009ApJ...690.1764B} {690, 1764}

\bibitem[\protect\citeauthoryear{{Bouwens} et~al.,}{{Bouwens}
  et~al.}{2011}]{Bouwens2011}
{Bouwens} R.~J.,  et~al., 2011, \mn@doi [\nat] {10.1038/nature09717}, \href
  {http://adsabs.harvard.edu/abs/2011Natur.469..504B} {469, 504}

\bibitem[\protect\citeauthoryear{{Bouwens} et~al.,}{{Bouwens}
  et~al.}{2012}]{Bouwens2012}
{Bouwens} R.~J.,  et~al., 2012, \mn@doi [\apj] {10.1088/0004-637X/754/2/83},
  \href {http://adsabs.harvard.edu/abs/2012ApJ...754...83B} {754, 83}

\bibitem[\protect\citeauthoryear{{Bouwens} et~al.,}{{Bouwens}
  et~al.}{2014}]{Bouwens2014}
{Bouwens} R.~J.,  et~al., 2014, \mn@doi [\apj] {10.1088/0004-637X/795/2/126},
  \href {http://adsabs.harvard.edu/abs/2014ApJ...795..126B} {795, 126}

\bibitem[\protect\citeauthoryear{{Bouwens} et~al.,}{{Bouwens}
  et~al.}{2015}]{Bouwens2015}
{Bouwens} R.~J.,  et~al., 2015, \mn@doi [\apj] {10.1088/0004-637X/803/1/34},
  \href {http://adsabs.harvard.edu/abs/2015ApJ...803...34B} {803, 34}

\bibitem[\protect\citeauthoryear{{Bouwens} et~al.,}{{Bouwens}
  et~al.}{2016}]{Bouwens2016}
{Bouwens} R.~J.,  et~al., 2016, \mn@doi [\apj] {10.3847/0004-637X/830/2/67},
  \href {http://adsabs.harvard.edu/abs/2016ApJ...830...67B} {830, 67}

\bibitem[\protect\citeauthoryear{{Bouwens}, {Illingworth}, {Oesch}, {Atek},
  {Lam}  \& {Stefanon}}{{Bouwens} et~al.}{2017a}]{Bouwens2017}
{Bouwens} R.~J.,  {Illingworth} G.~D.,  {Oesch} P.~A.,  {Atek} H.,  {Lam} D.,
  {Stefanon} M.,  2017a, \mn@doi [\apj] {10.3847/1538-4357/aa74e4}, \href
  {http://adsabs.harvard.edu/abs/2017ApJ...843...41B} {843, 41}

\bibitem[\protect\citeauthoryear{{Bouwens}, {Oesch}, {Illingworth}, {Ellis}  \&
  {Stefanon}}{{Bouwens} et~al.}{2017b}]{Bouwens2017b}
{Bouwens} R.~J.,  {Oesch} P.~A.,  {Illingworth} G.~D.,  {Ellis} R.~S.,
  {Stefanon} M.,  2017b, \mn@doi [\apj] {10.3847/1538-4357/aa70a4}, \href
  {http://adsabs.harvard.edu/abs/2017ApJ...843..129B} {843, 129}

\bibitem[\protect\citeauthoryear{{Brada{\v c}} et~al.,}{{Brada{\v c}}
  et~al.}{2009}]{Bradac2009}
{Brada{\v c}} M.,  et~al., 2009, \mn@doi [\apj] {10.1088/0004-637X/706/2/1201},
  \href {http://adsabs.harvard.edu/abs/2009ApJ...706.1201B} {706, 1201}

\bibitem[\protect\citeauthoryear{{Bradley} et~al.,}{{Bradley}
  et~al.}{2014}]{Bradley2014}
{Bradley} L.~D.,  et~al., 2014, \mn@doi [\apj] {10.1088/0004-637X/792/1/76},
  \href {http://adsabs.harvard.edu/abs/2014ApJ...792...76B} {792, 76}

\bibitem[\protect\citeauthoryear{{Brammer}, {van Dokkum}  \& {Coppi}}{{Brammer}
  et~al.}{2008}]{Brammer2008}
{Brammer} G.~B.,  {van Dokkum} P.~G.,   {Coppi} P.,  2008, \mn@doi [\apj]
  {10.1086/591786}, \href {http://adsabs.harvard.edu/abs/2008ApJ...686.1503B}
  {686, 1503}

\bibitem[\protect\citeauthoryear{{Brammer} et~al.,}{{Brammer}
  et~al.}{2016}]{Brammer2016}
{Brammer} G.~B.,  et~al., 2016, \mn@doi [\apjs] {10.3847/0067-0049/226/1/6},
  \href {http://adsabs.harvard.edu/abs/2016ApJS..226....6B} {226, 6}

\bibitem[\protect\citeauthoryear{{Bruzual} \& {Charlot}}{{Bruzual} \&
  {Charlot}}{2003}]{Bruzual2003}
{Bruzual} G.,  {Charlot} S.,  2003, \mn@doi [\mnras]
  {10.1046/j.1365-8711.2003.06897.x}, \href
  {http://adsabs.harvard.edu/abs/2003MNRAS.344.1000B} {344, 1000}

\bibitem[\protect\citeauthoryear{{Calzetti}, {Kinney}  \&
  {Storchi-Bergmann}}{{Calzetti} et~al.}{1994}]{Calzetti1994}
{Calzetti} D.,  {Kinney} A.~L.,   {Storchi-Bergmann} T.,  1994, \mn@doi [\apj]
  {10.1086/174346}, \href {http://adsabs.harvard.edu/abs/1994ApJ...429..582C}
  {429, 582}

\bibitem[\protect\citeauthoryear{{Calzetti}, {Armus}, {Bohlin}, {Kinney},
  {Koornneef}  \& {Storchi-Bergmann}}{{Calzetti} et~al.}{2000}]{Calzetti2000}
{Calzetti} D.,  {Armus} L.,  {Bohlin} R.~C.,  {Kinney} A.~L.,  {Koornneef} J.,
   {Storchi-Bergmann} T.,  2000, \mn@doi [\apj] {10.1086/308692}, \href
  {http://adsabs.harvard.edu/abs/2000ApJ...533..682C} {533, 682}

\bibitem[\protect\citeauthoryear{{Caminha} et~al.,}{{Caminha}
  et~al.}{2017}]{Caminha2017}
{Caminha} G.~B.,  et~al., 2017, \mn@doi [\aap] {10.1051/0004-6361/201629297},
  \href {http://adsabs.harvard.edu/abs/2017A%26A...600A..90C} {600, A90}

\bibitem[\protect\citeauthoryear{{Castellano} et~al.,}{{Castellano}
  et~al.}{2016}]{Castellano2016}
{Castellano} M.,  et~al., 2016, \mn@doi [\aap] {10.1051/0004-6361/201527514},
  \href {http://adsabs.harvard.edu/abs/2016A%26A...590A..31C} {590, A31}

\bibitem[\protect\citeauthoryear{{Chabrier}}{{Chabrier}}{2003}]{chabrier2003}
{Chabrier} G.,  2003, \mn@doi [\pasp] {10.1086/376392}, \href
  {http://adsabs.harvard.edu/abs/2003PASP..115..763C} {115, 763}

\bibitem[\protect\citeauthoryear{{Coe} et~al.,}{{Coe} et~al.}{2013}]{Coe2013}
{Coe} D.,  et~al., 2013, \mn@doi [\apj] {10.1088/0004-637X/762/1/32}, \href
  {http://adsabs.harvard.edu/abs/2013ApJ...762...32C} {762, 32}

\bibitem[\protect\citeauthoryear{{Conselice}, {Wilkinson}, {Duncan}  \&
  {Mortlock}}{{Conselice} et~al.}{2016}]{Conselice2016}
{Conselice} C.~J.,  {Wilkinson} A.,  {Duncan} K.,   {Mortlock} A.,  2016,
  \mn@doi [\apj] {10.3847/0004-637X/830/2/83}, \href
  {http://adsabs.harvard.edu/abs/2016ApJ...830...83C} {830, 83}

\bibitem[\protect\citeauthoryear{{Dahlen} et~al.,}{{Dahlen}
  et~al.}{2013}]{Dahlen2013}
{Dahlen} T.,  et~al., 2013, \mn@doi [\apj] {10.1088/0004-637X/775/2/93}, \href
  {http://adsabs.harvard.edu/abs/2013ApJ...775...93D} {775, 93}

\bibitem[\protect\citeauthoryear{{Dav{\'e}}}{{Dav{\'e}}}{2008}]{Dave2008}
{Dav{\'e}} R.,  2008, \mn@doi [\mnras] {10.1111/j.1365-2966.2008.12866.x},
  \href {http://adsabs.harvard.edu/abs/2008MNRAS.385..147D} {385, 147}

\bibitem[\protect\citeauthoryear{{De Santis}, {Grazian}, {Fontana}  \&
  {Santini}}{{De Santis} et~al.}{2007}]{DeSantis2007}
{De Santis} C.,  {Grazian} A.,  {Fontana} A.,   {Santini} P.,  2007, \mn@doi
  [\na] {10.1016/j.newast.2006.10.004}, \href
  {http://adsabs.harvard.edu/abs/2007NewA...12..271D} {12, 271}

\bibitem[\protect\citeauthoryear{{Dekel}, {Zolotov}, {Tweed}, {Cacciato},
  {Ceverino}  \& {Primack}}{{Dekel} et~al.}{2013}]{Dekel2013}
{Dekel} A.,  {Zolotov} A.,  {Tweed} D.,  {Cacciato} M.,  {Ceverino} D.,
  {Primack} J.~R.,  2013, \mn@doi [\mnras] {10.1093/mnras/stt1338}, \href
  {http://adsabs.harvard.edu/abs/2013MNRAS.435..999D} {435, 999}

\bibitem[\protect\citeauthoryear{{Duncan} \& {Conselice}}{{Duncan} \&
  {Conselice}}{2015}]{Duncan2015}
{Duncan} K.,  {Conselice} C.~J.,  2015, \mn@doi [\mnras]
  {10.1093/mnras/stv1049}, \href
  {http://adsabs.harvard.edu/abs/2015MNRAS.451.2030D} {451, 2030}

\bibitem[\protect\citeauthoryear{{Duncan} et~al.,}{{Duncan}
  et~al.}{2014}]{Duncan2014}
{Duncan} K.,  et~al., 2014, \mn@doi [\mnras] {10.1093/mnras/stu1622}, \href
  {http://adsabs.harvard.edu/abs/2014MNRAS.444.2960D} {444, 2960}

\bibitem[\protect\citeauthoryear{{Eales}}{{Eales}}{1993}]{Eales1993}
{Eales} S.,  1993, \mn@doi [\apj] {10.1086/172257}, \href
  {http://adsabs.harvard.edu/abs/1993ApJ...404...51E} {404, 51}

\bibitem[\protect\citeauthoryear{{Erb}, {Pettini}, {Shapley}, {Steidel}, {Law}
  \& {Reddy}}{{Erb} et~al.}{2010}]{Erb2010}
{Erb} D.~K.,  {Pettini} M.,  {Shapley} A.~E.,  {Steidel} C.~C.,  {Law} D.~R.,
  {Reddy} N.~A.,  2010, \mn@doi [\apj] {10.1088/0004-637X/719/2/1168}, \href
  {http://adsabs.harvard.edu/abs/2010ApJ...719.1168E} {719, 1168}

\bibitem[\protect\citeauthoryear{{Ferguson} et~al.,}{{Ferguson}
  et~al.}{2004}]{Ferguson2004}
{Ferguson} H.~C.,  et~al., 2004, \mn@doi [\apjl] {10.1086/378578}, \href
  {http://adsabs.harvard.edu/abs/2004ApJ...600L.107F} {600, L107}

\bibitem[\protect\citeauthoryear{{Finkelstein} et~al.,}{{Finkelstein}
  et~al.}{2012}]{Finkelstein2012}
{Finkelstein} S.~L.,  et~al., 2012, \mn@doi [\apj]
  {10.1088/0004-637X/756/2/164}, \href
  {http://adsabs.harvard.edu/abs/2012ApJ...756..164F} {756, 164}

\bibitem[\protect\citeauthoryear{{Finkelstein} et~al.,}{{Finkelstein}
  et~al.}{2015}]{Finkelstein2015}
{Finkelstein} S.~L.,  et~al., 2015, \mn@doi [\apj]
  {10.1088/0004-637X/810/1/71}, \href
  {http://adsabs.harvard.edu/abs/2015ApJ...810...71F} {810, 71}

\bibitem[\protect\citeauthoryear{{Fioc} \& {Rocca-Volmerange}}{{Fioc} \&
  {Rocca-Volmerange}}{1997}]{Fioc1997}
{Fioc} M.,  {Rocca-Volmerange} B.,  1997, \aap, \href
  {http://adsabs.harvard.edu/abs/1997A%26A...326..950F} {326, 950}

\bibitem[\protect\citeauthoryear{{Foreman-Mackey}, {Hogg}, {Lang}  \&
  {Goodman}}{{Foreman-Mackey} et~al.}{2013}]{Foreman2013}
{Foreman-Mackey} D.,  {Hogg} D.~W.,  {Lang} D.,   {Goodman} J.,  2013, \mn@doi
  [\pasp] {10.1086/670067}, \href
  {http://adsabs.harvard.edu/abs/2013PASP..125..306F} {125, 306}

\bibitem[\protect\citeauthoryear{{Galametz} et~al.,}{{Galametz}
  et~al.}{2013}]{Galametz2013}
{Galametz} A.,  et~al., 2013, \mn@doi [\apjs] {10.1088/0067-0049/206/2/10},
  \href {http://adsabs.harvard.edu/abs/2013ApJS..206...10G} {206, 10}

\bibitem[\protect\citeauthoryear{{Gardner} et~al.,}{{Gardner}
  et~al.}{2006}]{Gardner2006}
{Gardner} J.~P.,  et~al., 2006, \mn@doi [\ssr] {10.1007/s11214-006-8315-7},
  \href {http://adsabs.harvard.edu/abs/2006SSRv..123..485G} {123, 485}

\bibitem[\protect\citeauthoryear{{Gonz{\'a}lez}, {Labb{\'e}}, {Bouwens},
  {Illingworth}, {Franx}, {Kriek}  \& {Brammer}}{{Gonz{\'a}lez}
  et~al.}{2010}]{Gonzalez2010}
{Gonz{\'a}lez} V.,  {Labb{\'e}} I.,  {Bouwens} R.~J.,  {Illingworth} G.,
  {Franx} M.,  {Kriek} M.,   {Brammer} G.~B.,  2010, \mn@doi [\apj]
  {10.1088/0004-637X/713/1/115}, \href
  {http://adsabs.harvard.edu/abs/2010ApJ...713..115G} {713, 115}

\bibitem[\protect\citeauthoryear{{Gonz{\'a}lez}, {Labb{\'e}}, {Bouwens},
  {Illingworth}, {Franx}  \& {Kriek}}{{Gonz{\'a}lez}
  et~al.}{2011}]{Gonzalez2011}
{Gonz{\'a}lez} V.,  {Labb{\'e}} I.,  {Bouwens} R.~J.,  {Illingworth} G.,
  {Franx} M.,   {Kriek} M.,  2011, \mn@doi [\apjl]
  {10.1088/2041-8205/735/2/L34}, \href
  {http://adsabs.harvard.edu/abs/2011ApJ...735L..34G} {735, L34}

\bibitem[\protect\citeauthoryear{{Gonz{\'a}lez}, {Bouwens}, {Labb{\'e}},
  {Illingworth}, {Oesch}, {Franx}  \& {Magee}}{{Gonz{\'a}lez}
  et~al.}{2012}]{Gonzalez2012}
{Gonz{\'a}lez} V.,  {Bouwens} R.~J.,  {Labb{\'e}} I.,  {Illingworth} G.,
  {Oesch} P.,  {Franx} M.,   {Magee} D.,  2012, \mn@doi [\apj]
  {10.1088/0004-637X/755/2/148}, \href
  {http://adsabs.harvard.edu/abs/2012ApJ...755..148G} {755, 148}

\bibitem[\protect\citeauthoryear{{Gonz{\'a}lez}, {Bouwens}, {Illingworth},
  {Labb{\'e}}, {Oesch}, {Franx}  \& {Magee}}{{Gonz{\'a}lez}
  et~al.}{2014}]{Gonzalez2014}
{Gonz{\'a}lez} V.,  {Bouwens} R.,  {Illingworth} G.,  {Labb{\'e}} I.,  {Oesch}
  P.,  {Franx} M.,   {Magee} D.,  2014, \mn@doi [\apj]
  {10.1088/0004-637X/781/1/34}, \href
  {http://adsabs.harvard.edu/abs/2014ApJ...781...34G} {781, 34}

\bibitem[\protect\citeauthoryear{{Goodman} \& {Weare}}{{Goodman} \&
  {Weare}}{2010}]{Goodman2010}
{Goodman} J.,  {Weare} J.,  2010, \mn@doi [Communications in Applied
  Mathematics and Computational Science, Vol.~5, No.~1, p.~65-80, 2010]
  {10.2140/camcos.2010.5.65}, \href
  {http://adsabs.harvard.edu/abs/2010CAMCS...5...65G} {5, 65}

\bibitem[\protect\citeauthoryear{{Grazian} et~al.,}{{Grazian}
  et~al.}{2011}]{Grazian2011}
{Grazian} A.,  et~al., 2011, \mn@doi [\aap] {10.1051/0004-6361/201015754},
  \href {http://adsabs.harvard.edu/abs/2011A%26A...532A..33G} {532, A33}

\bibitem[\protect\citeauthoryear{{Grazian} et~al.,}{{Grazian}
  et~al.}{2012}]{Grazian2012}
{Grazian} A.,  et~al., 2012, \mn@doi [\aap] {10.1051/0004-6361/201219669},
  \href {http://adsabs.harvard.edu/abs/2012A%26A...547A..51G} {547, A51}

\bibitem[\protect\citeauthoryear{{Grazian} et~al.,}{{Grazian}
  et~al.}{2015}]{Grazian2015}
{Grazian} A.,  et~al., 2015, \mn@doi [\aap] {10.1051/0004-6361/201424750},
  \href {http://adsabs.harvard.edu/abs/2015A%26A...575A..96G} {575, A96}

\bibitem[\protect\citeauthoryear{{Grillo} et~al.,}{{Grillo}
  et~al.}{2015}]{Grillo2015}
{Grillo} C.,  et~al., 2015, \mn@doi [\apj] {10.1088/0004-637X/800/1/38}, \href
  {http://adsabs.harvard.edu/abs/2015ApJ...800...38G} {800, 38}

\bibitem[\protect\citeauthoryear{{Gu}, {Ho}, {Peng}  \& {Huang}}{{Gu}
  et~al.}{2013}]{Gu2013}
{Gu} M.,  {Ho} L.~C.,  {Peng} C.~Y.,   {Huang} S.,  2013, \mn@doi [\apj]
  {10.1088/0004-637X/773/1/34}, \href
  {http://adsabs.harvard.edu/abs/2013ApJ...773...34G} {773, 34}

\bibitem[\protect\citeauthoryear{{Guo} et~al.,}{{Guo} et~al.}{2013}]{Guo2013}
{Guo} Y.,  et~al., 2013, \mn@doi [\apjs] {10.1088/0067-0049/207/2/24}, \href
  {http://adsabs.harvard.edu/abs/2013ApJS..207...24G} {207, 24}

\bibitem[\protect\citeauthoryear{{Ilbert} et~al.,}{{Ilbert}
  et~al.}{2005}]{Ilbert2005}
{Ilbert} O.,  et~al., 2005, \mn@doi [\aap] {10.1051/0004-6361:20041961}, \href
  {http://adsabs.harvard.edu/abs/2005A%26A...439..863I} {439, 863}

\bibitem[\protect\citeauthoryear{{Ishigaki}, {Kawamata}, {Ouchi}, {Oguri},
  {Shimasaku}  \& {Ono}}{{Ishigaki} et~al.}{2015}]{Ishigaki2015}
{Ishigaki} M.,  {Kawamata} R.,  {Ouchi} M.,  {Oguri} M.,  {Shimasaku} K.,
  {Ono} Y.,  2015, \mn@doi [\apj] {10.1088/0004-637X/799/1/12}, \href
  {http://adsabs.harvard.edu/abs/2015ApJ...799...12I} {799, 12}

\bibitem[\protect\citeauthoryear{{Ishigaki}, {Kawamata}, {Ouchi}, {Oguri},
  {Shimasaku}  \& {Ono}}{{Ishigaki} et~al.}{2018}]{Ishigaki2017}
{Ishigaki} M.,  {Kawamata} R.,  {Ouchi} M.,  {Oguri} M.,  {Shimasaku} K.,
  {Ono} Y.,  2018, \mn@doi [\apj] {10.3847/1538-4357/aaa544}, \href
  {http://adsabs.harvard.edu/abs/2018ApJ...854...73I} {854, 73}

\bibitem[\protect\citeauthoryear{{Jauzac} et~al.,}{{Jauzac}
  et~al.}{2014}]{Jauzac2014}
{Jauzac} M.,  et~al., 2014, \mn@doi [\mnras] {10.1093/mnras/stu1355}, \href
  {http://adsabs.harvard.edu/abs/2014MNRAS.443.1549J} {443, 1549}

\bibitem[\protect\citeauthoryear{{Johnson}, {Sharon}, {Bayliss}, {Gladders},
  {Coe}  \& {Ebeling}}{{Johnson} et~al.}{2014}]{Johnson2014}
{Johnson} T.~L.,  {Sharon} K.,  {Bayliss} M.~B.,  {Gladders} M.~D.,  {Coe} D.,
   {Ebeling} H.,  2014, \mn@doi [\apj] {10.1088/0004-637X/797/1/48}, \href
  {http://adsabs.harvard.edu/abs/2014ApJ...797...48J} {797, 48}

\bibitem[\protect\citeauthoryear{{Jullo} \& {Kneib}}{{Jullo} \&
  {Kneib}}{2009}]{JulloKneib2009}
{Jullo} E.,  {Kneib} J.-P.,  2009, \mn@doi [\mnras]
  {10.1111/j.1365-2966.2009.14654.x}, \href
  {http://adsabs.harvard.edu/abs/2009MNRAS.395.1319J} {395, 1319}

\bibitem[\protect\citeauthoryear{{Jullo}, {Kneib}, {Limousin},
  {El{\'{\i}}asd{\'o}ttir}, {Marshall}  \& {Verdugo}}{{Jullo}
  et~al.}{2007}]{Jullo2007}
{Jullo} E.,  {Kneib} J.-P.,  {Limousin} M.,  {El{\'{\i}}asd{\'o}ttir} {\'A}.,
  {Marshall} P.~J.,   {Verdugo} T.,  2007, \mn@doi [New Journal of Physics]
  {10.1088/1367-2630/9/12/447}, \href
  {http://adsabs.harvard.edu/abs/2007NJPh....9..447J} {9, 447}

\bibitem[\protect\citeauthoryear{{Kalirai}}{{Kalirai}}{2018}]{Kalirai2018}
{Kalirai} J.,  2018, \mn@doi [Contemporary Physics]
  {10.1080/00107514.2018.1467648}, \href
  {http://adsabs.harvard.edu/abs/2018ConPh..59..251K} {59, 251}

\bibitem[\protect\citeauthoryear{{Kawamata}, {Ishigaki}, {Shimasaku}, {Oguri}
  \& {Ouchi}}{{Kawamata} et~al.}{2015}]{Kawamata2015}
{Kawamata} R.,  {Ishigaki} M.,  {Shimasaku} K.,  {Oguri} M.,   {Ouchi} M.,
  2015, \mn@doi [\apj] {10.1088/0004-637X/804/2/103}, \href
  {http://adsabs.harvard.edu/abs/2015ApJ...804..103K} {804, 103}

\bibitem[\protect\citeauthoryear{{Kennicutt}}{{Kennicutt}}{1998}]{Kennicutt1998}
{Kennicutt} Jr. R.~C.,  1998, \mn@doi [\araa] {10.1146/annurev.astro.36.1.189},
  \href {http://adsabs.harvard.edu/abs/1998ARA%26A..36..189K} {36, 189}

\bibitem[\protect\citeauthoryear{{Kneib}}{{Kneib}}{1993}]{Kneib1993}
{Kneib} J.-P.,  1993, PhD thesis, Ph.~D.~thesis, Universit{\'e} Paul Sabatier,
  Toulouse, (1993)

\bibitem[\protect\citeauthoryear{{Kneib} \& {Natarajan}}{{Kneib} \&
  {Natarajan}}{2011}]{Kneib2011}
{Kneib} J.-P.,  {Natarajan} P.,  2011, \mn@doi [\aapr]
  {10.1007/s00159-011-0047-3}, \href
  {http://adsabs.harvard.edu/abs/2011A%26ARv..19...47K} {19, 47}

\bibitem[\protect\citeauthoryear{{Labb{\'e}} et~al.,}{{Labb{\'e}}
  et~al.}{2010}]{Labbe2010}
{Labb{\'e}} I.,  et~al., 2010, \mn@doi [\apjl] {10.1088/2041-8205/716/2/L103},
  \href {http://adsabs.harvard.edu/abs/2010ApJ...716L.103L} {716, L103}

\bibitem[\protect\citeauthoryear{{Laidler} et~al.,}{{Laidler}
  et~al.}{2007}]{Laidler2007}
{Laidler} V.~G.,  et~al., 2007, \mn@doi [\pasp] {10.1086/523898}, \href
  {http://adsabs.harvard.edu/abs/2007PASP..119.1325L} {119, 1325}

\bibitem[\protect\citeauthoryear{{Laporte} et~al.,}{{Laporte}
  et~al.}{2015}]{Laporte2015}
{Laporte} N.,  et~al., 2015, \mn@doi [\aap] {10.1051/0004-6361/201425040},
  \href {http://adsabs.harvard.edu/abs/2015A%26A...575A..92L} {575, A92}

\bibitem[\protect\citeauthoryear{{Laporte} et~al.,}{{Laporte}
  et~al.}{2016}]{Laporte2016}
{Laporte} N.,  et~al., 2016, \mn@doi [\apj] {10.3847/0004-637X/820/2/98}, \href
  {http://adsabs.harvard.edu/abs/2016ApJ...820...98L} {820, 98}

\bibitem[\protect\citeauthoryear{{Livermore}, {Finkelstein}  \&
  {Lotz}}{{Livermore} et~al.}{2017}]{Livermore2017}
{Livermore} R.~C.,  {Finkelstein} S.~L.,   {Lotz} J.~M.,  2017, \mn@doi [\apj]
  {10.3847/1538-4357/835/2/113}, \href
  {http://adsabs.harvard.edu/abs/2017ApJ...835..113L} {835, 113}

\bibitem[\protect\citeauthoryear{{Lotz} et~al.,}{{Lotz}
  et~al.}{2017}]{Lotz2017}
{Lotz} J.~M.,  et~al., 2017, \mn@doi [\apj] {10.3847/1538-4357/837/1/97}, \href
  {http://adsabs.harvard.edu/abs/2017ApJ...837...97L} {837, 97}

\bibitem[\protect\citeauthoryear{{Madau}}{{Madau}}{1995}]{Madau1995}
{Madau} P.,  1995, \mn@doi [\apj] {10.1086/175332}, \href
  {http://adsabs.harvard.edu/abs/1995ApJ...441...18M} {441, 18}

\bibitem[\protect\citeauthoryear{{McLeod}, {McLure}  \& {Dunlop}}{{McLeod}
  et~al.}{2016}]{Mcleod2016}
{McLeod} D.~J.,  {McLure} R.~J.,   {Dunlop} J.~S.,  2016, \mn@doi [\mnras]
  {10.1093/mnras/stw904}, \href
  {http://adsabs.harvard.edu/abs/2016MNRAS.459.3812M} {459, 3812}

\bibitem[\protect\citeauthoryear{{McLure} et~al.,}{{McLure}
  et~al.}{2011}]{McLure2011}
{McLure} R.~J.,  et~al., 2011, \mn@doi [\mnras]
  {10.1111/j.1365-2966.2011.19626.x}, \href
  {http://adsabs.harvard.edu/abs/2011MNRAS.418.2074M} {418, 2074}

\bibitem[\protect\citeauthoryear{{McLure} et~al.,}{{McLure}
  et~al.}{2013}]{McLure2013}
{McLure} R.~J.,  et~al., 2013, \mn@doi [\mnras] {10.1093/mnras/stt627}, \href
  {http://adsabs.harvard.edu/abs/2013MNRAS.432.2696M} {432, 2696}

\bibitem[\protect\citeauthoryear{{Merlin} et~al.,}{{Merlin}
  et~al.}{2015}]{Merlin2015}
{Merlin} E.,  et~al., 2015, \mn@doi [\aap] {10.1051/0004-6361/201526471}, \href
  {http://adsabs.harvard.edu/abs/2015A%26A...582A..15M} {582, A15}

\bibitem[\protect\citeauthoryear{{Merlin} et~al.,}{{Merlin}
  et~al.}{2016}]{Merlin2016}
{Merlin} E.,  et~al., 2016, \mn@doi [\aap] {10.1051/0004-6361/201527513}, \href
  {http://adsabs.harvard.edu/abs/2016A%26A...590A..30M} {590, A30}

\bibitem[\protect\citeauthoryear{{Merten} et~al.,}{{Merten}
  et~al.}{2011}]{Merten2011}
{Merten} J.,  et~al., 2011, \mn@doi [\mnras]
  {10.1111/j.1365-2966.2011.19266.x}, \href
  {http://adsabs.harvard.edu/abs/2011MNRAS.417..333M} {417, 333}

\bibitem[\protect\citeauthoryear{{Meurer}, {Heckman}  \& {Calzetti}}{{Meurer}
  et~al.}{1999}]{Meurer1999}
{Meurer} G.~R.,  {Heckman} T.~M.,   {Calzetti} D.,  1999, \mn@doi [\apj]
  {10.1086/307523}, \href {http://adsabs.harvard.edu/abs/1999ApJ...521...64M}
  {521, 64}

\bibitem[\protect\citeauthoryear{{Mobasher} et~al.,}{{Mobasher}
  et~al.}{2015}]{Mobasher2015}
{Mobasher} B.,  et~al., 2015, \mn@doi [\apj] {10.1088/0004-637X/808/1/101},
  \href {http://adsabs.harvard.edu/abs/2015ApJ...808..101M} {808, 101}

\bibitem[\protect\citeauthoryear{{Oesch} et~al.,}{{Oesch}
  et~al.}{2010}]{Oesch2010}
{Oesch} P.~A.,  et~al., 2010, \mn@doi [\apjl] {10.1088/2041-8205/709/1/L21},
  \href {http://adsabs.harvard.edu/abs/2010ApJ...709L..21O} {709, L21}

\bibitem[\protect\citeauthoryear{{Oesch} et~al.,}{{Oesch}
  et~al.}{2013}]{Oesch2013}
{Oesch} P.~A.,  et~al., 2013, \mn@doi [\apj] {10.1088/0004-637X/773/1/75},
  \href {http://adsabs.harvard.edu/abs/2013ApJ...773...75O} {773, 75}

\bibitem[\protect\citeauthoryear{{Oesch} et~al.,}{{Oesch}
  et~al.}{2014}]{Oesch2014}
{Oesch} P.~A.,  et~al., 2014, \mn@doi [\apj] {10.1088/0004-637X/786/2/108},
  \href {http://adsabs.harvard.edu/abs/2014ApJ...786..108O} {786, 108}

\bibitem[\protect\citeauthoryear{{Oesch}, {Bouwens}, {Illingworth}, {Labb{\'e}}
   \& {Stefanon}}{{Oesch} et~al.}{2018}]{Oesch2018}
{Oesch} P.~A.,  {Bouwens} R.~J.,  {Illingworth} G.~D.,  {Labb{\'e}} I.,
  {Stefanon} M.,  2018, \mn@doi [\apj] {10.3847/1538-4357/aab03f}, \href
  {http://adsabs.harvard.edu/abs/2018ApJ...855..105O} {855, 105}

\bibitem[\protect\citeauthoryear{{Oguri}}{{Oguri}}{2010}]{Oguri2010}
{Oguri} M.,  2010, \mn@doi [\pasj] {10.1093/pasj/62.4.1017}, \href
  {http://adsabs.harvard.edu/abs/2010PASJ...62.1017O} {62, 1017}

\bibitem[\protect\citeauthoryear{{Oke} \& {Gunn}}{{Oke} \&
  {Gunn}}{1983}]{Oke1983}
{Oke} J.~B.,  {Gunn} J.~E.,  1983, \mn@doi [\apj] {10.1086/160817}, \href
  {http://adsabs.harvard.edu/abs/1983ApJ...266..713O} {266, 713}

\bibitem[\protect\citeauthoryear{{Ono}, {Ouchi}, {Shimasaku}, {Dunlop},
  {Farrah}, {McLure}  \& {Okamura}}{{Ono} et~al.}{2010}]{Ono2010}
{Ono} Y.,  {Ouchi} M.,  {Shimasaku} K.,  {Dunlop} J.,  {Farrah} D.,  {McLure}
  R.,   {Okamura} S.,  2010, \mn@doi [\apj] {10.1088/0004-637X/724/2/1524},
  \href {http://adsabs.harvard.edu/abs/2010ApJ...724.1524O} {724, 1524}

\bibitem[\protect\citeauthoryear{{Ono} et~al.,}{{Ono} et~al.}{2013}]{Ono2013}
{Ono} Y.,  et~al., 2013, \mn@doi [\apj] {10.1088/0004-637X/777/2/155}, \href
  {http://adsabs.harvard.edu/abs/2013ApJ...777..155O} {777, 155}

\bibitem[\protect\citeauthoryear{{Peng}, {Ho}, {Impey}  \& {Rix}}{{Peng}
  et~al.}{2002}]{Peng2002}
{Peng} C.~Y.,  {Ho} L.~C.,  {Impey} C.~D.,   {Rix} H.-W.,  2002, \mn@doi [\aj]
  {10.1086/340952}, \href {http://adsabs.harvard.edu/abs/2002AJ....124..266P}
  {124, 266}

\bibitem[\protect\citeauthoryear{{Ravindranath} et~al.,}{{Ravindranath}
  et~al.}{2006}]{Ravindranath2006}
{Ravindranath} S.,  et~al., 2006, \mn@doi [\apj] {10.1086/507016}, \href
  {http://adsabs.harvard.edu/abs/2006ApJ...652..963R} {652, 963}

\bibitem[\protect\citeauthoryear{{Schaerer} \& {de Barros}}{{Schaerer} \& {de
  Barros}}{2009}]{Schaerer2009}
{Schaerer} D.,  {de Barros} S.,  2009, \mn@doi [\aap]
  {10.1051/0004-6361/200911781}, \href
  {http://adsabs.harvard.edu/abs/2009A%26A...502..423S} {502, 423}

\bibitem[\protect\citeauthoryear{{Schaerer} \& {de Barros}}{{Schaerer} \& {de
  Barros}}{2010}]{Schaerer2010}
{Schaerer} D.,  {de Barros} S.,  2010, \mn@doi [\aap]
  {10.1051/0004-6361/200913946}, \href
  {http://adsabs.harvard.edu/abs/2010A%26A...515A..73S} {515, A73}

\bibitem[\protect\citeauthoryear{{Schaerer} \& {de Barros}}{{Schaerer} \& {de
  Barros}}{2012}]{Schaerer2012}
{Schaerer} D.,  {de Barros} S.,  2012, in {Tuffs} R.~J.,  {Popescu} C.~C.,
  eds,  IAU Symposium Vol. 284, The Spectral Energy Distribution of Galaxies -
  SED 2011. pp 20--25 (\mn@eprint {arXiv} {1111.6373}),
  \mn@doi{10.1017/S1743921312008630}

\bibitem[\protect\citeauthoryear{{Schechter}}{{Schechter}}{1976}]{Schechter1976}
{Schechter} P.,  1976, \mn@doi [\apj] {10.1086/154079}, \href
  {http://adsabs.harvard.edu/abs/1976ApJ...203..297S} {203, 297}

\bibitem[\protect\citeauthoryear{{Schenker} et~al.,}{{Schenker}
  et~al.}{2013}]{Schenker2013}
{Schenker} M.~A.,  et~al., 2013, \mn@doi [\apj] {10.1088/0004-637X/768/2/196},
  \href {http://adsabs.harvard.edu/abs/2013ApJ...768..196S} {768, 196}

\bibitem[\protect\citeauthoryear{{Schmidt}}{{Schmidt}}{1968}]{Schmidt1968}
{Schmidt} M.,  1968, \mn@doi [\apj] {10.1086/149446}, \href
  {http://adsabs.harvard.edu/abs/1968ApJ...151..393S} {151, 393}

\bibitem[\protect\citeauthoryear{{Schmidt} et~al.,}{{Schmidt}
  et~al.}{2014}]{Schmidt2014}
{Schmidt} K.~B.,  et~al., 2014, \mn@doi [\apj] {10.1088/0004-637X/786/1/57},
  \href {http://adsabs.harvard.edu/abs/2014ApJ...786...57S} {786, 57}

\bibitem[\protect\citeauthoryear{{Sersic}}{{Sersic}}{1968}]{Sersic1968}
{Sersic} J.~L.,  1968, {Atlas de Galaxias Australes}

\bibitem[\protect\citeauthoryear{{Shipley} et~al.,}{{Shipley}
  et~al.}{2018}]{Shipley2018}
{Shipley} H.~V.,  et~al., 2018, \mn@doi [\apjs] {10.3847/1538-4365/aaacce},
  \href {http://adsabs.harvard.edu/abs/2018ApJS..235...14S} {235, 14}

\bibitem[\protect\citeauthoryear{{Song} et~al.,}{{Song}
  et~al.}{2016}]{Song2016}
{Song} M.,  et~al., 2016, \mn@doi [\apj] {10.3847/0004-637X/825/1/5}, \href
  {http://adsabs.harvard.edu/abs/2016ApJ...825....5S} {825, 5}

\bibitem[\protect\citeauthoryear{{Stark}, {Ellis}, {Bunker}, {Bundy},
  {Targett}, {Benson}  \& {Lacy}}{{Stark} et~al.}{2009}]{Stark2009}
{Stark} D.~P.,  {Ellis} R.~S.,  {Bunker} A.,  {Bundy} K.,  {Targett} T.,
  {Benson} A.,   {Lacy} M.,  2009, \mn@doi [\apj]
  {10.1088/0004-637X/697/2/1493}, \href
  {http://adsabs.harvard.edu/abs/2009ApJ...697.1493S} {697, 1493}

\bibitem[\protect\citeauthoryear{{Stark}, {Schenker}, {Ellis}, {Robertson},
  {McLure}  \& {Dunlop}}{{Stark} et~al.}{2013}]{Stark2013}
{Stark} D.~P.,  {Schenker} M.~A.,  {Ellis} R.,  {Robertson} B.,  {McLure} R.,
  {Dunlop} J.,  2013, \mn@doi [\apj] {10.1088/0004-637X/763/2/129}, \href
  {http://adsabs.harvard.edu/abs/2013ApJ...763..129S} {763, 129}

\bibitem[\protect\citeauthoryear{{Steidel}, {Giavalisco}, {Pettini},
  {Dickinson}  \& {Adelberger}}{{Steidel} et~al.}{1996}]{Steidel1996}
{Steidel} C.~C.,  {Giavalisco} M.,  {Pettini} M.,  {Dickinson} M.,
  {Adelberger} K.~L.,  1996, \mn@doi [\apjl] {10.1086/310029}, \href
  {http://adsabs.harvard.edu/abs/1996ApJ...462L..17S} {462, L17}

\bibitem[\protect\citeauthoryear{{Trenti} \& {Stiavelli}}{{Trenti} \&
  {Stiavelli}}{2008}]{Trenti2008}
{Trenti} M.,  {Stiavelli} M.,  2008, \mn@doi [\apj] {10.1086/528674}, \href
  {http://adsabs.harvard.edu/abs/2008ApJ...676..767T} {676, 767}

\bibitem[\protect\citeauthoryear{{Trenti} et~al.,}{{Trenti}
  et~al.}{2011}]{Trenti2011}
{Trenti} M.,  et~al., 2011, \mn@doi [\apjl] {10.1088/2041-8205/727/2/L39},
  \href {http://adsabs.harvard.edu/abs/2011ApJ...727L..39T} {727, L39}

\bibitem[\protect\citeauthoryear{{Weinmann}, {Neistein}  \& {Dekel}}{{Weinmann}
  et~al.}{2011}]{Weinmann2011}
{Weinmann} S.~M.,  {Neistein} E.,   {Dekel} A.,  2011, \mn@doi [\mnras]
  {10.1111/j.1365-2966.2011.19440.x}, \href
  {http://adsabs.harvard.edu/abs/2011MNRAS.417.2737W} {417, 2737}

\bibitem[\protect\citeauthoryear{{Zheng} et~al.,}{{Zheng}
  et~al.}{2012}]{Zheng2012}
{Zheng} W.,  et~al., 2012, \mn@doi [\nat] {10.1038/nature11446}, \href
  {http://adsabs.harvard.edu/abs/2012Natur.489..406Z} {489, 406}

\bibitem[\protect\citeauthoryear{{Zitrin} et~al.,}{{Zitrin}
  et~al.}{2013}]{Zitrin2013}
{Zitrin} A.,  et~al., 2013, \mn@doi [\apjl] {10.1088/2041-8205/762/2/L30},
  \href {http://adsabs.harvard.edu/abs/2013ApJ...762L..30Z} {762, L30}

\bibitem[\protect\citeauthoryear{{Zitrin} et~al.,}{{Zitrin}
  et~al.}{2014}]{Zitrin2014}
{Zitrin} A.,  et~al., 2014, \mn@doi [\apjl] {10.1088/2041-8205/793/1/L12},
  \href {http://adsabs.harvard.edu/abs/2014ApJ...793L..12Z} {793, L12}

\makeatother
\end{thebibliography}




\begin{appendix}

\section{Catalogs}
\label{sec:catalogs}

\onecolumn
\begin{table}
	 \centering
	 \renewcommand{\arraystretch}{1.3}
	 \setlength{\tabcolsep}{10pt}
	  \caption{Catalog of the $z\sim8$ sample. Column (1) lists the source IDs, column (2) and (3) lists their coordinates, column (4) gives the photometric redshift, column (5) lists the median magnification, column (6) is the absolute magnitude at 1500 angstrom and column (7) is the logarithm of the stellar mass. Both absolute magnitude and stellar mass values are corrected for magnification.}
	 \label{tab:uvlf_table}
	 \begin{tabular}{lcccccr} 
		 \hline
		 ID & RA  & Dec. & $z_{\mathrm{phot}}$ & Magnification & $M_{1500}$ & log $M_{*}$ \\ 
		 	& & & & & 													   & ($M_{\odot}$)\\
		 \hline
		 MACS0416 cluster & & & & & & \\
		 \hline
		 2095 & $64.0342$ & $-24.0699$ & $8.43_{-0.66}^{+0.27}$ & $3.12$ & $-18.78_{-0.32}^{+0.45}$ & $8.21_{-0.24}^{+0.34}$\\
		 257  & $64.0336$ & $-24.0900$ & $7.71_{-0.71}^{+0.16}$ & $2.21$ & $-18.60_{-0.36}^{+0.28}$ & $8.29_{-0.29}^{+0.45}$ \\
		 2365 & $64.0476$ & $-24.0667$ & $8.06_{-1.05}^{+0.18}$ & $11.92$ & $-15.55_{-0.86}^{+0.61}$ & $7.84_{-1.7}^{+0.79}$\\
		 2613 & $64.0603$ & $-24.0649$ & $8.06_{-0.36}^{+0.22}$ & $1.61$ & $-18.47_{-0.31}^{+0.42}$ & $8.36_{-0.35}^{+0.44}$\\
		 1288 & $64.0327$ & $-24.0790$ & $8.15_{-0.36}^{+0.49}$ & $9.67$ & $-17.26_{-0.26}^{+0.29}$ & $7.82_{-0.25}^{+0.28}$\\
		 \hline
		 Parallel field & & & & & & \\
		 \hline
		 154 & $64.1319$ & $-24.1311$ & $7.71_{-1.09}^{+0.10}$ & - & $-18.76_{-0.56}^{+0.22}$ & $8.27_{-0.23}^{+0.32}$\\
		 2674 & $64.1268$ & $-24.1000$ & $8.25_{-0.62}^{+0.12}$ & - & $-19.00_{-0.35}^{+0.27}$ & $8.51_{-0.31}^{+0.49}$\\
		 495 & $64.1213$ & $-24.1258$ & $7.88_{-0.59}^{+0.10}$ & - & $-19.82_{-0.32}^{+0.28}$ & $8.62_{-0.21}^{+0.27}$\\
		 1237 & $64.1196$ & $-24.1169$ & $7.71_{-0.66}^{+0.27}$ & - & $-18.78_{-0.66}^{+0.27}$ & $8.64_{-0.66}^{+0.27}$\\
		 1084 & 64.1545 & -24.1188 & $8.25_{-0.66}^{+0.27}$ & - & $-20.13_{-0.66}^{+0.27}$ & $8.91_{-0.66}^{+0.27}$\\
		 \hline
		 \hline
	 \end{tabular}
\end{table}

\begin{table}
	 \centering
	 \renewcommand{\arraystretch}{1.2}
	 \setlength{\tabcolsep}{10pt}
	  \caption{Catalog of the $z\sim9$ sample. Column (1) lists the source IDs, column (2) and (3) lists their coordinates, column (4) gives the photometric redshift, column (5) lists the median magnification, column (6) is the absolute magnitude at 1500 angstrom and column (7) is the logarithm of the stellar mass. Both absolute magnitude and stellar mass values are corrected for magnification.}
	 \label{tab:uvlf_table}
	 \begin{tabular}{lcccccr} 
		 \hline
		 ID & RA  & Dec. & $z_{\mathrm{phot}}$ & Magnification & $M_{1500}$ & log $M_{*}$ \\ 
		 	& & & & & 													   & ($M_{\odot}$)\\
		 \hline
		 MACS0416 cluster & & & & & & \\
		 \hline
		 126 & $64.0391$ & $-24.0931$ & $8.62_{-0.18}^{+0.22}$ & $1.72$ & $-20.55_{-0.36}^{+0.47}$ & $8.84_{-0.18}^{+0.20}$\\
		 1040  & $64.0479$ & $-24.0816$ & $8.72_{-0.23}^{+0.12}$ & $1.61$ & $-20.12_{-0.16}^{+0.27}$ & $8.71_{-0.21}^{+0.28}$ \\
		 1065 & $64.0480$ & $-24.0814$ & $8.72_{-0.22}^{+0.52}$ & $1.56$ & $-20.25_{-0.22}^{+0.18}$ & $9.44_{-0.37}^{+0.42}$\\
		 393 & $64.0375$ & $-24.0881$ & $8.62_{-0.55}^{+0.14}$ & $1.58$ & $-18.78_{-0.26}^{+0.27}$ & $8.28_{-0.29}^{+0.42}$\\
		 \hline
		 Parallel field & & & & & & \\
		 \hline
		 1524 & $64.1498$ & $-24.1133$ & $9.42_{-0.76}^{+0.23}$ & - & $-19.59_{-0.21}^{+0.23}$ & $8.63_{-0.39}^{+0.51}$\\
		 1957 & $64.1517$ & $-24.1084$ & $9.01_{-0.35}^{+0.52}$ & - & $-18.96_{-0.18}^{+0.20}$ & $8.93_{-0.48}^{+0.60}$\\
		 2660 & $64.1267$ & $-24.1003$ & $8.53_{-0.62}^{+0.17}$ & - & $-18.97_{-0.32}^{+0.27}$ & $8.50_{-0.31}^{+0.46}$\\
		 1646 & $64.1479$ & $-24.1117$ & $8.72_{-0.94}^{+0.09}$ & - & $-18.84_{-0.18}^{+0.22}$ & $8.31_{-0.43}^{+0.51}$\\
		 \hline
		 \hline
	 \end{tabular}
\end{table}

 
 \end{appendix}


\bsp	
\label{lastpage}
\end{document}